\documentclass{amsart}
\usepackage{amsmath}
\usepackage{amssymb, latexsym}
\usepackage{amsfonts}
\usepackage{amsthm,amscd}

\newcommand{\beqa}{\begin{eqnarray*}}
\newcommand{\eeqa}{\end{eqnarray*}}
\newcommand{\beqn}{\begin{eqnarray}}
\newcommand{\eeqn}{\end{eqnarray}}

\newcommand{\iy}{\infty}

\newcommand{\R}{\mathbb R}

\newcommand{\N}{\mathbb N}

\newcommand{\mcH}{\mathcal H}

\newcommand{\la}{\lambda}

\newcounter{cnt1}
\newcounter{cnt2}
\newcounter{cnt3}
\newcommand{\blr}{\begin{list}{$($\roman{cnt1}$)$}
 {\usecounter{cnt1} \setlength{\topsep}{0pt}
 \setlength{\itemsep}{0pt}}}
\newcommand{\bla}{\begin{list}{$($\alph{cnt2}$)$}
 {\usecounter{cnt2} \setlength{\topsep}{0pt}
 \setlength{\itemsep}{0pt}}}
\newcommand{\bln}{\begin{list}{$($\arabic{cnt3}$)$}
 {\usecounter{cnt3} \setlength{\topsep}{0pt}
 \setlength{\itemsep}{0pt}}}
\newcommand{\el}{\end{list}}
\newtheorem{thm}{Theorem}[section]
\newtheorem{lem}[thm]{Lemma}
\newtheorem{cor}[thm]{Corollary}
\newtheorem{ex}[thm]{Example}

\newtheorem{Def}[thm]{Definition}

\newtheorem{rem}[thm]{Remark}
\newcommand{\Rem}{\begin{rem} \rm}
\newcommand{\bdfn}{\begin{Def} \rm}
\newcommand{\edfn}{\end{Def}}

\newcommand{\ba}{\begin{array}}
\newcommand{\ea}{\end{array}}

\newtheorem*{acknowledgements}{Acknowledgements}
\date{}
\begin{document}
\title{\bf Feynman Operator Calculus: The Constructive Theory }
\author[Gill]{T. L. Gill}
\address[Tepper L. Gill]{ Department of Electrical Engineering, Howard University\\
Washington DC 20059 \\ USA, {\it E-mail~:} {\tt tgill@howard.edu}}
\author[Zachary]{W. W. Zachary}
\address[Woodford W. Zachary]{ Department of Electrical Engineering \\ Howard
University\\ Washington DC 20059 \\ USA, {\it E-mail~:} {\tt
wwzachary@earthlink.net}}
\date{}
\subjclass{Primary (46T12), Secondary (47DO6, 28C20)}
\keywords{Feynman Operator Calculus, Path Integral, Hille-Yosida Theorem, Trotter-Kato Theorem}
\maketitle
\begin{abstract}In this paper, we  survey progress on the constructive foundations for the Feynman operator calculus in which operators acting at different times commute.   We begin with the development of an operator version of the Henstock-Kurzweil integral, and a new Hilbert space which allows us to construct the elementary path integral in the manner originally envisioned by Feynman.   After developing our time-ordered operator theory we extend a few of the important theorems of semigroup theory, including the Hille-Yosida theorem.  This means that our approach is a natural extension of basic operator theory to the time-ordered setting.   As an application, we unify and extend the theory of time-dependent parabolic and hyperbolic evolution equations.  We then develop a general perturbation theory and use it to prove that all theories generated by semigroups are asympotic in the operator-valued sense of Poincar\'{e}.  This allows us to provide a general theory for the interaction representation of relativistic quantum theory.   We then show that our theory can be reformulated as a physically motivated sum over paths, and use this version to extend the Feynman path integral to include more general interactions. Our approach is independent of the space measures and the space of  continuous functions and thus makes it clear that the need for a measure is more of a natural expectation based on past experience than a death blow to the foundations for the Feynman path integral.  

In addition, we provide a simple and direct solution to the problem of disentanglement, a method used by Feynman to relate his theory to conventional analysis.  Using our disentanglement approach, we extend the Trotter-Kato theory to include the case where the intersection of the domains of two operators may be empty 
\end{abstract}
\tableofcontents
\section{\bf{Introduction}}
At the end of his book on path integrals with Hibbs \cite{FH},  Feynman states: ``Nevertheless, many of the results and formulations of path integrals can be reexpressed by another mathematical system, a kind of ordered operator calculus.  In this form many of the results of the preceding chapters find an analogous but more general representation ... involving noncommuting variables."  Feynman is referring to his 1951 paper \cite{F}, in which he introduces his time-ordered operator calculus.  

Feynman's basic idea for this calculus is to first lay out spacetime as one would a photographic film and imagine that the evolution of a physical system appears as a picture on this film, in which one sees more and more of the future as more and more of the film comes into view.   From this point of view, we see that time takes on a special role in that it orders the flow of the spacetime events that appear.  Feynman then suggested that we let time take on this role in the manipulation of noncommuting variables in quantum field theory.  He went on to show that this approach allowed him to write down and compute highly complicated expressions in a very fast, efficient and effective manner. 

The paper by Feynman was written after Dyson had shown that, using Feynman's time-ordering ideas, he could relate the Feynman and Schwinger-Tomonaga theories of QED.  Indeed, it was the work of Dyson \cite{DY} that first brought the power of time-ordering to the larger community.  (A very nice introduction to the path integral side of this story along with the way that Feynman used path integral ideas to create his computational methods can be found in the recent survey by Pierre Cartier and C\'{e}cile DeWitt-Morette \cite{CDM}.)  In response to the importance of time-ordering in relating the Feynman and Schwinger-Tomonaga theories, Segal \cite{S} suggested that the provision of mathematical meaning for time-ordering is one of the major problems in the foundations for QED. 

A number of investigators have attempted to solve this problem using formal methods. Miranker and Weiss \cite{MW} showed how the ordering process could be done (in a restricted manner) using the theory of Banach algebras. Nelson \cite{N} also used Banach algebras to develop a theory of ÔÔoperantsÕÕ as an alternate (formal) approach. Araki \cite{AK}, motivated by the interesting paper by Fujiwara \cite{FW} (see below), used yet another formal approach to the problem.  Other workers include Maslov \cite{M}, who used the idea of a T-product as an approach to formally order the operators and developed an operational theory.  An idea that is closest to Feynman was developed by Johnson and Lapidus in a series of papers.  Their work can be found in their recent book on the subject \cite{JL}. (The recent paper by DeFacio, Johnson and Lapidus  \cite{DJL} should also be consulted.)  

A major difficulty with each approach (other than that of Johnson and Lapidus  \cite{JL}) is the problem of disentanglement, the method  proposed by Feynman to relate his results to conventional analysis.  Johnson and Lapidus  develop a general ordering approach via a probability measure on the parameter space.  This approach is also constructive and offers a different perspective on possible frameworks  for disentanglement in the Feynman program. 

Cartier and DeWitt-Morette \cite{CDM} point out that, during the early years, few researchers in mathematics or physics investigated the path integral.  The same is true with respect to the number of researchers investigating the Feynman operator calculus.  To our knowledge, the paper by Fujiwara \cite{FW} is the only one by a physicist in the early literature.  Fujiwara  agrees with the ideas and results of Feynman with respect to the operator calculus, but is critical of what he calls notational ambiguities, and introduces a different approach. ``What is wanted, and what I have striven after, is a logical well-ordering of the main ideas concerning the operator calculus.  The present study is entirely free from ambiguities in Feynman's notation, which might obscure the fundamental concepts of the operator calculus and hamper the rigorous organization of the disentanglement technique."   Fujiwara's main idea was that the Feynman program should be implemented  using a sheet of unit operators at every point except at time t, where the true operator should be placed.  He called the exponential of such an operator an expansional to distinguish it from the normal exponential so that, loosely speaking, disentanglement becomes the process of going from an expansional to an exponential. 

\noindent
{\textbf{Purpose}}\paragraph{} 
The purpose of this review is to provide a survey of recent progress on the constructive implementation of Feynman's program for the operator calculus [F].  The theory is constructive in that we use a sheet of unit operators at every point except at time t, where the true operator is placed, so that operators acting at different times actually commute. Thus, our approach is the mathematical embodiment of Fujiwara's suggestion.  More importantly, the structure developed allows us to lift all of analysis and operator theory to the time-ordered setting (see also Gill and Zachary \cite{GZ}).  The work in \cite{GZ} was primarily written for researchers concerned with the theoretical and/or mathematical foundations for quantum field theory.  (A major objective was to prove two important conjectures of Dyson for quantum electrodynamics; namely that, in general, we can only expect the perturbation expansion to be asymptotic, and that the ultraviolet divergence is caused by a violation of the Heisenberg uncertainty relation at each point in time.)  

As suggested in \cite{GZ}, it is our contention that the correct mathematical formulation of the Feynman operator calculus should at least have the following desirable features: 
\begin{itemize}
\item 
It should provide a transparent generalization of current analytic methods without sacrificing the physically intuitive and computationally useful ideas of Feynman [F].
\item
It should provide a clear approach to some of the mathematical problems of relativistic quantum theory. 
\item 
It should explain the connection with path integrals.
\end{itemize}

Although we shall obtain a general theory for path integrals, our approach is distinct from the methods of functional integration, so this work does not discuss that subject directly.  However, since functional integration represents an important approach to path integrals,   a few brief remarks on this subject are in order.  The methods of functional differentiation and integration were major tools for the Schwinger  program in quantum electrodynamics, which was developed in parallel with the Feynman theory (see \cite{DY}). Thus, these methods were not developed for the study of path integrals.  However, historically, path integrals have been studied from the functional integration point of view, and many authors have sought to restrict consideration to the space of continuous functions or related function spaces in their definition of the path integral.  The best known example is undoubtedly the Wiener integral \cite{WSRM}.  However, from the time-ordering point of view, such a restriction is not natural nor desirable.   Thus, our approach does not encourage attempts at standard measure theoretic formulations with countably additive measures.  In fact, we take the view that integration theory, as contrasted with measure theory, is the appropriate vehicle to use for path integration.   Indeed, as shown in \cite{GZ1},  there is a one-to-one mapping between path integrals and semigroups of operators that have a kernel representation.   In this case, the semigroup operation generates the reproducing property of the kernel (see Section 6.2).

In their recent (2000) review of functional integration, Cartier and  DeWitt-Morette \cite{CDM} discuss three of the most fruitful and important applications of functional integration to the construction of path integrals.  In 1995, the  Journal of Mathematical Physics devoted a special issue to this subject, Vol. {\bf{36}} No. 5 (edited by Cartier and  DeWitt-Morette).  Thus, those with interest in the functional integration approach will find ample material in the above references.  It should  be noted that one remark in \cite{CDM} could be misleading.  They suggest that a function space is richer than/or less constrained than $\mathbb{R}^{\infty}$.  This is not completely correct in the sense that  $\mathbb{R}^{\infty}$ is a separable Fr\'{e}chet space and every separable Banach space can be isometrically embedded in it.  This is obvious if the space has a Schauder basis, for example, $\mathbf{C}[0,1]$, or 
$\bf{L}^2(\mathbb{R})$.   More important is the fact that the construction of path integrals over $[0,t]$ by time-slicing is done on $\mathbb{R}^{[0,t]}$, which clearly includes all function spaces.  They seem to imply that this construction is done on the limit of $\mathbb{R}^{m}$, as $m \to \infty$. (Other than this minor criticism, the review is excellent on many levels, in addition to the historical information that could only come from one with first-hand information on the evolution of the subject.)  

\subsection*{Objective}
This paper is written for those in the larger research community including applied and pure mathematics, biology, chemistry, engineering  and physics, who may not be aware of this approach to the theory of evolution equations and its relationship to path integrals.  With this in mind, and in order to make the paper self contained, we have provided a number of results and ideas that may not be normal fare.  We assume the standard mathematics background of an aggressive graduate student in engineering or science, and have provided proofs for all nonstandard material.  

\subsection*{\bf{Summary}}               
	In Section 2 we introduce the Henstock-Kurzweil integral (HK-integral).  This integral is easier to understand (and learn) compared to the Lebesgue or Bochner integrals, and provides useful variants of the same theorems that have made those integrals so important.  Furthermore, it arises from a simple (transparent) generalization of the Riemann integral that is taught in elementary calculus.  Its usefulness in the construction of Feynman path integrals was first shown by Henstock \cite{HS}, and has been further explored in the book by Muldowney \cite{MD}.
	
	In Section 2.1, we construct a new Hilbert space that contains the class of HK-integrable functions.   In order to show that this space has all the properties required and for our later use,  Section 3 is devoted to a substantial  review of operator theory, including some recently published results and some new results on operator extensions that have not appeared elsewhere.  As an application, we show that the Fourier transform and the convolution operator have bounded extensions to our new Hilbert space. In Section 3.1 we review the basics of semigroup theory and, in Section 3.2, we apply our results to provide a rigorous proof that the elementary Feynman integral exists on the new Hilbert space.	
	
	 In Section 4, we construct the continuous tensor product Hilbert space of von Neumann, which we use to construct our version of Feynman's film.   In Section 5 we define what we mean by time-ordering, prove our fundamental theorem on the existence of time-ordered integrals and extend basic semigroup theory to the time-ordered setting, providing, among other results,  a time-ordered version of the Hille-Yosida Theorem.  In Section 6 we construct time-ordered evolution operators and prove that they have all the expected properties.  As an application, we unify and extend the theory of time-dependent parabolic and hyperbolic evolution equations.   
	 
	 In Section 7 we define what is meant by the phrase "asymptotic in the sense of Poincar\'{e}" for operators.  We then develop a general perturbation theory and use it to prove that all theories generated by semigroups are asympotic in the operator-valued sense of Poincar\'{e}.  This result allows us to extend the Dyson expansion and provide a general theory for the interaction representation of relativistic quantum theory.  Finally,  we show that Feynman's approach to disentanglement can be implemented in a direct manner, which allows us to extend the Trotter-Kato theory.

In Section 8 we return to the Feynman path integral. First, we show that our theory can be reformulated as a physically motivated sum over paths.  We use this version to extend the Feynman path integral in a very general manner and  prove a generalized version of the well-known Feynman-Kac theorem. The theory is independent of the space of continuous functions and hence makes the question of the existence of measures more of a desire than a requirement.  (Whenever a measure exists, our theory can be easily restricted to the space of continuous paths.)
 In this section, we also consider a number of examples so that one can see how the time-ordering ideas appear in concrete cases.  We then use some results due to Maslov and Shishmarev (see Shishmarev \cite{SH}) on hypoelliptic pseudodifferential operators to construct a general class of path integrals generated by Hamitonians that are not perturbations of Laplacians. 

\section{\bf{Henstock-Kurzweil integral}}
The standard university analysis courses tend to produce a natural bias and unease concerning the use of finitely additive set functions as a basis for the general theory of integration (despite the efforts of Alexandroff \cite{AX}, Bochner \cite{BO}, Blackwell and Dubins \cite{BD}, Dunford and Schwartz \cite{DS}, de Finetti \cite{DFN} and Yosida and Hewitt \cite{YH}).    

Without denying an important place for countable additivity, Blackwell and Dubins, and Dubins and Prikry (see \cite{BD}, \cite{DUK}, and \cite{DU}) argue forcefully for the intrinsic advantages in using finite additivity in the basic axioms of probability theory.  (The penetrating analysis of the foundations of probability theory by de Finetti \cite{DFN} also supports this position.)   In a very interesting paper \cite{DU}, Dubins shows that the Wiener process has a number of "cousins", related processes all with the same finite dimensional distributions as the Wiener process.  For example, there is one cousin with polynomial paths and another with piecewise linear paths.   Since the Wiener measure is unique, these cousins must necessarily have finitely additive limiting distributions. 

In this section, we give an introduction to the class of HK-integrable functions on $\R$, while providing a generalization to the operator-valued case.   The integral is well-defined for operator-valued functions that may not be separably valued (where both the Bochner and Pettis integrals are undefined).  Loosely speaking, one uses a version of the Riemann integral with the interior points chosen first, while the size of the base rectangle around any interior point is determined by an arbitrary positive function defined at that point.  This integral was discovered independently by Henstock \cite{HS} and Kurzweil \cite{KW}.  In order to make the conceptual and technical simplicity of the HK-integral available to all, we prove all except the elementary or well-known results.  

The extension to $\R^n$ follows the same basic approach (see Henstock \cite{HS} and Pfeffer \cite{PF}).  In his latest book, \cite{PF1}, Pfeffer presents a  nice exposition of a relatively new invariant multidimensional process of recovering a function from its derivative, that also extends the HK-integral to Euclidean spaces.

Let ${\mathcal{H}}$ be a separable Hilbert space and let $L({\mathcal{H}})$ be the algebra of bounded linear operators on ${\mathcal{H}}$.   Let $[a,b] \subset \mathbb{R}$ and, for each $t \in [a,b]$, let $A(t) \in L({\mathcal{H}})$ be a given family of operators. 
\begin{Def} Let $\delta (t)$ map $[a,b] \to (0,\infty )$, and let ${\mathbf{P}} = \{ t_0 ,\tau _1 ,t_1 ,\tau _2 , \cdots ,\tau _n ,t_n \},$ where $a = t_0  \leqslant \tau _1  \leqslant t_1  \leqslant  \cdots  \leqslant \tau _n  \leqslant t_n  = b$.  We call ${\mathbf{P}}$ a HK-partition for $\delta$ (or HK-partition when $\delta$ is understood) provided that, for $0 \leqslant i \leqslant n - 1$, $
t_i ,t_{i + 1}  \in (\tau _{i + 1}  - \delta (\tau _{i + 1} ),\tau _{i + 1}  + \delta (\tau _{i + 1} )).$
\end{Def}
\begin{lem} (Cousin's Lemma) If  $\delta (t)$ is a mapping of $[a,b] \to (0,\infty )$, then a HK-partition exists for $\delta$. 
\end{lem}
\begin{lem} Let $
\delta _1 (t)$ and $\delta _2 (t)$ map $[a,b] \to (0,\infty )$, and suppose that $
\delta _1 (t) \leqslant \delta _2 (t).$ Then, if  ${\mathbf{P}}$ is a HK-partition for $
\delta _1 (t)$, it is also one for $\delta _2 (t)$.
\end{lem}
\begin{Def} The family $A(t) $, $t \in [a,b] $, is said to have a (uniform) HK-integral if there is an operator $Q[a,b] $ in $L({\mathcal{H}})$ such that, for each $\varepsilon  > 0$, there exists a function $\delta$ from $
[a,b] \to (0,\infty ) $ such that, whenever ${\mathbf{P}}$ is a HK-partition for $\delta$, then
\[
\left\| {\sum\nolimits_{i = 1}^n {\Delta t_i A(\tau _i ) - Q[a,b]} } \right\| < \varepsilon. 
\]
In this case, we write
\[
Q[a,b] = (HK)\,\int_a^b {A(t)dt}. 
\]
\end{Def}
\begin{thm}For $t \in [a,b] $, suppose the operators $A_1 (t) $ and $A_2 (t) $ both have HK-integrals, then so does their sum and 
\[
(HK)\int_a^b {[A_1 (t) + A_2 (t)]dt}  = (HK)\,\int_a^b {A_1 (t)dt}  + (HK)\,\int_a^b {A_2 (t)dt}. 
\]
\end{thm}
\begin{thm} Suppose $\{ A_k (t)\left| {} \right.\,k \in {\mathbb{N}}\}$ is a family of operator-valued functions in $L[{\mathcal{H}}]$, converging uniformly to $A(t) $ on $[a,b] $, and $
A_k (t) $ has a HK-integral $Q_k [a,b] $ for each $k$; then $A(t)$ has a HK-integral $
Q[a,b] $ and $Q_k [a,b] \to Q[a,b] $ uniformly.
\end{thm}
\begin{thm} Suppose $A(t)$ is Bochner integrable on $[a,b] $, then $A(t) $ has a HK-integral $
Q[a,b]$ and:
\beqn
(B)\,\,\int_a^b {A(t)dt}  = (HK)\,\int_a^b {A(t)dt}. 
\eeqn
\end{thm}
\begin{proof} First, let $E$ be a measurable subset of $[a,b] $ and assume that $A(t) = A\chi _E (t) $, where $\chi _E (t) $ is the characteristic function of $E$.  In this case, we show that $
Q[a,b] = A\la(E) $, where $\la(E)$ is the Lebesgue measure of $E$.  Let $\varepsilon  > 0$ be given and let $D$ be a compact subset of $E$.  Let $F \subset [a,b] $ be an open set containing $E$ such that $\la(F\backslash D) < {\varepsilon  \mathord{\left/
 {\vphantom {\varepsilon  {\left\| A \right\|}}} \right.
 \kern-\nulldelimiterspace} {\left\| A \right\|}}$; and define $\delta \,:[a,b] \to (0,\infty ) $ such that:
\[
\delta (t) = \left\{ {\begin{array}{*{20}c}
   {d(t,[a,b]\backslash F),\;t \in E}  \\
   {d(t,D),\;t \in [a,b]\backslash E},  \\
\end{array} } \right.
\]
where $d(x\,,\,y) = \left| {x - y} \right|$ is the distance function.  Let $
{\mathbf{P}} = \{ t_0 ,\tau _1 ,t_1 ,\tau _2 , \cdots ,\tau _n ,t_n \}$ be a HK-partition for $
\delta$ ; for $1 \leqslant i \leqslant n$, if $\tau _i  \in E$ then $(t_{i - 1} ,t_i ) \subset F$ so that
\beqn
\left\| {\sum\nolimits_{i = 1}^n {\Delta t_i A(\tau _i ) - A\la(F)} } \right\| = \left\| A \right\|\left[ {\la(F) - \sum\nolimits_{\tau _i  \in E} {\Delta t_i } } \right].
\eeqn
On the other hand, if  $\tau _i  \notin E$ then $(t_{i - 1} ,t_i ) \cap D = \emptyset$ (empty set), and it follows that:
\beqn
\left\| {\sum\nolimits_{i = 1}^n {\Delta t_i A(\tau _i ) - A\la(D)} } \right\| = \left\| A \right\|\left[ {\sum\nolimits_{\tau _i  \notin E} {\Delta t_i }  - \la(D)} \right].
\eeqn
Combining equations (1.2) and (1.3), we have that
\beqa
\begin{gathered}
  \left\| {\sum\nolimits_{i = 1}^n {\Delta t_i A(\tau _i ) - A\la(E)} } \right\| = \left\| A \right\|\left[ {\sum\nolimits_{\tau _i  \in E} {\Delta t_i }  - \la(E)} \right] \hfill \\
  {\text{                      }} \leqslant \left\| A \right\|\left[ {\la(F) - \la(E)} \right] \leqslant \left\| A \right\|\left[ {\la(F) - \la(D)} \right] \leqslant \left\| A \right\| \la(F\backslash D) < \varepsilon . \hfill \\ 
\end{gathered} 
\eeqa
Now suppose that $A(t) = \sum\nolimits_{k = 1}^\infty  {A_k \chi _{E_k } (t)} $ .  By definition, $A(t) $ is Bochner integrable if and only if  $\left\| {A(t)} \right\|$ is Lebesgue integrable with: 
\[
(B)\int_a^b {A(t)} dt = \sum\nolimits_{k = 1}^\infty  {A_k \la(E_k )}, 
\]
and (cf. Hille and Phillips \cite{HP})
\[
(L)\int_a^b {\left\| {A(t)} \right\|} dt = \sum\nolimits_{k = 1}^\infty  {\left\| {A_k } \right\|\la(E_k )}. 
\]
As the partial sums converge uniformly by Theorem 7, $Q[a,b] $ exists and 
\[
Q[a,b] \equiv (HK)\int_a^b {A(t)} dt = (B)\int_a^b {A(t)} dt.
\]
Now let $A(t) $ be an arbitrary Bochner integrable operator-valued function in $ L({\mathcal{H}})$, uniformly measurable and defined on $[a,b] $.  By definition, there exists a sequence $\{ A_k (t)\} $ of countably-valued operator-valued functions in $L({\mathcal{H}})
$ which converges to $A(t) $ in the uniform operator topology such that:
\[
\mathop {\lim }\limits_{k \to \infty } (L)\int_a^b {\left\| {A_k (t) - A(t)} \right\|dt}  = 0,
\]
and
\[
(B)\int_a^b {A(t)dt}  = \mathop {\lim }\limits_{k \to \infty } (B)\int_a^b {A_k (t)dt}. 
\]
Since the $A_k (t) $ are countably-valued, 
\[
(HK)\int_a^b {A_k (t)dt}  = (B)\int_a^b {A_k (t)dt}, 
\]
so
\[
(B)\int_a^b {A(t)dt}  = \mathop {\lim }\limits_{k \to \infty } (HK)\int_a^b {A_k (t)dt}. 
\]
We are done if we show that $Q[a,b] $ exists.  First, by the basic result of Henstock, every L-integral is a HK-integral, so that $f_k (t) = \left\| {A_k (t) - A(t)} \right\|$ has a HK-integral.  The above means that $\mathop {\lim }\limits_{k \to \infty } (HK)\int_a^b {f_k (t)dt}  = 0$.
 Let $\varepsilon  > 0$ and choose $m$ so large that
\[
\left\| {(B)\int_a^b {A(t)dt}  - (HK)\int_a^b {A_m (t)dt} } \right\| < {\varepsilon  \mathord{\left/
 {\vphantom {\varepsilon  4}} \right.
 \kern-\nulldelimiterspace} 4}
\]
and
\[
(HK)\int_a^b {f_k (t)dt}  < {\varepsilon  \mathord{\left/
 {\vphantom {\varepsilon  4}} \right.
 \kern-\nulldelimiterspace} 4}.
\]
Choose $\delta _1$ so that, if $\{ t_0 ,\tau _1 ,t_1 ,\tau _2 , \cdots ,\tau _n ,t_n \}$ is a HK-partition for $\delta _1$, then 
\[
\left\| {(HK)\int_a^b {A_m (t)dt}  - \sum\nolimits_{i = 1}^n {\Delta t_i A_m (\tau _i )} } \right\| < {\varepsilon  \mathord{\left/
 {\vphantom {\varepsilon  4}} \right.
 \kern-\nulldelimiterspace} 4}.
\]
Now choose $\delta _2 $ so that, whenever $\{ t_0 ,\tau _1 ,t_1 ,\tau _2 , \cdots ,\tau _n ,t_n \}$ is a HK-partition for $\delta _2$,
\[
\left\| {(HK)\int_a^b {f_m (t)dt}  - \sum\nolimits_{i = 1}^n {\Delta t_i f_m (\tau _i )} } \right\| < {\varepsilon  \mathord{\left/
 {\vphantom {\varepsilon  4}} \right.
 \kern-\nulldelimiterspace} 4}.
\]
Set $\delta  = \delta _1  \wedge \delta _2$ so that, by Lemma 3, if $\{ t_0 ,\tau _1 ,t_1 ,\tau _2 , \cdots ,\tau _n ,t_n \}$ is a HK-partition for $\delta$, it is also one for $\delta _1$ and $
\delta _2$, so that:
\[
\begin{gathered}
  \left\| {(B)\int_a^b {A(t)dt}  - \sum\limits_{i = 1}^n {\Delta t_i A(\tau _i )} } \right\| \leqslant \left\| {(B)\int_a^b {A(t)dt}  - (HK)\int_a^b {A_m (t)dt} } \right\| \hfill \\
  {\text{       }} + \left\| {(HK)\int_a^b {A_m (t)dt}  - \sum\nolimits_{i = 1}^n {\Delta t_i A_m (\tau _i )} } \right\| + \left| {(HK)\int_a^b {f_m (t)dt}  - \sum\nolimits_{i = 1}^n {\Delta t_i f_m (\tau _i )} } \right| \hfill \\
  {\text{                                                                                 }} + (HK)\int_a^b {f_m (t)dt}  < \varepsilon . \hfill \\ 
\end{gathered} 
\]
\end{proof} 
\section{\bf{Operator Theory}}\paragraph{}
\subsection{{Semigroups of Operators}}\paragraph{}
In this section, we introduce some basic results from the theory of semigroups of operators, which will be used throughout the remainder of the paper.   
\begin{Def} A family of bounded linear operators $\{ S(t), 0 \leqslant t< \infty \}$, defined on a Hilbert space ${\mcH}$, is a semigroup if
\begin{enumerate}
\item $S{\text{(}}t{\text{ + }}s{\text{)}} \varphi {\text{ = }}S{\text{(}}t{\text{)}}S{\text{(}}s{\text{)}} \varphi $ for $\varphi  \in {\mcH}$.  
\item The semigroup is said to be strongly continuous if $
\mathop {\lim }\limits_{\tau  \to 0} S{\text{(}}t + \tau {\text{)}} \varphi {\text{ = }}S{\text{(}}t{\text{)}} \varphi $ for  $ t > 0$.
\item It is said to be a $C_0$-semigroup if it is strongly continuous, $S{\text{(}}0{\text{) = }}I$, and $
\mathop {\lim }\limits_{t \to 0} S{\text{(}}t{\text{)}} \varphi {\text{ = }} \varphi $ for all $ \varphi  \in {\mathcal{H}}$.
\item $S{\text{(}}t{\text{)}}$ is a $C_0$-contraction semigroup if $\left\| {S{\text{(}}t{\text{)}}} \right\| \leqslant 1$. 
\item $S{\text{(}}t{\text{)}}$ is a $C_0$-unitary group if  $S{\text{(}}t{\text{)}}S{\text{(}}t{\text{)}}^ *   = S{\text{(}}t{\text{)}}^ *  S{\text{(}}t{\text{) = }}I$, and $\left\| {S{\text{(}}t{\text{)}}} \right\| = 1$.   
\end{enumerate}
\end{Def}
\begin{Def} A closed densely defined operator $A$ is said to be m-dissipative if  $
{\text{Re}}\left\langle {A \varphi, \varphi  } \right\rangle  \leqslant {\text{0}}$ for all
$ \varphi  \in {\text{D(}}A{\text{)}}$, and $
{\text{Ran(}}I - A{\text{) = }}{\mathcal{H}}$  (range of $
{\text{(}}I - A{\text{)}}$).
\end{Def}
\begin{thm}[see Goldstein \cite{GS} or Pazy \cite{PZ}] Let $S(t)$ be a $C_0$-semigroup of contraction operators on ${\mathcal{H}}$.  Then
\begin{enumerate}
\item 
$
A \varphi = \mathop{lim} \limits_{t \to 0} {{\left[ {S{\text{(}}t{\text{)}}\varphi  - \varphi } \right]} \mathord{\left/
 {\vphantom {{\left[ {S{\text{(}}t{\text{)}}\varphi  - \varphi } \right]} t}} \right.
 \kern-\nulldelimiterspace} t}
$ exists for 
$\varphi$ in a dense set, and 
$
R{\text{(}}\lambda {\text{, }}A{\text{) = (}}\lambda I - A{\text{)}}^{{\text{ - }}1} $ 
(the resolvent) exists for $\lambda  > 0$ and 
$
\left\| {R{\text{(}}\lambda {\text{, }}A{\text{)}}} \right\| \leqslant \lambda ^{ - 1}$.
\item The closed densely defined operator $A$ generates a $C_0$-semigroup of contractions on 
${\mathcal{H}}$, $
\left\{ {S(t), 0 \leqslant t < \infty } \right\}$, if and only if  $A$ is m-dissipative.
\item  If $A$ is closed and densely defined with both $A$ and $A^*$ dissipative, then $A$ is m-dissipative.
\item If $A_{\la}= \la AR(\la,A)$, then $A_{\la}$ generates an uniformly continuous contraction semigroup, $\varphi \in D(A) \Rightarrow AA_{\la} \varphi = A_{\la}A \varphi$ and, for $\varphi \in D(A), \ \mathop{lim} \limits_{t \to \iy} A_{\la} \varphi = \varphi$.  (The operator $A_{\la}$ is called the Yosida approximator for $A$.) 
\end{enumerate}
\end{thm}
\section{\textbf{Continuous Tensor Product Hilbert Space}}
In this section, we study the continuous tensor product Hilbert space of von Neumann.  This space contains a class of subspaces that we will use for our constructive representation of the Feynman operator calculus.  Although von Neumann \cite{VN2} did not develop his theory for our purpose, it will be clear that the theory is  natural for our approach.  Some might object that these spaces are too big (nonseparable) for physics.  However, we observe that past objections to nonseparable spaces do not apply to a theory which lays out all of spacetime from past to present to future as required by Feynman.  (It should be noted that the theory presented is formulated so that the basic space is separable at each instant of time, which is all that is required by quantum theory.)   Since von Neumann's approach is central to our theory and this subject is not discussed in the standard  analysis/functional analysis programs, we have provided a fairly complete exposition.   In addition, we have included a number of new and/or simplifed proofs from the literature. 
\par Let $I=[a,b]$, $0\leq a<b\leq \iy$ and, in order to avoid trivialities, we always assume that, in any product, all terms are nonzero. 
\begin{Def} If $\{z_\nu \}$ is a sequence of complex numbers indexed by $\nu \in I$,  
\begin{enumerate}
\item  We say that the product $\prod\nolimits_{\nu \in I} {z_\nu }$ is convergent with limit $z$ if, for every $\varepsilon  > 0$, there is a finite set $J(\varepsilon)$ such that, for all finite sets $J \subset I$, with   $J(\varepsilon)\subset J$, we have $\left| {\prod\nolimits_{\nu \in J} {z_\nu }  - z} \right| < \varepsilon$.
\item We say that the product $\prod\nolimits_{\nu  \in I} {z_\nu  }$ is quasi-convergent if $
\prod\nolimits_{\nu  \in I} {\left| {z_\nu  } \right|}$ is convergent.  (If the product is quasi-convergent, but not convergent, we assign it the value zero.)
\end{enumerate}
\end{Def}
Since $I$ is not countable, we note that 
\beqn
0 < \left| {\prod\nolimits_{\nu \in I} {z_\nu } } \right| < \infty \; {\text {if and only if}} \; \sum\nolimits_{\nu \in I} {\left| {1 - z_\nu } \right|}  < \infty.  
\eeqn
Thus, it follows that convergence implies that at most a countable number of the 
${z_\nu} \neq{1}$. 

Let ${\mathcal{H}}_\nu  = \mathcal{H}$ be a fixed Hilbert space for each $\nu \in I$ and, for $
\{ \phi _\nu \}  \in \prod _{\nu \in I}{\mathcal{H}}_\nu$, let $\Delta _I $
 be those sequences $\{\phi_\nu \}$ such that $\sum\nolimits_{\nu \in I} {\left| {\left\| {\varphi _\nu } \right\|_\nu  - 1} \right|}  < \infty $. Define a functional on $\Delta _I $
 by 
\beqn
\Phi (\psi ) = \sum\nolimits_{k = 1}^n {\prod _{\nu \in I} \left\langle {\varphi _{\nu}^k ,\psi _\nu } \right\rangle _\nu } ,
\eeqn
where $\psi  = \{ \psi _\nu \} ,\{ \varphi _{\nu}^k \}  \in \Delta _I $, for $1 \leq k \leq n$.  It is easy to see that this functional is linear in each component.  Denote $\Phi$ by
\[
\Phi  = \sum\nolimits_{k = 1}^n { \otimes _{\nu \in I} \varphi _{\nu}^k } .
\]
Define the algebraic tensor product, 
$
 \otimes _{{\nu} \in I} {\mathcal{H}}_{\nu}
 $,
 by
\beqn
 \;\;\;\; \otimes _{{\nu}  \in I} {\mathcal{H}}_{\nu}   = \left\{ {\sum\nolimits_{k = 1}^n { \otimes _{{\nu}  \in I} \varphi _{\nu} ^k } \left| {\{ \varphi _{\nu} ^k \}  \in \Delta _I ,1 \leq k \leq n,n \in \mathbb{N}} \right.} \right\}.
\eeqn
We define a linear functional on $\otimes _{{\nu} \in I} {\mathcal{H}}_{\nu}$ by 
\beqn
\;\;\;\;\;\; \left( {\sum\nolimits_{k = 1}^n { \otimes _{{\nu} \in I} \varphi _{\nu}^k } ,\sum\nolimits_{l = 1}^m { \otimes _{{\nu} \in I} \psi _{\nu}^l } } \right)_ \otimes   = \sum\nolimits_{l = 1}^m {\sum\nolimits_{k = 1}^n {\prod _{{\nu} \in I} \left\langle {\varphi _{\nu}^k ,\psi _{\nu}^l } \right\rangle _{\nu} } }. 
\eeqn
\begin{lem}
 The functional $\left( { \cdot {\text{,}} \cdot } \right)_ \otimes  $
 is a well-defined mapping on $\otimes _{{\nu} \in I}{\mathcal{H}}_{\nu} $.
\end{lem}
 \begin{proof} It suffices to show that, if $\Phi  = 0$,
 then $\left( {\Phi ,\Psi } \right)_ \otimes   = 0$.  If 
$\Phi  = \sum\nolimits_{k = 1}^n { \otimes _{{\nu} \in I} \varphi _{\nu}^k }$ and 
$\Psi  = \sum\nolimits_{l = 1}^m { \otimes _{{\nu} \in I} \psi _{\nu}^l }$ , then with
$\psi _l  = \{ \psi _{\nu}^l \}$,
\beqn
\begin{gathered}\left( {\Phi ,\Psi } \right)_ \otimes   = \sum\nolimits_{l = 1}^m {\sum\nolimits_{k = 1}^n {\prod _{{\nu} \in I} \left\langle {\varphi _{\nu}^k ,\psi _{\nu}^l } \right\rangle _{\nu} } }  = \sum\nolimits_{l = 1}^m {\Phi (\psi _l )}  = 0.
\end{gathered} 
\eeqn
\end{proof}
	Before continuing our discussion of the above functional, we first need to look a little more closely at the structure of the algebraic tensor product space, $
 \otimes _{{\nu} \in I} {\mathcal{H}}_{\nu} $. 
\begin{Def}
 Let $\phi  = \mathop  \otimes \limits_{{\nu} \in I} \phi _{\nu}$ and $
\psi  = \mathop  \otimes \limits_{{\nu} \in I} \psi _{\nu} $ be in $
 \otimes _{{\nu} \in I} {\mathcal{H}}_{\nu} $. 
\begin{enumerate}
\item We say that $\phi$ is strongly equivalent to $\psi$ ($\phi  \equiv^s \psi $), if and only if $
\sum\limits_{{\nu} \in I} {\left| {1 - \left\langle {\phi _{\nu} ,\psi _{\nu} } \right\rangle _{\nu} } \right|}  < \infty \;.$
\item We say that $\phi$ is weakly equivalent to $\psi $  ($\phi  \equiv ^w \psi $), if and only if $
\sum\limits_{\nu  \in I} {\left| {1 - \left| {\left\langle {\phi _\nu  ,\psi _\nu  } \right\rangle _\nu  } \right|\,} \right|}  < \infty. $
\end{enumerate}
\end{Def}
\begin{lem} We have $\phi  \equiv ^w \psi$ if and only if there exist $z_\nu  ,\;|\,z_\nu  \,| = 1$, such that $
\mathop  \otimes \limits_{\nu  \in I} z_\nu  \phi _\nu   \equiv ^s \mathop  \otimes \limits_{\nu  \in I} \psi _\nu  $.
\end{lem}
\begin{proof} Suppose that $
\mathop  \otimes \limits_{\nu  \in I} z_\nu  \phi _\nu   \equiv ^s \mathop  \otimes \limits_{\nu  \in I} \psi _\nu $.  Then we have:
\[
\sum\limits_{\nu  \in I} {\left| {1 - \left| {\left\langle {\phi _\nu  ,\psi _\nu  } \right\rangle _\nu  } \right|} \right|}  = \sum\limits_{\nu  \in I} {\left| {1 - \left| {\left\langle {z_\nu  \phi _\nu  ,\psi _\nu  } \right\rangle _\nu  } \right|} \right|}  \leqslant \sum\limits_{\nu  \in I} {\left| {1 - \left\langle {z_\nu  \phi _\nu  ,\psi _\nu  } \right\rangle _\nu  } \right|}  < \infty. 
\]
If $\phi  \equiv ^w \psi $, set \[
z_\nu   = {{\left| {\left\langle {\phi _\nu  ,\psi _\nu  } \right\rangle _\nu  } \right|} \mathord{\left/
 {\vphantom {{\left| {\left\langle {\phi _\nu  ,\psi _\nu  } \right\rangle _\nu  } \right|} {\left\langle {\phi _\nu  ,\psi _\nu  } \right\rangle _\nu  }}} \right.
 \kern-\nulldelimiterspace} {\left\langle {\phi _\nu  ,\psi _\nu  } \right\rangle _\nu  }}
\]
 for $
\left\langle {\phi _\nu  ,\psi _\nu  } \right\rangle _\nu   \ne 0$, and set $z_\nu   = 1$ otherwise. It follows that 
\[
\sum\limits_{\nu  \in I} {\left| {1 - \left\langle {z_\nu  \phi _\nu  ,\psi _\nu  } \right\rangle _\nu  } \right|}  = \sum\limits_{\nu  \in I} {\left| {1 - \left| {\left\langle {\phi _\nu  ,\psi _\nu  } \right\rangle _\nu  } \right|} \right|}  < \infty ,
\]
so that $\mathop  \otimes \limits_{\nu  \in I} z_\nu  \phi _\nu   \equiv ^s \mathop  \otimes \limits_{\nu  \in I} \psi _\nu $. 
\end{proof}  
\begin{thm}
 The relations defined above are equivalence relations on $\otimes _{{\nu} \in I}{\mathcal{H}}_{\nu} $, which decomposes $\otimes _{{\nu} \in I} {\mathcal{H}}_{\nu}$ into disjoint equivalence classes.
\end{thm}  
\begin{proof} 
Suppose $
\mathop  \otimes \limits_{{\nu} \in I} \phi _{\nu}  \equiv^s \mathop  \otimes \limits_{{\nu} \in I} \psi _{\nu}$.  First note that the relation is clearly reflexive.   Thus, we need only prove that it is symmetric and  transitive. To prove that the first relation is symmetric, observe that $
\left| {1 - \left\langle {\psi _\nu  ,\phi _\nu  } \right\rangle _\nu  } \right| = \left| {1 - \overline {\left\langle {\phi _\nu  ,\psi _\nu  } \right\rangle _\nu  } } \right| = \left| {\overline {\left[ {1 - \left\langle {\phi _\nu  ,\psi _\nu  } \right\rangle _\nu  } \right]} } \right| = \left| {1 - \left\langle {\phi _\nu  ,\psi _\nu  } \right\rangle _\nu  } \right|.$ To show that it is transitive, without loss, we can assume that $
\left\| {\psi _{\nu} } \right\|_{\nu}  = \left\| {\phi _{\nu} } \right\|_{\nu}  = 1$.  It is then easy to see that, if $
\mathop  \otimes \limits_{{\nu} \in I} \phi _{\nu}  \equiv^s \mathop  \otimes \limits_{{\nu} \in I} \psi _{\nu}$ and $
\mathop  \otimes \limits_{{\nu} \in I} \psi _{\nu}  \equiv^s \mathop  \otimes \limits_{{\nu} \in I} \rho _{\nu}$, then 
\[
1 - \left\langle {\phi _\nu  ,\rho _\nu  } \right\rangle _\nu   = \left[ {1 - \left\langle {\phi _\nu  ,\psi _\nu  } \right\rangle _\nu  } \right] + \left[ {1 - \left\langle {\psi _\nu  ,\rho _\nu  } \right\rangle _\nu  } \right] + \left\langle {\phi _\nu   - \psi _\nu  ,\psi _\nu   - \rho _\nu  } \right\rangle _\nu.  
\]
Now $
\left\langle {\phi _\nu   - \psi _\nu  ,\phi _\nu   - \psi _\nu  } \right\rangle _\nu   = 2\left[ {1 - \operatorname{Re} \left\langle {\phi _\nu  ,\psi _\nu  } \right\rangle _\nu  } \right] \leqslant 2\left| {1 - \left\langle {\phi _\nu  ,\psi _\nu  } \right\rangle _\nu  } \right|$, so that  $
\sum\limits_{\nu} {\left\| {\phi _{\nu}  - \psi _{\nu} } \right\|_{_{\nu} }^{_2 } }  < \infty 
$ and, by the same observation, $
\sum\limits_{\nu} {\left\| {\psi _{\nu}  - \rho _{\nu} } \right\|_{_{\nu} }^{_2 } }  < \infty.
$ It now follows from Schwartz's inequality that
$
\sum\limits_{\nu} {\left\| {\phi _{\nu}  - \psi _{\nu} } \right\|_{\nu} } \left\| {\psi _{\nu}  - \rho _{\nu} } \right\|_{\nu}  < \infty $.  Thus we have that 
\beqa
\begin{gathered}
  \sum _{{\nu} \in I} \left| {1 - \left\langle {\phi _{\nu} ,\rho _{\nu} } \right\rangle _{\nu} } \right| \leq \sum _{{\nu} \in I} \left| {1 - \left\langle {\phi _{\nu} ,\psi _{\nu} } \right\rangle _{\nu} } \right| + \sum _{{\nu} \in I} \left| {1 - \left\langle {\psi _{\nu} ,\rho _{\nu} } \right\rangle _{\nu} } \right| \hfill \\
  {\text{                            }} + \sum _{{\nu} \in I} \left\| {\phi _{\nu}  - \psi _{\nu} } \right\|_{\nu}  \left\| {\psi _{\nu}  - \rho _{\nu} } \right\|_{\nu}   < \infty . \hfill \\ 
\end{gathered} 
\eeqa
This proves the first case.  The proof of the second case (weak equivalence) now follows from the above lemma.
\end{proof}
\begin{thm} 
Let $ \otimes _{\nu  \in I} \varphi _\nu  $ be in $
 \otimes _{\nu  \in I} {\mathcal{H}}_\nu $.  Then: 
\begin{enumerate} 
\item
The product $\prod _{\nu  \in I} \left\| {\varphi _\nu  } \right\|_\nu $ converges if and only if $\prod _{\nu  \in I} \left\| {\varphi _\nu  } \right\|_\nu ^2$ converges. 
\item 
If  $\prod _{\nu  \in I} \left\| {\varphi _\nu  } \right\|_\nu$  and $
\prod _{\nu  \in I} \left\| {\psi _\nu  } \right\|_\nu$  converge, then $
\prod _{\nu  \in I} \left\langle {\varphi _\nu  ,\psi _\nu  } \right\rangle _\nu$  is quasi-convergent.
\item 
If  $\prod _{\nu  \in I} \left\langle {\varphi _\nu  ,\psi _\nu  } \right\rangle _\nu$  is quasi-convergent, then there exist complex numbers $\{ z_\nu  \} ,\;|z_\nu  {\kern 1pt} |\, = 1$, such that $\prod _{\nu  \in I} \left\langle {z_\nu  \varphi _\nu  ,\psi _\nu  } \right\rangle _\nu$   converges.
\end{enumerate}
\end{thm}
\begin{proof} For the first case, convergence of either term implies that $
\{ \left\| {\varphi _\nu  } \right\|_\nu  ,\;\nu  \in I\}$ has a finite upper bound $M > 0$.  Hence
\[
\left| {1 - \left\| {\varphi _\nu  } \right\|_\nu  } \right| \leqslant \left| {1 + \left\| {\varphi _\nu  } \right\|_\nu  } \right|\left| {1 - \left\| {\varphi _\nu  } \right\|_\nu  } \right| = \left| {1 - \left\| {\varphi _\nu  } \right\|_\nu ^2 } \right| \leqslant (1 + M)\left| {1 - \left\| {\varphi _\nu  } \right\|_\nu  } \right|.
\]

To prove (2), note that, if $J \subset I$ is any finite subset, 
\[
0 \leqslant \left| {\prod _{\nu  \in J} \left\langle {\varphi _\nu  ,\psi _\nu  } \right\rangle _\nu  } \right| \leqslant \prod _{\nu  \in J} \left\| {\varphi _\nu  } \right\|_\nu  \prod _{\nu  \in J} \left\| {\psi _\nu  } \right\|_\nu   < \infty. 
\]
Therefore, $
0 \leqslant \left| {\prod _{\nu  \in I} \left\langle {\varphi _\nu  ,\psi _\nu  } \right\rangle _\nu  } \right| < \infty$ so that $
\prod _{\nu  \in I} \left\langle {\varphi _\nu  ,\psi _\nu  } \right\rangle _\nu$  is quasi-convergent and, if $0 < \left| {\prod _{\nu  \in I} \left\langle {\varphi _\nu  ,\psi _\nu  } \right\rangle _\nu  } \right| < \infty $, it is convergent.  
The proof of  (3) now follows directly from the above lemma.
\end{proof}
\begin{Def} 
For $
\varphi  = \mathop  \otimes \limits_{{\nu} \in I} \varphi _{\nu}  \in  \otimes _{{\nu} \in I} {\mathcal{H}}_{\nu} $, we define ${\kern 1pt} {\mathcal{H}}_ \otimes ^2 (\varphi )$ to be the closed subspace generated by the span of all $
\psi  \equiv^s \varphi $ and we call it the strong partial tensor product space generated by the vector $\varphi $.
\end{Def}
\begin{thm}
For the partial tensor product spaces, we have the following:  
\begin{enumerate}
\item 
If $\psi _{\nu}  \ne \varphi _{\nu} $ occurs for at most a finite number of ${\nu}$, then $
\psi  = \mathop  \otimes \limits_{{\nu} \in I} \psi _{\nu}  \equiv^s \varphi  = \mathop  \otimes \limits_{{\nu} \in I} \varphi _{\nu} $.
\item 
The space ${\kern 1pt} {\mathcal{H}}_ \otimes ^2 (\varphi )$ is the closure of the linear span of $\psi  = \mathop  \otimes \limits_{{\nu} \in I} \psi _{\nu} $ such that $\psi _{\nu}  \ne \varphi _{\nu} $ occurs for at most a finite number of ${\nu}$.
\item 
If $\Phi  =  \otimes _{{\nu}  \in I} \varphi _{\nu}  $
 and $\Psi  =  \otimes _{{\nu}  \in I} \psi _{\nu}  $
 are in different equivalence classes of $
 \otimes _{{\nu}  \in I} {\mathcal{H}}_{\nu}  $, then $
\left( {\Phi ,\Psi } \right)_ \otimes   = \prod _{{\nu} \in I} \left\langle {\varphi _{\nu} ,\psi _{\nu} } \right\rangle _{\nu}  = 0$.
\item
$
{\kern 1pt} {\mathcal{H}}_ \otimes ^2 (\varphi )^w  = \mathop  \oplus \limits_{\psi  \equiv ^w \phi } {\kern 1pt} \left[ {\mathcal{H}}_ \otimes ^2 (\psi )^s  \right].
$
\end{enumerate}
\end{thm}  
\begin{proof} 
To prove (1), let $J$ be the finite set of ${\nu}$ for which $\psi _{\nu}  \ne \varphi _{\nu} $.  Then 
\beqa
\sum\limits_{{\nu} \in I} {\left| {1 - \left\langle {\varphi _{\nu} ,\psi _{\nu} } \right\rangle _{\nu} } \right|}  = \sum\limits_{{\nu} \in J} {\left| {1 - \left\langle {\varphi _{\nu} ,\psi _{\nu} } \right\rangle _{\nu} } \right|}  + \sum\limits_{{\nu} \in I\backslash J} {\left| {1 - \left\langle {\varphi _{\nu} ,\varphi _{\nu} } \right\rangle _{\nu} } \right|}  \leq c + \sum\limits_{{\nu} \in I} {\left| {1 - \left\| {\varphi _{\nu} } \right\|_{\nu}^2 } \right|}  < \infty, 
\eeqa
so that $
\mathop  \otimes \limits_{{\nu} \in I} \psi _{\nu}  \equiv \mathop  \otimes \limits_{{\nu} \in I} \varphi _{\nu} $.

To prove (2), let ${\kern 1pt} {\mathcal{H}}_ \otimes ^2 (\varphi )^\# $ be the closure of the linear span of all $\psi  =  \otimes _{{\nu} \in I} \psi _{\nu} $
 such that $\psi _{\nu}  \ne \varphi _{\nu} $ occurs for at most a finite number of ${\nu}$.  There is no loss in assuming that $\left\| {\varphi _{\nu} } \right\|_{\nu}  = 1$ for all ${\nu} \in I$.  It is clear from (1) that ${\kern 1pt} {\mathcal{H}}_ \otimes ^2 (\varphi )^\#  {\kern 1pt}  \subseteq {\kern 1pt} \,{\mathcal{H}}_ \otimes ^2 (\varphi )$.  Thus, we are done if we can show that $
{\mathcal{H}}_ \otimes ^2 (\varphi )^\#  \, \supseteq {\kern 1pt} \,{\mathcal{H}}_ \otimes ^2 (\varphi )$.  For any vector $\psi  =  \otimes _{{\nu} \in I} \psi _{\nu} $
 in ${\kern 1pt} {\mathcal{H}}_ \otimes ^2 (\varphi )$,  $\varphi  \equiv \psi$ so that $
\sum\limits_{{\nu} \in I} {\left| {1 - \left\langle {\varphi _{\nu} ,\psi _{\nu} } \right\rangle _{\nu} } \right|}  < \infty $.   If  $\left\| \psi  \right\|_ \otimes ^2  = 0$ then $\psi  \in {\kern 1pt} {\mathcal{H}}_ \otimes ^2 (\varphi )^\#  $, so we can assume that $\left\| \psi  \right\|_ \otimes ^2  \ne 0$.  This implies that $\left\| {\psi _{\nu} } \right\|_{\nu}  \ne 0$ for all ${\nu} \in I$ and 
$0 \ne \prod _{{\nu} \in I} ({1 \mathord{\left/ {\vphantom {1 {\left\| {\psi _{\nu} } \right\|_{\nu} }}} \right. \kern-\nulldelimiterspace} {\left\| {\psi _{\nu} } \right\|_{\nu} }}) < \infty $;
 hence, by scaling if necessary, we may also assume that $\left\| {\psi _{\nu} } \right\|_{\nu}  = 1$ for all ${\nu} \in I$.  Let $0 < \varepsilon  < 1$ be given, and choose $\delta $ so that $0 < \sqrt {2\delta e}  < \varepsilon$ ($e$ is the base for the natural log).  Since $
\sum\limits_{{\nu} \in I} {\left| {1 - \left\langle {\varphi _{\nu} ,\psi _{\nu} } \right\rangle _{\nu} } \right|}  < \infty $, there is a finite set of distinct values $J = \{ {\nu}_1 , \cdots ,{\nu}_n \} $ such that $\sum\limits_{{\nu} \in I - J} {\left| {1 - \left\langle {\varphi _{\nu} ,\psi _{\nu} } \right\rangle _{\nu} } \right|}  < \delta$ .  Since, for any finite set of numbers $z_1 , \cdots ,z_n $, it is easy to see that $
\left| {\prod _{k = 1}^n z_k  - 1} \right| = \left| {\prod _{k = 1}^n \left[ {1 + (z_k  - 1)} \right] - 1} \right| \leq \left( {\prod _{k = 1}^n e^{\left| {z_k  - 1} \right|}  - 1} \right)$, we have that
\beqa
\left| {\prod _{{\nu} \in I\backslash J} \left\langle {\varphi _{\nu} ,\psi _{\nu} } \right\rangle _{\nu}  - 1} \right| \leq (\exp \{ \sum\limits_{{\nu} \in I\backslash J} {\left| {\left\langle {\varphi _{\nu} ,\psi _{\nu} } \right\rangle _{\nu}  - 1} \right|} \}  - 1) \leq e^\delta   - 1 \leq e\delta. 
\eeqa
Now, define $
\phi _{\nu}  = \psi _{\nu} {\text{ if }}{\nu} \in J,{\text{ and }}\phi _{\nu}  = \varphi _{\nu} {\text{ if }}{\nu} \in {I \backslash J}$, and set $\phi _J  =  \otimes _{{\nu} \in I} \phi _{\nu}$ so that $
\phi _J  \in {\kern 1pt} {\mathcal{H}}_ \otimes ^2 (\varphi )^\#  
$ and
\[
\begin{gathered}
  \left\| {\psi  - \phi _J } \right\|_ \otimes ^2  = 2 - 2\operatorname{Re} \left[ {\prod _{\nu  \in J} \left\langle {\varphi _\nu  ,\psi _\nu  } \right\rangle _\nu   \cdot \prod _{\nu  \in I - J} \left\langle {\varphi _\nu  ,\psi _\nu  } \right\rangle _\nu  } \right] \hfill \\
   = 2 - 2\operatorname{Re} \left[ {\prod _{\nu  \in I} \left\| {\psi _\nu  } \right\|_\nu ^2  \cdot \prod _{\nu  \in I - J} \left\langle {\varphi _\nu  ,\psi _\nu  } \right\rangle _\nu  } \right] = 2\operatorname{Re} \left[ {1 - \prod _{\nu  \in I - J} \left\langle {\varphi _\nu  ,\psi _\nu  } \right\rangle _\nu  } \right] \leqslant 2e\delta  < \varepsilon ^2 . \hfill \\ 
\end{gathered} 
\]
Since $\varepsilon $ is arbitrary, $\psi$ is in the closure of $
{\kern 1pt} {\mathcal{H}}_ \otimes ^2 (\varphi )^\#  $, so $
{\kern 1pt} {\kern 1pt} {\mathcal{H}}_ \otimes ^2 (\varphi )^\#  \, = {\kern 1pt} \,{\kern 1pt} {\mathcal{H}}_ \otimes ^2 (\varphi )$.  

To prove (3), first note that, if  $\prod _{{\nu}  \in I} \left\| {\varphi _{\nu}  } \right\|_{\nu}  $
and  $\prod _{{\nu}  \in I} \left\| {\psi _{\nu}  } \right\|_{\nu} $ converge, then, for any finite subset $J \subset I$, 
$
0 \leq \left| {\prod _{{\nu}  \in J} \left\langle {\varphi _{\nu}  ,\psi _{\nu}  } \right\rangle _{\nu}  } \right| \leq \prod _{{\nu}  \in J} \left\| {\varphi _{\nu}  } \right\|_{\nu}  \prod _{{\nu}  \in J} \left\| {\psi _{\nu}  } \right\|_{\nu}   < \infty 
$. Therefore,
$
0 \leq \left| {\prod _{{\nu}  \in I} \left\langle {\varphi _{\nu}  ,\psi _{\nu}  } \right\rangle _{\nu}  } \right| = \left| {\left( {\Phi ,\Psi } \right)_ \otimes  } \right| < \infty 
$ so that 
$\prod _{{\nu}  \in I} \left\langle {\varphi _{\nu}  ,\psi _{\nu}  } \right\rangle _{\nu} $ is convergent or zero. If $
0 < \left| {\left( {\Phi ,\Psi } \right)_ \otimes  } \right| < \infty $, then $\sum\limits_{{\nu}  \in I} {\left| {1 - \left\langle {\phi _{\nu}  ,\psi _{\nu}  } \right\rangle _{\nu}  } \right|}  < \infty $ and, by definition, $\Phi $ and $\Psi $ are in the same equivalence class, so we must have $\left| {\left( {\Phi ,\Psi } \right)_ \otimes  } \right| = 0$. The proof of (4) follows from the definition of weakly equivalent spaces. 
\end{proof}
\begin{thm} 
$\left( {\Phi ,\Psi } \right)_ \otimes$  is a conjugate bilinear positive definite functional.
\end{thm}
\begin{proof} 
The first part is trivial.  To prove that it is positive definite, let $
\Phi  = \sum\nolimits_{k = 1}^n { \otimes _{{\nu}  \in I} \varphi _{\nu} ^k } $, and assume that the vectors $ \otimes _{{\nu}  \in I} \varphi _{\nu} ^k ,1 \leq k \leq n$, are in distinct equivalence classes.  This means that, with $\Phi _k  =  \otimes _{{\nu}  \in I} \varphi _{\nu} ^k $, we have
\[
\left( {\Phi ,\Phi } \right)_ \otimes   = \left( {\sum\nolimits_{k = 1}^n {\Phi _k } ,\sum\nolimits_{k = 1}^n {\Phi _k } } \right)_ \otimes   =      \sum\nolimits_{k = 1}^n {\sum\nolimits_{j = 1}^n {\left( {\Phi _k ,\Phi _j } \right)_ \otimes  } }  = \sum\nolimits_{k = 1}^n {\left( {\Phi _k ,\Phi _k } \right)_ \otimes  }. 
\]
Note that, from Theorem 4.8 (3), $
k \ne j$ implies $\left( {\Phi _k ,\Phi _j } \right)_ \otimes   = 0$.  Thus, it suffices to assume that $ \otimes _{{\nu}  \in I} \varphi _{\nu} ^k ,{\text{ }}1 \leq k \leq n$, are all in the same equivalence class.  In this case, we have that
\[
\left( {\Phi ,\Phi } \right)_ \otimes   = \sum\nolimits_{k = 1}^n {\sum\nolimits_{j = 1}^n {\prod _{{\nu}  \in I} \left\langle {\varphi _{\nu} ^k ,\varphi _{\nu} ^j } \right\rangle _{\nu} } }, 
\]
where each product is convergent.  It follows that the above will be positive definite if we can show that, for all possible finite sets $J = \{ {\nu} _1 , {\nu} _2  \cdots , {\nu} _m \} ,m \in \mathbb{N}$,
\[
\sum\nolimits_{k = 1}^n {\sum\nolimits_{j = 1}^n {\prod _{{\nu}  \in J} \left\langle {\varphi _{\nu} ^k ,\varphi _{\nu} ^j } \right\rangle _{\nu}  } }  \geq 0.
\]
This is equivalent to showing that the above defines a positive definite functional on $ \otimes _{{\nu} \in J} {\mathcal{H}}_{\nu} $, which follows from the standard result for finite tensor products of Hilbert spaces (see Reed and Simon, \cite{RS}).
\end{proof} 
\begin{Def}
We define ${\mathcal{H}}_ \otimes ^2  = \hat  \otimes _{{\nu} \in I} {\mathcal{H}}_{\nu} $ to be the completion of the linear space $ \otimes _{{\nu}  \in I} {\mathcal{H}}_{\nu}  $, relative to the inner product $\left( { \cdot {\text{,}} \cdot } \right)_ \otimes $.
\end{Def}

\subsection{{Orthonormal Basis for} $\mathcal{H}_ \otimes ^2 (\varphi )$}\paragraph{}

We now construct an orthonormal basis for each ${\kern 1pt} {\mathcal{H}}_ \otimes ^2 (\varphi )$.  Let ${\mathbf{N}}$ be the natural numbers, and let $
\{ e_n^{\nu} ,\;n \in \mathbb{N} = {\mathbf{N}} \cup \{ 0\} \}$ be a complete orthonormal basis for ${\kern 1pt} {\mathcal{H}}_{\nu} $.  Let $e_0^{\nu}$ be a fixed unit vector in $
{\kern 1pt} {\mathcal{H}}_{\nu}$ and set $E =  \otimes _{{\nu} \in I} e_0^{\nu} $.  Let $
{\mathbf{F}}$ be the set of all functions $f:\;I \to \mathbb{N}$ such that $f({\nu}) = 0$ for all but a finite number of ${\nu}$.  Let $F(f)$ be the image of $
f \in {\mathbf{F}}$ (e.g., $F(f) = \{ f({\nu}),\; {\nu} \in I\}$), and set $
E_{F(f)}  =  \otimes _{{\nu} \in I} e_{{\nu},f({\nu})} $, where $
f({\nu}) = 0$ implies that $ e_{{\nu},0}  = e_0^{\nu} $ and $
f({\nu}) = n$ implies $e_{{\nu}{\text{,}}n}  = e_n^{\nu} $.  
\begin{thm}
The set $\{ E_{F(f)} ,\;f \in {\mathbf{F}}\}$ is a complete orthonormal basis for $
{\mathcal{H}}_ \otimes ^2 (E)$.  
\end{thm}
\begin{proof} 
First, note that $E \in \{ E_{F(f)} ,\;f \in {\mathbf{F}}\}$ and each $E_{F(f)}$ is a unit vector.   Also, we have $E_{F(f)}  {\equiv}^s E$ and $
\left\langle {E_{F(f{\text{)}}} ,E_{F(g{\text{)}}} } \right\rangle  = \prod _{{\nu} \in I} \left\langle {e_{{\nu}{\text{,}}f{\text{(}}{\nu}{\text{)}}} ,e_{{\nu}{\text{,}}g{\text{(}}{\nu}{\text{)}}} } \right\rangle  = 0$ unless $f{\text{(}}{\nu}{\text{)}} = g{\text{(}}{\nu}{\text{)}}$ for all ${\nu}$.  Hence, the family $\{ E_{F(f)} ,\;f \in {\mathbf{F}}\}$ is an orthonormal set of vectors in ${\mathcal{H}}_ \otimes ^2 (E)$.  Let ${\mathcal{H}}_ \otimes ^2 (E)^\# $ be the completion of the linear span of this set of vectors.  Clearly $
{\mathcal{H}}_ \otimes ^2 (E)^\#   \subseteq {\mathcal{H}}_ \otimes ^2 (E)$, so we only need prove that every vector in ${\mathcal{H}}_ \otimes ^2 (E) \subset {\kern 1pt} {\mathcal{H}}_ \otimes ^2 (E )^\#  $.  By Theorem 4.8 (2), it suffices to prove that ${\mathcal{H}}_ \otimes ^2 (E)^\# $ contains the closure of the set of all $\varphi  =  \otimes _{{\nu} \in I} \varphi _{\nu}$ such that $\varphi _{\nu}  \ne e_0^{\nu}$ occurs for only a finite number of ${\nu}$.  Let $\varphi  =  \otimes _{{\nu} \in I} \varphi _{\nu}$ be any such vector, and let $J = \{ {\nu}_1 , \cdots, {\nu}_k \}$ be the finite set of distinct values of ${\nu}$ for which $
\varphi _{\nu}  \ne e_0^{\nu}$ occurs.  Since $\{ e_n^{\nu} ,\;n \in \mathbb{N}\}$ is a basis for $
{\kern 1pt} {\mathcal{H}}_{\nu} $, for each $
{\nu}_i $ there exist constants $a_{{\nu}_i ,n}$ such that $
\sum\nolimits_{n \in \mathbb{N}} {a_{{\nu}_i ,n} } e_n^{{\nu}_i }  = \varphi _{{\nu}_i }$ for $
1 \leq i \leq k$.  Let $\varepsilon  > 0$ be given.  Then, for each ${\nu}_i$ there exists a finite subset $\mathbb{N}_i  \subset \mathbb{N}$ such that $
\left\| {\varphi _{{\nu}_i }  - \sum\nolimits_{n \in \mathbb{N}_i } {a_{{\nu}_i ,n} } e_n^{{\nu}_i } } \right\|_ \otimes   < \tfrac{1}
{n}({\varepsilon  \mathord{\left/
 {\vphantom {\varepsilon  {\left\| \varphi  \right\|_ \otimes  )}}} \right.
 \kern-\nulldelimiterspace} {\left\| \varphi  \right\|_ \otimes  )}}$.  Let $
\vec {\mathbb{N}} = (\mathbb{N}_1 , \cdots \mathbb{N}_k )$ and set $
\varphi _{{\nu}_i }^{\mathbb{N}_i }  = \sum\nolimits_{n \in \mathbb{N}_i } {a_{{\nu}_i ,n} } e_n^{{\nu}_i }$ so that $
\varphi ^{\vec {\mathbb{N}}}  =  \mathop  \otimes \limits _{{\nu}_i  \in J} \varphi _{{\nu}_i }^{\mathbb{N}_i }  \otimes (\mathop  \otimes \limits _{{\nu} \in I\backslash J} e_0^{\nu} )$ and $
\varphi  =  \mathop  \otimes \limits_{{\nu}_i \in J}  \varphi _{{\nu}_i }  \otimes (  \mathop  \otimes \limits _{{\nu} \in I\backslash J} e_0^{\nu} )$.  It follows that: 
\[
\begin{gathered}
  \left\| \varphi  - \varphi ^{\vec {\mathbb{N}}} \right\|_ \otimes   = \left\| {\left[ {\mathop  \otimes \limits _{{\nu}_i  \in J} \varphi _{{\nu}_i }  -  \mathop  \otimes \limits _{{\nu}_i  \in J} \varphi _{{\nu}_i }^{\mathbb{N}_i } } \right] \otimes ( \mathop  \otimes \limits _{ {\nu} \in I\backslash J} e_0^\nu  )} \right\|_ \otimes   \hfill \\
  {\text{               }} = \left\| { \mathop  \otimes \limits _{{\nu}_i  \in J} \varphi _{{\nu}_i }  -  \mathop  \otimes \limits _{{\nu}_i  \in J} \varphi _{{\nu}_i }^{\mathbb{N}_i } } \right\|_ \otimes.   \hfill \\ 
\end{gathered} 
\]
We can rewrite this as:
\[
\begin{gathered}
  \left\| { \mathop  \otimes \limits _{{\nu}_i  \in J} \varphi _{{\nu}_i }  -  \mathop  \otimes \limits _{{\nu}_i  \in J} \varphi _{{\nu}_i }^{\mathbb{N}_i } } \right\|_ \otimes   = \left\| {\varphi _{{\nu}_1 } \, \otimes \varphi _{{\nu}_2 }  \cdots  \otimes \varphi _{{\nu}_k } } \right. - \varphi _{{\nu}_1 }^{\mathbb{N}_1 } \, \otimes \varphi _{{\nu}_2 }  \cdots  \otimes \varphi _{{\nu}_k }  \hfill \\
  {\text{                                    + }}\varphi _{{\nu}_1 }^{\mathbb{N}_1 } \, \otimes \varphi _{{\nu}_2 }  \cdots  \otimes \varphi _{{\nu}_k }  - \varphi _{{\nu}_1 }^{\mathbb{N}_1 } \, \otimes \varphi _{{\nu}_2 }^{\mathbb{N}_2 }  \cdots  \otimes \varphi _{{\nu}_k }  \hfill \\
  {\text{                                                           }} \vdots  \hfill \\
  {\text{ + }}\left. {\varphi _{{\nu}_1 }^{\mathbb{N}_1 } \, \otimes \varphi _{{\nu}_2 }^{\mathbb{N}_2 }  \cdots  \otimes \varphi _{{\nu}_{k - 1} }^{\mathbb{N}_{k - 1} }  \otimes \varphi _{{\nu}_k }  - \varphi _{{\nu}_1 }^{\mathbb{N}_1 } \, \otimes \varphi _{{\nu}_2 }^{\mathbb{N}_2 }  \cdots  \otimes \varphi _{{\nu}_k }^{\mathbb{N}_k } } \right\|_ \otimes  \hfill \\
   {\text{                                                          }}   \leq \sum\nolimits_{i = 1}^n {\left\| {\varphi _{{\nu}_i }  - \varphi _{{\nu}_i }^{\mathbb{N}_i } } \right\|} _ \otimes  \left\| \varphi  \right\|_ \otimes   \leq \varepsilon.  \hfill \\ 
\end{gathered} 
\]
Now, as the tensor product is multilinear and continuous in any finite number of variables, we have: 
\beqa
\begin{gathered}
  {\text{ }}\varphi ^{\vec {\mathbb{N}}}  =  \mathop  \otimes \limits _{{\nu}_i  \in J} \varphi _{{\nu}_i }^{\mathbb{N}_i }  \otimes ( \mathop  \otimes \limits _{{\nu} \in I\backslash J} e_0^{\nu} ) = \varphi _{{\nu}_1 }^{\mathbb{N}_1 } \, \otimes \varphi _{{\nu}_2 }^{\mathbb{N}_2 }  \cdots  \otimes \varphi _{{\nu}_k }^{\mathbb{N}_k }  \otimes ( \mathop  \otimes \limits _{{\nu} \in I\backslash J} e_0^{\nu} ) \hfill \\
   = \left[ {\sum\nolimits_{n_1  \in \mathbb{N}_1 } {a_{{\nu}_1 ,n_1 } e_{n_1 }^{{\nu}_1 } } } \right] \otimes \left[ {\sum\nolimits_{n_2  \in \mathbb{N}_2 } {a_{{\nu}_2 ,n_2 } e_{n_2 }^{{\nu}_2 } } } \right] \cdots  \otimes \left[ {\sum\nolimits_{n_k  \in \mathbb{N}_k } {a_{{\nu}_k ,n_k } e_{n_k }^{{\nu}_k } } } \right] \otimes ( \mathop  \otimes \limits _{ {\nu} \in I\backslash J} e_0^{\nu}  ) \hfill \\
   = \sum\nolimits_{\gamma _1  \in N_1  \cdots \gamma _n  \in N_n } {a_{{\nu}_1 ,n_1 } a_{{\nu}_2 ,n_2 }  \cdots a_{{\nu}_k ,n_k } } \left[ {e_{n_1 }^{{\nu}_1 }  \otimes e_{n_2 }^{{\nu}_2 }  \cdots  \otimes e_{n_k }^{{\nu}_k }  \otimes ( \mathop  \otimes \limits_{ {\nu} \in I\backslash J} e_0^{\nu}  )} \right]. \hfill \\ 
\end{gathered} 
\eeqa
It is now clear that, by definition of ${\mathbf{F}}$, for each fixed set of indices $n_1 ,\;n_2 , \cdots \;n_k$ there exists a function $f:\;I \to \mathbb{N}$ such that $
f{\text{(}}{\nu}_i {\text{)}} = n_i$ for ${\nu}_i  \in J$ and $f({\nu}) = 0$ for ${\nu} \in I\backslash J$.  Since each $\mathbb{N}_i$  is finite, $
\vec {\mathbb{N}} = (\mathbb{N}_1 , \cdots \mathbb{N}_k )$ is also finite, so that only a finite number of functions are needed.  It follows that $\varphi ^{\vec {\mathbb{N}}}$ is in 
${\mathcal{H}}_ \otimes ^2 (E)^\#  $, so that $
\varphi $ is a limit point and $
{\mathcal{H}}_ \otimes ^2 (E)^\#   = {\mathcal{H}}_ \otimes ^2 (E)$. 
\end{proof}

\subsection{{Tensor Product Semigroups}}\paragraph{}
Let $S_i (t),\;i = 1,2$, be $C_0$-contraction semigroups with generators $A_i $ defined on ${\mathcal{H}}$, so that $\left\| {S_i (t)} \right\|_{\mathcal{H}}  \leqslant 1$.  Define operators ${\mathbf{S}}_1 (t) = S_1 (t)\hat  \otimes {\kern 1pt} {\mathbf{I}}_2 $, ${\mathbf{S}}_2 (t) = {\mathbf{I}}_1 \hat  \otimes {\kern 1pt} S_2 (t)$ and ${\mathbf{S}}(t) = S_1 (t)\hat  \otimes {\kern 1pt} S_2 (t)$ on ${\mathcal{H}}\hat  \otimes {\kern 1pt}{\mathcal{H}}$.  The proof of the next result is easy. 
\begin{thm} The operators ${\mathbf{S}}(t),{\text{ }}{\mathbf{S}}_i (t){\text{ }},i = 1,2$, are $C_0$-contraction semigroups with generators $\mathcal{A} = \overline {A_1 \hat  \otimes {\kern 1pt} {\mathbf{I}}_2  + {\mathbf{I}}_1 \hat  \otimes A_2 } $, ${\mathcal{A}}_1  = A_1 \hat  \otimes {\kern 1pt} {\mathbf{I}}_2 $, ${\mathcal{A}}_2  = {\kern 1pt} {\mathbf{I}}_1 \hat  \otimes A_2$, and ${\mathbf{S}}(t)={{\mathbf{S}}_1 (t)}{{\mathbf{S}}_2 (t)}={{\mathbf{S}}_2 (t)}{{\mathbf{S}}_1 (t)}$.  
\end{thm} 
Let $S_i (t),\;1 \leqslant i \leqslant n$, be a family of $C_0$-contraction semigroups with generators $A_i $ defined on ${\mathcal{H}}$.  
\begin{cor} ${\mathbf{S}}(t) = \hat  \otimes _{i = 1}^n {\kern 1pt} S_i (t)$ is a $C_0$-contraction semigroup on $\hat  \otimes _{i = 1}^n {\kern 1pt} {\mathcal{H}}$ and the closure of $
A_1 \hat  \otimes {\kern 1pt} {\mathbf{I}}_2 \hat  \otimes  \cdots \hat  \otimes {\kern 1pt} {\mathbf{I}}_n  + {\mathbf{I}}_1 \hat  \otimes {\kern 1pt} A_2 \hat  \otimes  \cdots \hat  \otimes {\kern 1pt} {\mathbf{I}}_n  +  \cdots {\mathbf{I}}_1 \hat  \otimes {\kern 1pt} {\mathbf{I}}_2 \hat  \otimes  \cdots \hat  \otimes A_n$  is the generator ${\mathcal{A}}$ of ${\mathbf{S}}(t)$.
\end{cor}

\section{\textbf{Time-Ordered Operators}}
For the remainder of the paper, our index set $I=[a,b]$ is a subset of the reals, $\mathbf{R}$, and we replace
${\mathcal{H}}_ \otimes ^2 = \hat  \otimes _{\nu  \in I} \mathcal{H}_\nu$ by  
$\hat  \otimes _{t \in I} \mathcal{H}(t)$. Let $L({\mathcal{H}}_ \otimes ^2 )$ be the set of bounded operators on ${\mathcal{H}}_ \otimes ^2 $, and define $L({\mathcal{H}}(t)) \subset L({\mathcal{H}}_ \otimes ^2 )$ by: 
\beqn
\;\;\;\;\; L({\mathcal{H}}(t)) = \left\{ {{\text{  }}\mathcal{A}(t) = (\mathop {\hat  \otimes }\limits_{b \geqslant s > t} {\text{I}}_s)  \otimes A{\text{(}}t{\text{)}} \otimes (\mathop  \otimes \limits_{t > s \geqslant a} {\text{I}}_s ),\forall A{\text{(}}t{\text{)}} \in L({\mathcal{H}})} \right\},
\eeqn
where ${\text{I}}_{\text{s}}$ is the identity operator.  Let $L^\#  ({\mathcal{H}}_ \otimes ^2 )$ be the uniform closure of the algebra generated by $\{ L({\mathcal{H}}(t)),\;t \in I\}$.  If the family $\left\{ {A{\text{(}}t{\text{), }}t \in I} \right\}$ is in $L({\mathcal{H}})$, then the operators $\left\{ {\mathcal{A}(t),{\text{ }}t \in I} \right\} \in L^\#  ({\mathcal{H}}_ \otimes ^2 )$ commute when acting at different times: 
\[
\mathcal{A}(t)\mathcal{A}(\tau ) = \mathcal{A}(\tau )\mathcal{A}(t)
\; {\text{for}}\; t \ne \tau.
\]
Let ${\mathbf{P}}_\varphi $ denote the projection from ${\mathcal{H}}_ \otimes ^2$ onto ${\mathcal{H}}_ \otimes ^2 (\varphi )$.
\begin{thm} If ${\mathbf{T}} \in L^\#  [{\mathcal{H}}_ \otimes ^2 ]$, then ${\mathbf{P}}_\varphi  {\mathbf{T}} = {\mathbf{TP}}_\varphi$.
\end{thm}
\begin{proof} Since vectors of the form $\Phi  = \sum\nolimits_{i = 1}^L { \otimes _{s \in I} \varphi _s^i } $, with $\varphi _s^i  = \varphi _s $ for all but a finite number of $s$, are dense in ${\mathcal{H}}_ \otimes ^2 (\varphi )$; it suffices to show that $
{\mathbf{T}} \in L^\#  [{\mathcal{H}}_ \otimes ^2 ]$ implies ${\mathbf{T}}\Phi  \in {\mathcal{H}}_ \otimes ^2 (\varphi )$.  Now, ${\mathbf{T}} \in L^\#  [{\mathcal{H}}_ \otimes ^2 ]$ implies that there exists a sequence of operators ${\mathbf{T}}_n$ such that $\left\| {{\mathbf{T}} - {\mathbf{T}}_n } \right\|_ \otimes   \to 0$ as $n \to \infty$, where each ${\mathbf{T}}_n$ is of the form: ${\mathbf{T}}_n  = \sum\nolimits_{k = 1}^{N_n } {a_k^n T_k^n }$, with $a_k^n$ a scalar, $N_n  < \infty$, and each $T_k^n  = \hat  \otimes _{s \in J_k } T_{ks}^n \hat  \otimes _{s \in I\backslash J_k } I_s $ for some finite set of $s$-values $J_k $.  Hence,
\[
{\mathbf{T}}_n \Phi  = \sum\nolimits_{i = 1}^L {\sum\nolimits_{k = 1}^{N_n } {a_k^n  \otimes _{s \in J_k } T_{ks}^n \varphi _s^i  \otimes _{s \in I\backslash J_k } \varphi _s^i } }. 
\]
It is easy to see that, for each $i$, $ \otimes _{s \in J_k } T_{ks}^n \varphi _s^i  \otimes _{s \in I\backslash J_k } \varphi _s^i  \equiv  \otimes _{s \in I} \varphi _s $.  It follows that ${\mathbf{T}}_n \Phi  \in {\mathcal{H}}_ \otimes ^2 (\varphi )$ for each $n$, so that ${\mathbf{T}}_n  \in L[{\mathcal{H}}_ \otimes ^2 (\varphi )]$.  As $L[{\mathcal{H}}_ \otimes ^2 (\varphi )]$ is a norm closed algebra, ${\mathbf{T}} \in L[{\mathcal{H}}_ \otimes ^2 (\varphi )]$ and it follows that ${\mathbf{P}}_\varphi  {\mathbf{T}} = {\mathbf{TP}}_\varphi$. 
\end{proof}
\begin{Def} We call $L^\#  ({\mathcal{H}}_ \otimes ^2 )$ the time-ordered von Neumann algebra over ${\mathcal{H}}_ \otimes ^2$.
\end{Def}
The following theorem is due to von Neumann \cite{VN2}.
\begin{thm} The mapping ${\mathbf{T}}_\theta ^t \;{\mathbf{:}}\,\;L({\mathcal{H}}) \to L({\mathcal{H}}(t))$ is an isometric isomorphism of algebras. (We call ${\mathbf{T}}_\theta ^t$ the time-ordering morphism.)
\end{thm}
\subsection {{Exchange Operator}}
\begin{Def}An exchange operator ${\mathbf{E}}[t,t']$ is a linear map defined for pairs $t, t'$ such that:
\begin{enumerate} \item
${\mathbf{E}}[t,t'] \,:\;L[{\mathcal{H}}(t)] \to L[{\mathcal{H}}( t')]$, (isometric isomorphism), \item $
{\mathbf{E}}[s, t'] {\mathbf{E}}[t,s]= {\mathbf{E}} [t,t'] $,
 \item
$
{\mathbf{E}}[t,t'] {\mathbf{E}}[t',t] = I,
$
\item  for $s \ne t,\;t'$,
${\mathbf{E}}[t,t'] {\mathcal{A}}(s) = {\mathcal{A}}(s)$, 
 for all ${\mathcal{A}}{\text{(}}s) \in L[{\mathcal{H}}(s)].
$
\end{enumerate}
\end{Def}
The exchange operator acts to exchange the time positions of a pair of operators in a more complicated expression.
\begin{thm} 
(Existence) There exists an exchange operator for $
L^\#  [{\mathcal{H}}_ \otimes ^2 ]$.
\end{thm}
\begin{proof} Define a map $
C[t,t']\,:\;{\mathcal{H}}_ \otimes ^2  \to {\mathcal{H}}_ \otimes ^2 $ (comparison operator) by its action on elementary vectors:
$$
C[t,t'] \otimes _{s \in I} \phi _s  =  (\otimes _{a \leqslant s < t'} \phi _s)  \otimes \phi _{t}  \otimes ( \otimes _{t' < s < t} \phi _s ) \otimes \phi _{t'}  \otimes ( \otimes _{t < s \leqslant b} \phi _s ),
$$
for all $
\phi  =  \otimes _{s \in I} \phi _s \, \in {\mathcal{H}}_ \otimes ^2 $.  Clearly, $
C[t,t']\, $ extends to an isometric isomorphism of ${\mathcal{H}}_ \otimes ^2 $.  For $
{\mathbf{U}}\, \in L^\#  [{\mathcal{H}}_ \otimes ^2 ] $, we define $
{\mathbf{E}}[t,t']\,{\mathbf{U}} = C[t,t']\,{\mathbf{U}}C[t',t] $.  It is easy to check that ${\mathbf{E}}{\text{[}} \cdot {\text{,}} \cdot {\text{]}}$ satisfies all the requirements for an exchange operator. 
\end{proof}
\subsection{{The Film}}
\paragraph{}
In the world view suggested by Feynman, physical reality is laid out as a three-dimensional motion picture in which we become aware of the future as more and more of the film comes into view. (The way the world appears to us in our consciousness.)   

In order to motivate our approach, let $\left\{ {e^i \,\left| {\,i \in \mathbb{N}} \right.} \right\}$ be a complete orthonormal basis for $\mathcal{H}$ and, for each $t \in I$ and $i \in \mathbb{N}$, let $e_t^i  = e^i$ and set $E^i  = \mathop  \otimes \nolimits_{t \in I} e_t^i$.  
Now notice that the Hilbert space ${\widehat{\mathcal{H}}}$ generated by the family of vectors $\{ E^i ,{\text{ }}i \in \mathbb{N}\} $  is isometrically isomorphic to ${\mathcal{H}}$.  
For later use, it should be noted that any vector in ${\mathcal{H}}$ of the form $\varphi  = \sum\nolimits_{k = 1}^\infty  {a_k e^k }$  has the corresponding representation in ${\widehat{\mathcal{H}}}$ as $\hat {\varphi}  = \sum\nolimits_{k = 1}^\infty  {a_k E^k }$.   The problem with using ${\widehat{\mathcal{H}}}$ to define our operator calculus is that this space is not invariant for any reasonable class of operators. 
We now construct a particular structure, which is our mathematical version of this film.  
\begin{Def} A film, ${\mathcal{F}\mathcal{D}}_ \otimes ^2 $, is the smallest subspace  containing ${\widehat{\mathcal{H}}}$ which is invariant for $L^\#  [{\mathcal{H}}_ \otimes ^2 ] $.  We call ${\mathcal{F}\mathcal{D}}_ \otimes ^2$ the  Feynman-Dyson space (FD-space) over $\mathcal{H}$.
\end{Def}
In order to construct our space, let ${\mathcal{F}\mathcal{D}}_2^i  = {\mathcal{H}}_ \otimes ^2 (E^i )$ be the strong partial tensor product space generated by the vector $E^{\text{i}}$.  It is clear  that ${\mathcal{F}\mathcal{D}}_2^i$ is the smallest space in  ${\mathcal{H}}_ \otimes ^2 $ which contains the vector $E^{\text{i}}$.  We now set $
{\mathcal{F}\mathcal{D}}_ \otimes ^2  = \mathop  \oplus \limits_{i = 1}^\infty {\mathcal{F}\mathcal{D}}_2^i $.  It is clear that the space $
{\mathcal{F}\mathcal{D}}_ \otimes ^2$ is a nonseparable Hilbert (space) bundle over $
I = [a,b] $.  However, by construction, it is not hard to see that the fiber at each time-slice is isomorphic to ${\mathcal{H}}$ almost everywhere.  

In order to facilitate the proofs in the next section, we need an explicit basis for each ${\mathcal{F}\mathcal{D}}_2^i $.   As in Section 4.1, let ${\mathbf{F}}$ be the set of all functions $f{\text{(}}\, \cdot \,{\text{):}}\;I \to   \mathbb{N} \cup \{ 0\}$ such that $
f{\text{(}}t{\text{)}}$ is zero for all but a finite number of $t$, and let $F(f{\text{)}}$ denote the image of the function $f( \cdot \, )$.  Set $ E_{F(f{\text{)}}}^i  =  \otimes _{t \in I} e_{t{\text{,}}f{\text{(t)}}}^i $ with $
e_{t{\text{,}}0}^i  = e_{}^i$ , and $f{\text{(}}t{\text{) = }}k $ implies 
$e_{t{\text{,}}k}^i  = e_{}^k $.    
\begin{lem} The set $\{ E_{F(f{\text{)}}}^i \left| {F(f{\text{)}} \in {\mathbf{F}}} \right.\} $ is a (c.o.b) for each ${\mathcal{F}\mathcal{D}}_2^i$.
 \end{lem}
If $\Phi ^{i} = \sum\nolimits_{F(f) \in {\mathbf{F}}} 
{a_{F(f)}^{i}} E_{F(f)}^i , \;\Psi ^{i}  
= \sum\nolimits_{F(f) \in {\mathbf{F}}} {b_{F(f)}^{i}} E_{F(f)}^i   \in {\mathcal{FD}_2^i} 
$, set $a_{F(f)}^{i}  = \left\langle {\Phi ^{\text{i}} ,E_{F(f{\text{)}}}^i }\right\rangle$ and $b_{F(f{\text{)}}}^{\text{i}}  = \left\langle {\Psi ^{\text{i}} ,E_{F(f{\text{)}}}^i } \right\rangle $, so that 
\[
\left\langle {\Phi ^{\text{i}} ,\Psi ^{\text{i}} } \right\rangle  = \sum\limits_{F(f{\text{),}}F(g{\text{)}} \in {\mathbf{F}}} {a_{F(f{\text{)}}}^{\text{i}} \bar b_{F(g{\text{)}}}^{\text{i}} \left\langle {E_{F(f{\text{)}}}^i ,E_{F(g{\text{)}}}^i } \right\rangle }, \; {\text{and}}\;
\left\langle {\Phi ^{\text{i}} ,\Psi ^{\text{i}} } \right\rangle  = \sum\limits_{F(f{\text{)}} \in {\mathbf{F}}} {a_{f{\text{(}}t{\text{)}}}^{\text{i}} \bar b_{f{\text{(}}t{\text{)}}}^{\text{i}} }. 
\]
 (Note that $
\left\langle {E_{F(f{\text{)}}}^i ,E_{F(g{\text{)}}}^i } \right\rangle  = \prod _{t \in I} \left\langle {e_{t{\text{,}}f{\text{(}}t{\text{)}}}^{\text{i}} ,e_{t{\text{,}}g{\text{(}}t{\text{))}}}^{\text{i}} } \right\rangle  = 0
$ unless $f(t) = g(t)$ for all  $t \in I$.)

The following notation will be used at various points of this section so we record the meanings here for reference. (The $t$ value referred to is in our fixed interval $I$.)
\begin{enumerate}
 \item (e.o.v): "except for at most one $t$ value"; 
\item (e.f.n.v): "except for an at most finite number of $t$ values"; and
\item  (a.s.c): "almost surely and the exceptional set is at most countable".
\end{enumerate}

\subsection{{Time-Ordered Integrals and Generation Theorems}}
\paragraph{} In this section, we assume that $I = [a,b] \subseteq [0,\infty ) $ and, for each $t \in I$, $ A(t)$ generates a $C_0$-semigroup on ${\mathcal{H}}$. 

To partially see the advantage of developing our theory on ${\mathcal{F}\mathcal{D}}_ \otimes ^2 $, suppose that $A(t)$ generates a $C_0$-semigroup for $t \in I$ and define $ {\mathbf{S}}_t (\tau ) $ by:  
\beqn
{\mathbf{S}}_t (\tau ) = \hat  \otimes _{s \in [a,t)} {\text{I}}_s  \otimes \left( {\exp \{ \tau A(t)\} } \right) \otimes \left( { \otimes _{s \in (t,b]} {\text{I}}_s } \right).
\eeqn
We briefly investigate the relationship between $S_t (\tau ) = \exp \{ \tau A(t)\} $ and $
{\mathbf{S}}_t (\tau ) = \exp \{ \tau {\mathcal{A}}(t)\} $. By Theorems 3.11 and 4.12, we know that ${\mathbf{S}}_t (\tau ) $ is a $C_0$-semigroup for $t \in I$ if and only if $S_t (\tau )$ is one also. 	For additional insight, we need a dense core for the family $\left\{ {\mathcal{A}(t)\left| t \right.} \right.\left. { \in I} \right\}$, so let $
{\bar{D}} = \mathop  \otimes \limits_{t \in I} {D(A(t))}$ and set $
{\text{D}}_0  = {\bar{D}} \cap {\mathcal{F}\mathcal{D}}_ \otimes ^2$.  Since $
{\bar{D}}$ is dense in ${\mathcal{H}}_ \otimes ^2 $, it follows that ${\text{D}}_0 $ is dense in ${\mathcal{F}\mathcal{D}}_ \otimes ^2 $.   Using our basis, if $\Phi ,\Psi  \in {\text{D}}_0 $, 
$
\Phi  = \sum\nolimits_{\text{i}} {\sum\nolimits_{F{\text{(}}f{\text{)}}} {a_{F{\text{(}}f{\text{)}}}^{\text{i}} E_{F{\text{(f)}}}^{\text{i}} } } ,
\Psi  = \sum\nolimits_{\text{i}} {\sum\nolimits_{F{\text{(}}g{\text{)}}} {b_{F{\text{(}}g{\text{)}}}^{\text{i}} E_{F{\text{(}}g{\text{)}}}^{\text{i}} } } 
$; then, as $\exp \{ \tau {\mathcal{A}}(t)\}$ is invariant on ${\mathcal{F}\mathcal{D}}_2^i$, we have 
\[
\left\langle {\exp \{ \tau {\mathcal{A}}(t)\} \Phi ,\Psi } \right\rangle  = \sum\nolimits_{\text{i}} {\sum\nolimits_{F{\text{(}}f{\text{)}}} {\sum\nolimits_{F{\text{(}}g{\text{)}}} {a_{F{\text{(}}f{\text{)}}}^{\text{i}} \bar b_{F{\text{(}}g{\text{)}}}^{\text{i}} \left\langle {\exp \{ \tau {\mathcal{A}}(t)\} E_{F{\text{(}}f{\text{)}}}^{\text{i}} ,E_{F{\text{(}}g{\text{)}}}^{\text{i}} } \right\rangle } } }, 
\]
and
\[
\begin{gathered}
  \left\langle {\exp \{ \tau {\mathcal{A}}(t)\} E_{F{\text{(}}f{\text{)}}}^{\text{i}} ,E_{F{\text{(}}g{\text{)}}}^{\text{i}} } \right\rangle  = \prod\limits_{s \ne t} {\left\langle {e_{s{\text{,}}f{\text{(}}s{\text{)}}}^{\text{i}} ,e_{s{\text{,}}g{\text{(}}s{\text{)}}}^{\text{i}} } \right\rangle } \left\langle {\exp \{ \tau A(t)\} e_{t{\text{,}}f{\text{(}}t{\text{)}}}^{\text{i}} ,e_{t{\text{,}}g{\text{(}}t{\text{)}}}^{\text{i}} } \right\rangle  \hfill \\
  {\text{                                       }} = \left\langle {\exp \{ \tau A(t)\} e_{t{\text{,}}f{\text{(}}t{\text{)}}}^{\text{i}} ,e_{t{\text{,}}f{\text{(}}t{\text{)}}}^{\text{i}} } \right\rangle {\text{ (e}}{\text{.o}}{\text{.v),}} \hfill \\
  {\text{                                       }} = \left\langle {\exp \{ \tau A(t)\} e_{}^{\text{i}} ,e_{}^{\text{i}} } \right\rangle {\text{ (e}}{\text{.f}}{\text{.n}}{\text{.v}}{\text{.) implies }} \hfill \\ 
\end{gathered} 
\]
\beqa
 \left\langle {\exp \{ \tau {\mathcal{A}}(t)\} \Phi ,\Psi } \right\rangle  = \sum\nolimits_{\text{i}} {\sum\nolimits_{F(f)} {a_{F(f)}^{\text{i}} \bar b_{F(f)}^{\text{i}} \left\langle {\exp \{ \tau A(t)\} e_{}^{\text{i}} ,e_{}^{\text{i}} } \right\rangle } } (a.s).  
\eeqa
Thus, by working on ${\mathcal{F}\mathcal{D}}_ \otimes ^2 $, we obtain a simple direct relationship between the conventional and time-ordered version of a semigroup.  This suggests that a parallel theory of semigroups of operators on ${\mathcal{F}\mathcal{D}}_ \otimes ^2 $ might make it possible for physical theories to be formulated in the intuitive and conceptually simpler time-ordered framework, offering substantial gain compared to the conventional mathematical structure.  Note that this approach would also obviate the need for the problematic process of disentanglement suggested by Feynman in order to relate the operator calculus to conventional mathematics.
Let ${\mathcal{A}}_z {\text{(}}t{\text{) = }}z {\mathcal{A}}{\text{(}}t{\text{)}}{\mathbf{R}}(z, {\mathcal{A}}{\text{(}}t{\text{)}})
$, where ${\mathbf{R}}(z,{\mathcal{A}}(t)) $, is the resolvent of 
${\mathcal{A}}(t)$. 

By Theorem 3.11(4), ${A_z}(t)$ generates a uniformly bounded semigroup and $
\mathop {\lim }\limits_{z \to \infty } A_z {\text{(}}t{\text{)}}\phi {\text{ = }}A{\text{(}}t{\text{)}}\phi$ for $\phi  \in {\text{D}}(A{\text{(t)}})
$.  
\begin{thm}
The operator ${{\mathcal{A}}_z}(t)$ satisfies
\begin{enumerate}
\item 
${\mathcal{A}}{\text{(}}t{\text{)}} {\mathcal{A}}_z {\text{(}}t{\text{)}}\Phi  = {{\mathcal{A}}}_z {\text{(}}t{\text{)}} {\mathcal{A}}{\text{(}}t{\text{)}}\Phi {\text{, }}\Phi  \in {\text{D}}
$, 
${\mathcal{A}}_z {\text{(}}t{\text{)}}$ 
generates a uniformly bounded contraction semigroup on ${\mathcal{F}\mathcal{D}}_ \otimes ^2 $ 
for each $t$, and 
$\mathop {\lim }\limits_{z \to \infty } {\mathcal{A}}_z {\text{(}}t{\text{)}}\Phi  = {\mathcal{A}}{\text{(}}t{\text{)}}\Phi {\text{, }}\Phi  \in {\text{D}}$. 
\item For each $n$, each set $\tau _1 , \cdots ,\tau _n  \in I$ and each set $a_1 , \cdots ,a_n ,\;\,a_i  \geqslant 0$; $\sum\nolimits_{i = 1}^n {a_i {\mathcal{A}}(\tau _i )} $ generates a $C_0$-semigroup on ${\mathcal{F}\mathcal{D}}_ \otimes ^2$.
\end{enumerate}
\end{thm}
\begin{proof} The proof of (1) follows from Theorem 3.13 and the relationship between  $\mathcal{A}(t)$ and  $A(t)$.  It is an easy computation to check that (2) follows from Theorem 4.12 and Corollary 4.1 3, with $
{\mathbf{S}}(t) = \prod _{i = 1}^n {\mathbf{S}}_{\tau _i } (a_i t) $. 
\end{proof}
We now assume that $A(t),\;t \in I$, is weakly continuous and that $D(A(t)) \supseteq D$, where $D$ is dense in ${\mathcal{H}}$ and independent of $t$.  It follows that this family has a weak KH-integral $Q[a,b] = \int_a^b {A(t)dt \in C({\mathcal{H}})}$ (the closed densely defined linear operators on ${\mathcal{H}}$).  Furthermore, it is not difficult to see that $A_z (t),\;t \in I$, is also weakly continuous and hence the family $
\left\{ {A_z (t)\left| {{\text{ }}t \in I} \right.} \right\} \subset L({\mathcal{H}})$ has a weak HK-integral $Q_z [a,b] = \int_a^b {A_z (t)dt \in L({\mathcal{H}})}$.  Let $
{\text{P}}_{\text{n}}$ be a sequence of HK-partitions for $\delta _n (t):\;[a,b] \to (0,\infty) $ with $\delta _{n + 1} (t) \le \delta _n (t) $ and $\lim _{n \to \infty } \delta _n (t) = 0$, so that the mesh $\mu _n  = \mu ({\text{P}}_{\text{n}} ) \to 0{\text{ as n}} \to \infty $.  Set $Q_{z,n}  = \sum\nolimits_{l = 1}^n {A_z (\bar{t}_l )\Delta t_l }$, 
$
Q_{z,m}  = \sum\nolimits_{q = 1}^m {A_z (\bar{s}_q )\Delta s_q }; $
${\mathbf{Q}}_{z,n}  = \sum\nolimits_{l = 1}^{\text{n}} {{\mathcal{A}}_z (\bar{t}_l )\Delta t_l }$, 
$
{\mathbf{Q}}_{z,m}  = \sum\nolimits_{q = 1}^{\text{m}} {{\mathcal{A}}_z (\bar{s}_q )\Delta s_q }$; and $\Delta Q_z  = Q_{z,n}  - Q_{z,m} {\text{,}}\;\,\,\Delta {\mathbf{Q}}_z  = {\mathbf{Q}}_{z,n}  - {\mathbf{Q}}_{z,m}$.  
Let 
$
\Phi ,\Psi  \in {\text{D}}_0 ;{\text{ }}\Phi  = \sum\nolimits_i^J {\Phi ^{\text{i}} }  = \sum\nolimits_i^J {\sum\nolimits_{F{\text{(}}f{\text{)}}}^K {a_{F{\text{(}}f{\text{)}}}^{\text{i}} E_{F{\text{(}}f{\text{)}}}^{\text{i}} } },\;\; 
$
$
\Psi  = \sum\nolimits_i^L {\Psi ^{\text{i}} }  = \sum\nolimits_i^L {\sum\nolimits_{F{\text{(}}g{\text{)}}}^M {b_{F{\text{(}}g{\text{)}}}^{\text{i}} E_{F{\text{(}}g{\text{)}}}^{\text{i}} } }. 
$
Then we have:
\begin{thm}\label{F: fun} (Fundamental Theorem for Time-Ordered Integrals)
\begin{enumerate}
\item The family $
\left\{ { {\mathcal{A}}_z ({\text{t}})\left| {{\text{ }}t \in I} \right.} \right\}$ has a weak KH-integral and
\beqn
\left\langle {\Delta {\mathbf{Q}}_z \Phi ,\Psi } \right\rangle  = \sum\nolimits_i^J {\sum\nolimits_{F{\text{(}}f{\text{)}}}^K {a_{F{\text{(}}f{\text{)}}}^{\text{i}} \bar b_{F{\text{(}}f{\text{)}}}^{\text{i}} \left\langle {\Delta Q_z e^i ,e^i } \right\rangle } } \,{\text{ (a}}{\text{.s}}{\text{.c)}}{\text{.}}
\eeqn
\item If, in addition, for each $i$
\beqn\label{CA: cona}
\sum\limits_{{\text{k,}}}^{\text{n}} {\Delta {\text{t}}_{\text{k}} \left\| {A_z {\text{(}}s_{\text{k}} {\text{)}}e^i  - \left\langle {A_z {\text{(}}s_{\text{k}} {\text{)}}e^i ,e^i } \right\rangle e^i } \right\|^2 }  \leqslant M\mu _n^{\delta  - 1}, 
\eeqn
where $M$ is a constant, $\mu _{\text{n}}$ is the mesh of $
{\text{P}}_{\text{n}}$, and $0 < \delta  < 1$, then the family $
\left\{ { {\mathcal{A}}_z ({\text{t}})\left| {{\text{ }}t \in I} \right.} \right\}$ has a strong integral, ${\mathbf{Q}}_z [t,a] = \int_a^t { {\mathcal{A}}_z (s)ds}$.
\item The linear operator ${\mathbf{Q}}_z [t,a] $ generates a uniformly continuous $C_0$-contraction semigroup.
\end{enumerate}
\end{thm}
\begin{rem}  
In general, the family 
$
\left\{ {A_z (t)\left| {{\text{ }}t \in I} \right.} \right\}
$ need not have a Bochner or Pettis integral.  (However, if it has a Bochner integral, our condition (\ref{CA: cona}) is automatically satisfied.) 
\end{rem}
\begin{proof}  To prove (1), note that 
\[
\left\langle {\Delta {\mathbf{Q}}_z \Phi ,\Psi } \right\rangle  = \sum\nolimits_i {\sum\nolimits_{F{\text{(}}f{\text{)}}} {\sum\nolimits_{F{\text{(}}g{\text{)}}} {a_{F{\text{(}}f{\text{)}}}^{\text{i}} \bar b_{F{\text{(}}g{\text{)}}}^{\text{i}} \left\langle {\Delta {\mathbf{Q}}_z E_{F{\text{(}}f{\text{)}}}^{\text{i}} ,E_{F{\text{(}}g{\text{)}}}^{\text{i}} } \right\rangle } } } 
\]
 (we omit the upper limit).  Now
\beqa
\begin{gathered}
  \left\langle {\Delta {\mathbf{Q}}_z E_{F{\text{(}}f{\text{)}}}^{\text{i}} ,E_{F{\text{(}}g{\text{)}}}^{\text{i}} } \right\rangle  = \sum\limits_{l = 1}^{\text{n}} {\Delta t_l } \prod\limits_{{\text{t}} \ne {\bar{t}}_l } {\left\langle {e_{t,f(t)}^{\text{i}} ,e_{t,g(t)}^{\text{i}} } \right\rangle } \left\langle {A_z (\bar{t}_l )e_{\bar{t}_l ,f(\bar{t}_l )}^{\text{i}} ,e_{\bar{t}_l ,g(\bar{t}_l )}^{\text{i}} } \right\rangle  \hfill \\
  {\text{   }} - \sum\limits_{q = 1}^{\text{m}} {\Delta s_q } \prod\limits_{{\text{t}} \ne {\bar{s}}_q } {\left\langle {e_{t,f(t)}^{\text{i}} ,e_{t,g(t)}^{\text{i}} } \right\rangle } \left\langle {A_z (\bar{s}_q )e_{\bar{s}_q ,f(\bar{s}_q )}^{\text{i}} ,e_{\bar{s}_q ,g(\bar{s}_q )}^{\text{i}} } \right\rangle  = \sum\limits_{l = 1}^{\text{n}} {\Delta t_l } \left\langle {A_z (\bar{t}_l )e_{\bar{t}_l ,f(\bar{t}_l )}^{\text{i}} ,e_{\bar{t}_l ,f(\bar{t}_l )}^{\text{i}} } \right\rangle  \hfill \\
  {\text{                                       }} - \sum\limits_{q = 1}^{\text{m}} {\Delta s_q } \left\langle {A_z (\bar{s}_q )e_{\bar{s}_q ,f(\bar{s}_q )}^{\text{i}} ,e_{\bar{s}_q ,f(\bar{s}_q )}^{\text{i}} } \right\rangle  = \left\langle {\Delta Q_z e^i ,e^i } \right\rangle {\text{ (e}}{\text{.f}}{\text{.n}}{\text{.v)}}{\text{.  }} \hfill \\ 
\end{gathered} 
\eeqa
This gives (5.3) and shows that the family $\left\{ { {\mathcal{A}}_z ({\text{t}})\left| {{\text{ }}t \in I} \right.} \right\}$ has a weak HK-integral if and only if the family $
\left\{ {A_z ({\text{t}})\left| {{\text{ }}t \in I} \right.} \right\}$ has one. 

To see that condition  (\ref{CA: cona}) makes ${\mathbf{Q}}_z$ a strong limit, let $\Phi  \in {\text{D}}_0 $.  Then 
\beqa
\begin{gathered}
  \left\langle {{\mathbf{Q}}_{z,n} \Phi ,{\mathbf{Q}}_{z,n} \Phi } \right\rangle  = \sum\nolimits_{\text{i}}^J {\sum\nolimits_{F(f),F(g)}^K {a_{F(f)}^{\text{i}} \bar a_{F(g)}^{\text{i}} \left( {\sum\limits_{{\text{k,m}}}^{\text{n}} {\sum\nolimits_{k = 1}^n {\Delta {\text{t}}_k \Delta {\text{t}}_m \left\langle { {\mathcal{A}}_z {\text{(}}s_k {\text{)}}E_{F(f)}^{\text{i}} , {\mathcal{A}}_z {\text{(}}s_m {\text{)}}E_{F(g)}^{\text{i}} } \right\rangle } } } \right)} }  \hfill \\
   = \sum\nolimits_{\text{i}}^J {\sum\nolimits_{F(f)}^K {\mathop {\left| {\mathop a\nolimits_{F(f)}^{\text{i}} } \right|}\nolimits^2 \left( {\sum\nolimits_{k \ne m}^n {\Delta {\text{t}}_k \Delta {\text{t}}_m \left\langle {A_z {\text{(}}s_k {\text{)}}e_{s_k ,f(s_k )}^{\text{i}} ,e_{s_k ,f(s_k )}^{\text{i}} } \right\rangle \left\langle {e_{s_m ,f(s_m )}^{\text{i}} ,A_z {\text{(}}s_m {\text{)}}e_{s_m ,f(s_m )}^{\text{i}} } \right\rangle } } \right)} }  \hfill \\
  {\text{                                      }} + \sum\nolimits_{\text{i}}^J {\sum\nolimits_{F(f)}^K {\mathop {\left| {\mathop a\nolimits_{F(f)}^{\text{i}} } \right|}\nolimits^2 \left( {\sum\nolimits_{k = 1}^n {(\Delta {\text{t}}_k )^2 \left\langle {A_z {\text{(}}s_k {\text{)}}e_{s_k ,f(s_k )}^{\text{i}} ,A_z {\text{(}}s_k {\text{)}}e_{s_k ,f(s_k )}^{\text{i}} } \right\rangle } } \right)} } . \hfill \\ 
\end{gathered} 
\eeqa
This can be rewritten as
\beqn
\begin{gathered}
  \left\| {{\mathbf{Q}}_{z,n} \Phi } \right\|_ \otimes ^2  = \sum\nolimits_i^J {\sum\nolimits_{F(f)}^K {\left| {a_{F(f)}^i } \right|^2 \left\{ {\left| {\left\langle {Q_{z,n} e^i ,e^i } \right\rangle } \right|^2 } \right.} }  \hfill \\
  {\text{                     }}\left. { + \sum\nolimits_{k = 1}^n {(\Delta t_k )^2 \left( {\left\| {A_z (s_k )e^i } \right\|^2  - \left| {\left\langle {A_z (s_k )e^i ,e^i } \right\rangle } \right|^2 } \right)} } \right\}\; (a.s.c). \hfill \\ 
\end{gathered} 
\eeqn
First note that: $$
\left\| {A_z (s_k )e^i } \right\|^2  - \left| {\left\langle {A_z (s_k )e^i ,e^i } \right\rangle } \right|^2  = \left\| {A_z (s_k )e^i  - \left\langle {A_z (s_k )e^i ,e^i } \right\rangle e^i } \right\|^2,$$ 
so that the last term in (5.5) can be written as
\beqa
\begin{gathered}
  \sum\nolimits_{k = 1}^n {(\Delta t_k )^2 \left( {\left\| {A_z (s_k )e^i } \right\|^2  - \left| {\left\langle {A_z (s_k )e^i ,e^i } \right\rangle } \right|^2 } \right)}  = \sum\nolimits_{k = 1}^n {(\Delta t_k )^2 \left\| {A_z (s_k )e^i  - \left\langle {A_z (s_k )e^i ,e^i } \right\rangle e^i } \right\|^2 }  \hfill \\
  {\text{                                                                }} \leqslant {\text{ }}\mu _n^\delta  M. \hfill \\
  {\text{                         }} \hfill \\ 
\end{gathered} 
\eeqa
We can now use the above result in (5.5) to get
\[
\left\| {{\mathbf{Q}}_{z,n} \Phi } \right\|_ \otimes ^2  \leqslant \sum\nolimits_i^J {\sum\nolimits_{F(f)}^K {\left| {a_{F(f)}^i } \right|^2 \left| {\left\langle {Q_{z,n} e^i ,e^i } \right\rangle } \right|^2 } }  + \mu _n^\delta  M\quad (a.s.c).
\]
Thus, ${\mathbf{Q}}_{z,n} [t,a] $ converges strongly to ${\mathbf{Q}}_z [t,a] $ on $
{\mathcal{F}\mathcal{D}}_ \otimes ^2 $.  To show that ${\mathbf{Q}}_z {\text{[}}t,a{\text{]}}$ generates a uniformly continuous contraction semigroup, it suffices to show that $Q_{\text{z}} {\text{[}}t,a{\text{]}}$ is dissipative.  For any $\Phi$ in $
{\mathcal{F}\mathcal{D}}_ \otimes ^2$, 
\[
\left\langle {{\mathbf{Q}}_z {\text{[}}t,a{\text{]}}\Phi ,\Phi } \right\rangle  = \sum\limits_i^J {\sum\limits_{F(f)}^K {\left| {a_{F(f)}^i } \right|^2 } } \left\langle {Q_z e^i ,e^i } \right\rangle {\text{ }}(a.s.c{\text{)}}
\]
and, for each n, we have
\[
\begin{gathered}
  \operatorname{Re} \left\langle {Q_z [t,a]e^i ,e^i } \right\rangle  = \operatorname{Re} \left\langle {Q_{z,n} [t,a]e^i ,e^i } \right\rangle  + \operatorname{Re} \left\langle {\left[ {Q_z [t,a] - Q_{z,n} [t,a]} \right]e^i ,e^i } \right\rangle  \hfill \\
  {\text{                          }} \leqslant \operatorname{Re} \left\langle {\left[ {Q_z [t,a] - Q_{z,n} [t,a]} \right]e^i ,e^i } \right\rangle, \hfill \\ 
\end{gathered} 
\]
since $Q_{z,n} [t,a]$ is dissipative. Letting $n \to \infty $ implies $
 \operatorname{Re} \left\langle {Q_z [t,a]e^i ,e^i } \right\rangle  \leqslant 0$, so that $
\operatorname{Re} \left\langle {{\mathbf{Q}}_z [t,a]\Phi ,\Phi } \right\rangle  \leqslant 0$.  Thus, ${\mathbf{Q}}_z {\text{[}}t,a{\text{]}}$ is a bounded dissipative linear operator on ${\mathcal{F}\mathcal{D}}_ \otimes ^2$, which completes our proof. 
\end{proof}
We can also prove Theorem \ref{F: fun} for the family 
$
\left\{ {\mathcal{A}}({\text{t}})\left| {\text{ }}t \in I\right. \right\}
$.  The same proof goes through, but now we restrict to $
{\text{D}}_0  = \mathop  \otimes \limits_{t \in I} {\text{D}}(A(t)) \cap {\mathcal{F}\mathcal{D}}_ \otimes ^2 $.   In this case  (\ref{CA: cona}) becomes: 
\beqn\label{CD: cond}
\sum\limits_{k,}^n {\Delta t_k \left\| {A(s_k )e^i  - \left\langle {A(s_k )e^i ,e^i } \right\rangle e^i } \right\|^2 }  \leqslant M\mu _n^{\delta  - 1}. 
\eeqn
From equation (5.5), we have the following important result: (set $
\sum\limits_{F(f)}^K {\left| {a_{F(f)}^{\text{i}} } \right|^{\text{2}} }  = \left| {b^i } \right|^2 $)
\beqn\label{B: basic}
\left\| {{\mathbf{Q}}_z {\text{[}}t,a{\text{]}}\Phi } \right\|_ \otimes ^2  = \sum\limits_{\text{i}}^J {\left| {b^i } \right|^2 } \left| {\left\langle {Q_z e^i ,e^i } \right\rangle } \right|^2 {\text{ }}(a.s.c).
\eeqn
The representation (\ref{B: basic}) makes it easy to prove the next theorem.  
\begin{thm} With the conditions of Theorem \ref{F: fun}, we have:
\begin{enumerate}
	\item  $
{\mathbf{Q}}_z [t,s] + {\mathbf{Q}}_z [s,a] = {\mathbf{Q}}_z [t,a]{\text{ (a}}{\text{.s}}{\text{.c),}} $
\item $
s{\text{ - }}\mathop {\lim }\limits_{h \to 0} \frac{{{\mathbf{Q}}_z [t + h,a] - {\mathbf{Q}}_z [t,a]}}
{h} = s{\text{ - }}\mathop {\lim }\limits_{h \to 0} \frac{{{\mathbf{Q}}_z [t + h,t]}}
{h}{\text{ }} = {\mathcal{A}}_z {\text{(}}t{\text{) (a}}{\text{.s}}{\text{.c),}}
$
\item $
s{\text{ - }}\mathop {\lim }\limits_{h \to 0} {\mathbf{Q}}_z [t + h,t]{\text{ = }}0{\text{ (a}}{\text{.s}}{\text{.c),}} $
\item $
s{\text{ - }}\mathop {\lim }\limits_{h \to 0} \exp \left\{ {\tau {\mathbf{Q}}_z [t + h,t]} \right\}{\text{ = I}}_ \otimes  {\text{ (a}}{\text{.s}}{\text{.c),}}\tau  \geqslant 0. $
\end{enumerate}
\end{thm}
\begin{proof} In each case, it suffices to prove the result for $
\Phi  \in {\text{D}}_0 $.  To prove (1), use 
\beqa
\begin{gathered}
  \left\| {\left[ {{\mathbf{Q}}_z [t,s] + {\mathbf{Q}}_z [s,a]} \right]\Phi } \right\|_ \otimes ^2  = \sum\nolimits_i^J {\left| {b^i } \right|^2 \left| {\left\langle {\left[ {Q_z [t,s] + Q_z [s,a]} \right]e^i ,e^i } \right\rangle } \right|^2 }  \hfill \\
  {\text{                           }} = \sum\nolimits_i^J {\left| {b^i } \right|^2 \left| {\left\langle {Q_z [t,a]e^i ,e^i } \right\rangle } \right|^2 } {\text{ = }}\left\| {{\mathbf{Q}}_z [t,a]\Phi } \right\|_ \otimes ^2 {\text{(a}}{\text{.s}}{\text{.c)}}. \hfill \\ 
\end{gathered} 
\eeqa
To prove (2), use (1) to get that $
{\mathbf{Q}}_z [t + h,a] - {\mathbf{Q}}_z [t,a] = {\mathbf{Q}}_z [t + h,t]{\text{ (a}}{\text{.s}}{\text{.),}}$ so that 
\[
\begin{gathered}
  \mathop {\lim }\limits_{h \to 0} \left\| {\frac{{{\mathbf{Q}}_z [t + h,t]}}
{h}\Phi } \right\|_ \otimes ^2  \hfill \\
  {\text{                  }} = \sum\limits_{\text{i}}^J {\left| {b^i } \right|^2 } \mathop {\lim }\limits_{h \to 0} \left| {\left\langle {\frac{{Q_z [t + h,t]}}
{h}e^i ,e^i } \right\rangle } \right|^2  = {\text{ }}\left\| { {\mathcal{A}}_z {\text{(}}t{\text{)}}\Phi } \right\|_ \otimes ^2 {\text{ (a}}{\text{.s}}{\text{.c}}{\text{.)}}{\text{.}} \hfill \\ 
\end{gathered} 
\]
The proof of (3) follows from (2) and the proof of (4) follows from (3). 
\end{proof}
	The results of the previous theorem are expected if ${\mathbf{Q}}_z [t,a] $ is an integral in the conventional sense.  The important point is that a weak integral on the base space gives a strong integral on ${\mathcal{F}\mathcal{D}}_ \otimes ^2$ (note that, by (2), we also get strong differentiability).  This clearly shows that our approach to time-ordering has more to offer than being simply a representation space to allow time to act as a place-keeper for operators in a product.  It should be observed that, in all results up to now, we have used the assumption that the family $A(t),t \in I$, is weakly continuous, satisfies equation (\ref{CD: cond}), and has a common dense domain $D \subseteq D(A(t)) $ in ${\mathcal{H}}$.  We now impose a condition that is equivalent to assuming that each $A(t) $ generates a $C_0$-contraction semigroup; namely, we assume that, for each $t$, $A(t) $ and $A^ * (t)$  (dual) are dissipative.  This form is an easier condition to check. 
\begin{thm} With the above assumptions, we have that $
\mathop {\lim }\limits_{z \to \infty } \left\langle {Q_{z} [t,a]\phi ,\psi } \right\rangle  = \left\langle {Q[t,a]\phi ,\psi } \right\rangle$ exists for all $
\phi  \in D[Q],\;\psi  \in D[Q^ *  ] $.  Furthermore:
\begin{enumerate}
\item the operator $Q[t,a]$ generates a $C_0$-contraction semigroup on $
{\mathcal{H}}$,
\item for $\Phi  \in {\text{D}}_0 $, 
\[
\mathop {\lim }\limits_{z \to \infty } {\mathbf{Q}}_z [t,a]\Phi  = {\mathbf{Q}}[t,a]\Phi, 
\]
and 
\item the operator ${\mathbf{Q}}[t,a]$ generates a $C_0$-contraction semigroup on ${\mathcal{F}\mathcal{D}}_ \otimes ^2 $,	
\item ${\mathbf{Q}}[t,s]\Phi  + {\mathbf{Q}}[s,a]\Phi  = {\mathbf{Q}}[t,a]\Phi {\text{ (a}}{\text{.s}}{\text{.c}}{\text{.)}}$,
\item 
\[
\mathop {\lim }\limits_{h \to 0} \left[ {{{\left( {{\mathbf{Q}}[t + h,a] - {\mathbf{Q}}[t,a]} \right)} \mathord{\left/
 {\vphantom {{\left( {{\mathbf{Q}}[t + h,a] - {\mathbf{Q}}[t,a]} \right)} h}} \right.
 \kern-\nulldelimiterspace} h}} \right]\Phi  = \mathop {\lim }\limits_{h \to 0} \left[ {{{\left( {{\mathbf{Q}}[t + h,t]} \right)} \mathord{\left/
 {\vphantom {{\left( {{\mathbf{Q}}[t + h,t]} \right)} h}} \right.
 \kern-\nulldelimiterspace} h}} \right]\Phi  = {\mathcal{A}}{\text{(}}t{\text{)}}\Phi {\text{ (a}}{\text{.s}}{\text{.c}}{\text{.)}},
\]
\item $\mathop {\lim }\limits_{h \to 0} {\mathbf{Q}}[t + h,t]\Phi {\text{ = }}0{\text{ (a}}{\text{.s}}{\text{.c}}{\text{.)}}$, and
\item $\mathop {\lim }\limits_{h \to 0} \exp \left\{ {\tau {\mathbf{Q}}[t + h,t]} \right\}\Phi {\text{ = }}\Phi {\text{ (a}}{\text{.s}}{\text{.c}}{\text{.),}}\tau  \geqslant 0$. 
\end{enumerate}
\end{thm}
\begin{proof} Since $A_{\text{z}} (t),\;A(t) $ are weakly continuous and $
A_z (t)\xrightarrow{s}A(t) $ for each $t \in I$, given $\varepsilon  > 0$ we can choose $Z$ such that, if $z > Z$, then 
\[
\sup _{s \in [a,b]} \left| {\left\langle {\left[ {A(s) - A_z (s)} \right]\varphi \,,\,\psi } \right\rangle } \right| < {\varepsilon  \mathord{\left/
 {\vphantom {\varepsilon  {3(b - a)}}} \right.
 \kern-\nulldelimiterspace} {3(b - a)}}.
\]
By uniform (weak) continuity, if $s,s' \in [a,b] $ we can also choose $\eta $ such that, if $
\left| {s - s'} \right| < \eta $, 
\[
\sup _{z > 0} \left| {\left\langle {\left[ {A_z (s) - A_z (s')} \right]\varphi \,,\,\psi } \right\rangle } \right| < {\varepsilon  \mathord{\left/
 {\vphantom {\varepsilon  {3(b - a)}}} \right.
 \kern-\nulldelimiterspace} {3(b - a)}}
\]
and
\[
\left| {\left\langle {\left[ {A(s) - A(s')} \right]\varphi \,,\,\psi } \right\rangle } \right| < {\varepsilon  \mathord{\left/
 {\vphantom {\varepsilon  {3(b - a)}}} \right.
 \kern-\nulldelimiterspace} {3(b - a)}}.
\]
Now choose $\delta (t):\;[a,b] \to (0,\infty ) $ so that, for any HK-partition ${\mathbf{P}}$ for $\delta$, we have that $\mu _n  < \eta $, where $\mu _n$ is the mesh of the partition.  If $Q_{z,n}  = \sum\nolimits_{j = 1}^n {A_z (\tau _j )\Delta t_j } $
 and $Q_n  = \sum\nolimits_{j = 1}^n {A(\tau _j )\Delta t_j }$, we have 
\beqa
\begin{gathered}
  \left| {\left\langle {\left[ {Q_z [t,a] - Q[t,a]} \right]\varphi \,,\,\psi } \right\rangle } \right| \leqslant \left| {\left\langle {\left[ {Q_n [t,a] - Q[t,a]} \right]\varphi \,,\,\psi } \right\rangle } \right| \hfill \\
  {\text{                           }} + \left| {\left\langle {\left[ {Q_{z,n} [t,a] - Q_z [t,a]} \right]\varphi \,,\,\psi } \right\rangle } \right| + \left| {\left\langle {\left[ {Q_n [t,a] - Q_{z,n} [t,a]} \right]\varphi \,,\,\psi } \right\rangle } \right| \hfill \\
  {\text{       }} \leqslant \sum\nolimits_{j = 1}^n {\int_{t_{j - 1} }^{t_j } {\left| {\left\langle {\left[ {A(\tau _j ) - A(\tau )} \right]\varphi \,,\,\psi } \right\rangle } \right|d\tau } }  + \sum\nolimits_{j = 1}^n {\int_{t_{j - 1} }^{t_j } {\left| {\left\langle {\left[ {A_z (\tau _j ) - A_z (\tau )} \right]\varphi \,,\,\psi } \right\rangle } \right|d\tau } }  \hfill \\
  {\text{                                 }} + \sum\nolimits_{j = 1}^n {\int_{t_{j - 1} }^{t_j } {\left| {\left\langle {\left[ {A(\tau _j ) - A_z (\tau _j )} \right]\varphi \,,\,\psi } \right\rangle } \right|d\tau } }  < \frac{\varepsilon }
{3} + \frac{\varepsilon }
{3} + \frac{\varepsilon }
{3} = \varepsilon . \hfill \\ 
\end{gathered} 
\eeqa
This proves that $
\mathop {\lim }\limits_{z \to \infty } \left\langle {Q_z [t,a]\phi ,\psi } \right\rangle  = \left\langle {Q[t,a]\phi ,\psi } \right\rangle $.  To prove (1), first note that $Q{\text{[}}t,a{\text{]}}$ is closable and use
\[
\begin{gathered}
  \operatorname{Re} \left\langle {Q[t,a]\phi ,\phi } \right\rangle  = \operatorname{Re} \left\langle {Q_z [t,a]\phi ,\phi } \right\rangle  + \operatorname{Re} \left\langle {\left[ {Q[t,a] - Q_z [t,a]} \right]\phi ,\phi } \right\rangle  \hfill \\
  {\text{                       }} \leqslant \operatorname{Re} \left\langle {\left[ {Q[t,a] - Q_z [t,a]} \right]\phi ,\phi } \right\rangle , \hfill \\ 
\end{gathered} 
\]
and let ${\text{z}} \to \infty $, to show that $Q[t,a]$ is dissipative.   Then do likewise for $\left\langle {\phi ,Q^ * [t,a] \phi } \right\rangle $ to show that the same is true for $Q^*[t,a]$, to complete the proof. (\emph{It is important to note that, although $Q[t,a]$ generates a contraction semigroup on $\mathcal{H}$, $exp\{Q[t,a]\}$ does not solve the original initial-value problem.})  

To prove (2), use (\ref{B: basic}) in the form 
\beqn
\;\;\;\;\;\;\; \left\| {\left[ {{\mathbf{Q}}_z {\text{[}}t,a{\text{]}} - {\mathbf{Q}}_{z'} {\text{[}}t,a{\text{]}}} \right]\Phi } \right\|_ \otimes ^2  = \sum\limits_{\text{i}}^J {\left| {b^i } \right|^2 } \left| {\left\langle {\left[ {Q_z {\text{[}}t,a{\text{]}} - Q_{z'} {\text{[}}t,a{\text{]}}} \right]e^i ,e^i } \right\rangle } \right|^2 \; .
\eeqn
This proves that ${\mathbf{Q}}_z {\text{[}}t,a{\text{]}}\xrightarrow{{\text{s}}}{\mathbf{Q}}{\text{[}}t,a{\text{]}}$.  Since ${\mathbf{Q}}[t,a] $ is densely defined, it is closable.  The same method as above shows that it is m-dissipative.   Proofs of the other results follow the methods of Theorem 5.12. 
\end{proof}
\subsection{{General Case}}
\paragraph{}
We relax the contraction condition and assume that $A(t),t \in I$, generates a $C_0$-semigroup on ${\mathcal{H}}$.  We can always  shift the spectrum (if necessary) so that $
\left\| {\exp \{ \tau A(t)\} } \right\|  \leqslant M(t) $.  We assume that $
\sup _J \prod _{i \in J}^{} \left\| {\exp \{ \tau A(t_i )\} } \right\|  \leqslant M$, where the sup is over all finite subsets $J \subset I$.
 \begin{thm}
Suppose that $A(t),t \in I$, generates a $C_0$-semigroup, satisfies (\ref{CD: cond}) and has a weak HK-integral, $Q[t,a] $, on a dense set $D$ in ${\mathcal{H}}$.  Then the family $ {\mathcal{A}}(t),t \in I$, has a strong HK-integral, $
{\mathbf{Q}}[t,a]$, which generates a $C_0$-semigroup on $ {\mathcal{F}\mathcal{D}}_ \otimes ^2 $
 (for each $t \in I$) and $\left\| {\exp \{ {\mathbf{Q}}[t,a] \} } \right\|_ \otimes   \leqslant M$.
\end{thm}
\begin{proof}  It is clear from part (2) of Theorem 5.9 that ${\mathbf{Q}}_n [t,a] = \sum\nolimits_{i = 1}^n { {\mathcal{A}}(\tau _i )} \Delta t_i $ generates a $C_0$-semigroup on ${\mathcal{F}\mathcal{D}}_ \otimes ^2$ and $
\left\| {\exp \{ {\mathbf{Q}}_n {\text{[}}t,a{\text{]}}\} } \right\|_ \otimes   \leqslant M$.  If $\Phi  \in D_0 $, let ${\mathbf{P}}_m $, ${\mathbf{P}}_n$ be arbitrary HK-partitions for $\delta _m ,\;  \delta _n $  (of order $m$ 
 and $n$ respectively) and set $\delta (s) = \delta _m (s) \wedge \delta _n (s) $.  Since any HK-partition for $\delta$ is one for $\delta _m$ and $\delta _n$, we have that 
\[
\begin{gathered}
  \left\| {\left[ {\exp \{ \tau {\mathbf{Q}}_n [t,a]\}  - \exp \{ \tau {\mathbf{Q}}_m [t,a]\} } \right]\Phi } \right\|_ \otimes   \hfill \\
  {\text{                }} = \left\| {\int_0^\tau  {\frac{d}
{{ds}}\left[ {{\text{exp}}\{ (\tau  - s){\mathbf{Q}}_n [t,a]\} {\text{exp}}\{ s{\mathbf{Q}}_m [t,a]\} } \right]} \Phi ds} \right\|_ \otimes   \hfill \\
  {\text{                }} \leqslant \int_0^\tau  {\left\| {\left[ {{\text{exp}}\{ (\tau  - s){\mathbf{Q}}_n [t,a]\} \left( {{\mathbf{Q}}_n [t,a] - {\mathbf{Q}}_m [t,a]} \right){\text{exp}}\{ s{\mathbf{Q}}_m [t,a]\} \Phi } \right]} \right\|} _ \otimes   \hfill \\
  {\text{                }} \leqslant M\int_0^\tau  {\left\| {\left( {{\mathbf{Q}}_n [t,a] - {\mathbf{Q}}_m [t,a]} \right)\Phi } \right\|} _ \otimes  ds \hfill \\
  {\text{                }} \leqslant M\tau \left\| {\left[ {{\mathbf{Q}}_n [t,a] - {\mathbf{Q}}[t,a]} \right]\Phi } \right\|_ \otimes   + M\tau \left\| {\left[ {{\mathbf{Q}}[t,a] - {\mathbf{Q}}_m [t,a]} \right]\Phi } \right\|_ \otimes  . \hfill \\ 
\end{gathered} 
\]
The existence of the weak HK-integral, $Q[t,a] $, on ${\mathcal{H}}$ satisfying equation (\ref{CD: cond}) implies that ${\mathbf{Q}}_n [t,a]\xrightarrow{s}{\mathbf{Q}}[t,a] $, so that $\exp \{ \tau {\mathbf{Q}}_n {\text{[}}t,a{\text{]}}\} \Phi$ converges as $n \to \infty$ for each fixed $t \in I$; and the convergence is uniform on bounded $\tau$ intervals.  As $\left\| {\exp \{ {\mathbf{Q}}_n {\text{[}}t,a{\text{]}}\} } \right\|_ \otimes   \leqslant M$, we have 
\[
\lim _{n \to \infty } \exp \{ \tau {\mathbf{Q}}_n [t,a]\} \Phi  = {\mathbf{S}}_t (\tau )\Phi ,{\text{ }}\Phi  \in {\mathcal{F}\mathcal{D}}_ \otimes ^2 .
\]
The limit is again uniform on bounded $\tau$ intervals.  It is easy to see that the limit $
{\mathbf{S}}_t (\tau ) $ satisfies the semigroup property, ${\mathbf{S}}_t (0) = I$, and 
$\left\| {{\mathbf{S}}_t (\tau )} \right\|_ \otimes   \leqslant M$.  Furthermore, as the uniform limit of continuous functions, we see that $\tau  \to {\mathbf{S}}_t (\tau )\Phi$ is continuous for $\tau  \geqslant 0$.  We are done if we show that ${\mathbf{Q}}[t,a] $ is the generator of ${\mathbf{S}}_t (\tau ) $.  For $\Phi  \in D_0 $, we have that
\[
\begin{gathered}
  {\mathbf{S}}_t (\tau )\Phi  - \Phi  = \lim _{n \to \infty } \exp \{ \tau {\mathbf{Q}}_n [t,a]\} \Phi  - \Phi  \hfill \\
  {\text{           }} = \lim _{n \to \infty } \int_0^\tau  {\exp \{ s{\mathbf{Q}}_n [t,a]\} } {\mathbf{Q}}_n [t,a]\Phi ds = \int_0^\tau  {{\mathbf{S}}_t (\tau )} {\mathbf{Q}}[t,a]\Phi ds. \hfill \\ 
\end{gathered} 
\]
Our result follows from the uniqueness of the generator, so that ${\mathbf{S}}_t (\tau ) = \exp \{ \tau {\mathbf{Q}}{\text{[}}t,a{\text{]}}\}$. 
\end{proof}
	The next result is the time-ordered version of the Hille-Yosida Theorem (see Pazy [PZ], pg. 8).    We assume that the family $A(t),t \in I$, is closed and densely defined.
\begin{thm} The family ${\mathcal{A}}(t),t \in I$, has a strong HK-integral, ${\mathbf{Q}}{\text{[}}t,a{\text{]}}$, which generates a $C_0$-contraction semigroup on $
{\mathcal{F}\mathcal{D}}_ \otimes ^2$ if and only if $\rho (A(t)) \supset (0,\infty ) $, $
\left\| {R\left( {\lambda \,:\,A(t)} \right)} \right\| < {1 \mathord{\left/
 {\vphantom {1 \lambda }} \right.
 \kern-\nulldelimiterspace} \lambda }$ for $\lambda  > 0$, $A(t),t \in I$, satisfies (\ref{CD: cond}) and has a densely defined weak HK-integral $Q[t,a] $ on ${\mathcal{H}}$. 
\end{thm}
\begin{proof} In the first direction, suppose ${\mathbf{Q}}{\text{[}}t,a{\text{]}}$ generates a $C_0$-contraction semigroup on ${\mathcal{F}\mathcal{D}}_ \otimes ^2 $.  Then 
${\mathbf{Q}}_n [t,a]\Phi \xrightarrow{s}{\mathbf{Q}}[t,a]\Phi $ for each $
\Phi  \in {\text{D}}_0 $ and each $t \in I$. Since ${\mathbf{Q}}{\text{[}}t,a{\text{]}}$ has a densely defined strong HK-integral, it follows from (\ref{CD: cond}) that $Q[t,a] $ must have a densely defined weak HK-integral.  Since ${\mathbf{Q}}_n {\text{[}}t,a{\text{]}}$ generates a $C_0$-contraction semigroup for each HK-partition of order $n$, it follows that $ {\mathcal{A}}(t) $ must generate a $C_0$-contraction semigroup for each $t \in I$.  From Theorem 4.12 and Theorem 5.9, we see that $A(t) $ must also generate a $C_0$-contraction semigroup for each $t \in I$.  From the conventional Hille-Yosida theorem, the resolvent condition follows.

	In the reverse direction, the conventional Hille-Yosida theorem along with the first part of Theorem 5.13 shows that $Q[t,a] $ generates a $C_0$-contraction semigroup for each $t \in I$.  From parts (2), (3) of Theorem 5.9 and Theorem 4.12, we have that, for each HK-partition of order $n$, ${\mathbf{Q}}_n {\text{[}}t,a{\text{]}}$ generates a $C_0$-contraction semigroup, ${\mathbf{Q}}_{\text{n}} [t,a]\Phi  \to {\mathbf{Q}}[t,a]\Phi $ for each $\Phi  \in {\text{D}}_0$ and each $t \in I$, and ${\mathbf{Q}}{\text{[}}t,a{\text{]}}$ generates a $C_0$-contraction semigroup on $
{\mathcal{F}\mathcal{D}}_ \otimes ^2 $. 
\end{proof}
The other generation theorems have a corresponding formulation in terms of time-ordered integrals.
\newpage
\section{\bf{Time-Ordered Evolutions}}
As ${\mathbf{Q}}{\text{[}}t,a{\text{]}}$ and ${\mathbf{Q}}_{\text{z}} {\text{[}}t,a{\text{]}}$ generate (uniformly bounded) $C_0$-semigroups, we can set $
{\mathbf{U}}[t,a] = exp\{ {\mathbf{Q}}[t,a]\} $, $
{\mathbf{U}}_z [t,a] = exp\{ {\mathbf{Q}}_z [t,a]\}$.  They are $C_0$-evolution operators and the following theorem generalizes a result due to Hille and Phillips [HP].
\begin{thm} For each $n$, and $\Phi  \in D\left[ {\left( {{\mathbf{Q}}[t,a]} \right)^{n + 1} } \right] $, we have: (w is positive and ${\mathbf{U}}^w [t,a] = exp\left\{ {w{\mathbf{Q}}[t,a]} \right\}$)
\beqa
{\mathbf{U}}^w [t,a]\Phi  = \left\{ {I_ \otimes   + \sum\limits_{k = 1}^n {\frac{{\left( {w{\mathbf{Q}}[t,a]} \right)^k }}
{{k!}}}  + \frac{1}
{{n!}}\int\limits_0^w {(w - \xi )^n } {\mathbf{Q}}[t,a]^{n + 1} {\mathbf{U}}_{}^\xi [t,a] d\xi } \right\}\Phi .
\eeqa
\end{thm}
\begin{proof}  The proof is easy.  Start with \[
\left[ {{\mathbf{U}}_{\text{z}}^w {\text{[}}t,a{\text{]}}\Phi  - I_ \otimes  } \right]\Phi  = \int\limits_0^w {{\mathbf{Q}}_{\text{z}} [t,a]{\mathbf{U}}_{\text{z}}^\xi  {\text{[}}t,a{\text{]}}d\xi \Phi } 
\]
 and use integration by parts to get that 
\[
\left[ {{\mathbf{U}}_z^w [t,a]\Phi  - I_ \otimes  } \right]\Phi  = w{\mathbf{Q}}_z [t,a]\Phi  + \int\limits_0^w {(w - \xi )\left[ {{\mathbf{Q}}_{\text{z}} [t,a]} \right]^2 {\mathbf{U}}_z^\xi [t,a] d\xi \Phi } .
\]
It is clear how to get the nth term.  Finally, let ${z} \to \infty$ to get the result.
\end{proof}
\begin{thm} If  $a < t< b$,
\begin{enumerate}
\item $
\mathop {\lim }\limits_{z \to \infty } {\mathbf{U}}_z [t,a]\Phi  = {\mathbf{U}}[t,a]\Phi {\text{,  }}\Phi  \in {\mathcal{F}\mathcal{D}}_ \otimes ^2. $
\item 
\[
\frac{\partial }
{{\partial t}}{\mathbf{U}}_z [t,a] \Phi  = {\mathcal{A}}_z (t){\mathbf{U}}_z [t,a]\Phi  = {\mathbf{U}}_z[t,a] {\mathcal{A}}_z (t)\Phi, 
\]
with $\Phi  \in {\mathcal{F}\mathcal{D}}_ \otimes ^2$, and 
\item \[
\frac{\partial }
{{\partial t}}{\mathbf{U}}[t,a]\Phi  = {\mathcal{A}}(t){\mathbf{U}}[t,a]\Phi  = {\mathbf{U}}[t,a]{\mathcal{A}}(t)\Phi {\text{, }}\Phi  \in D({\mathbf{Q}}[b,a]) \supset {\text{D}}_0.
\]
\end{enumerate}
\end{thm}
\begin{proof}  To prove (1), use the fact that ${\mathcal{A}}_z {\text{(}}t{\text{)}}$ and 
${\mathcal{A}}{\text{(}}t{\text{)}}$ commute, along with
\[
\begin{gathered}
  {\mathbf{U}}{\text{[}}t,a{\text{]}}\Phi  - {\mathbf{U}}_z {\text{[}}t,a{\text{]}}\Phi  = \int_0^1 {({d \mathord{\left/
 {\vphantom {d {ds}}} \right.
 \kern-\nulldelimiterspace} {ds}}} )\left( {e^{s{\mathbf{Q}}[t,a]} e^{(1 - s){\mathbf{Q}}_z [t,a]} } \right)\Phi ds \hfill \\
  {\text{            }} = \int_0^1 {s\left( {e^{s{\mathbf{Q}}[t,a]} e^{(1 - s){\mathbf{Q}}_{\text{z}} [t,a]} } \right)} \left( {{\mathbf{Q}}{\text{[}}t,a{\text{]}} - {\mathbf{Q}}_z {\text{[}}t,a{\text{]}}} \right)\Phi ds, \hfill \\ 
\end{gathered} 
\]
so that 
\[
\lim _{{\text{z}} \to {\text{0}}} \left\| {{\mathbf{U}}{\text{[}}t,a{\text{]}}\Phi  - {\mathbf{U}}_{\text{z}} {\text{[}}t,a{\text{]}}\Phi } \right\| \leqslant M\lim _{{\text{z}} \to {\text{0}}} \left\| {{\mathbf{Q}}{\text{[}}t,a{\text{]}}\Phi  - {\mathbf{Q}}_{\text{z}} {\text{[}}t,a{\text{]}}\Phi } \right\| = 0.
\]
To prove (2), use   
\[
{\mathbf{U}}_z {\text{[}}t + h,a{\text{]}} - {\mathbf{U}}_z {\text{[}}t,a{\text{]}} = {\mathbf{U}}_z {\text{[}}t,a{\text{]}}\left( {{\mathbf{U}}_z {\text{[}}t + h,t{\text{]}} - {\text{I}}} \right) = \left( {{\mathbf{U}}_z {\text{[}}t + h,t{\text{]}} - {\text{I}}} \right){\mathbf{U}}_z {\text{[}}t,a{\text{],}}
\]
so that 
\[
{{\left( {{\mathbf{U}}_z {\text{[}}t + h,a{\text{]}} - {\mathbf{U}}_z {\text{[}}t,a{\text{]}}} \right)} \mathord{\left/
 {\vphantom {{\left( {{\mathbf{U}}_z {\text{[}}t + h,a{\text{]}} - {\mathbf{U}}_z {\text{[}}t,a{\text{]}}} \right)} h}} \right.
 \kern-\nulldelimiterspace} h} = {\mathbf{U}}_z {\text{[}}t,a{\text{]}}\left[ {{{\left( {{\mathbf{U}}_z {\text{[}}t + h,t{\text{]}} - {\text{I}}} \right)} \mathord{\left/
 {\vphantom {{\left( {{\mathbf{U}}_z {\text{[}}t + h,t{\text{]}} - {\text{I}}} \right)} h}} \right.
 \kern-\nulldelimiterspace} h}} \right].
\]
Now set $\Phi _z^t  = {\mathbf{U}}_z {\text{[}}t,a{\text{]}}\Phi$ and use Theorem 6.1 with $
n = 1$ and $w = 1$ to get:
\[
{\mathbf{U}}_z {\text{[}}t + h,t{\text{]}}\Phi _z^t  = \left\{ {I_ \otimes   + {\mathbf{Q}}_z [t + h,t] + \int\limits_0^1 {(1 - \xi ){\mathbf{U}}_z^\xi  {\text{[}}t + h,t{\text{]}}} {\mathbf{Q}}_z [t + h,t]^2 d\xi } \right\}\Phi _z^t ,
\]
so 
\[
\begin{gathered}
  \frac{{\left( {{\mathbf{U}}_z {\text{[}}t + h,t{\text{]}} - {\text{I}}} \right)}}
{h}\Phi _z^t  - {\mathcal{A}}_z (t)\Phi _z^t  = \frac{{{\mathbf{Q}}_z {\text{[}}t + h,t{\text{]}}}}
{h}\Phi _z^t  - {\mathcal{A}}_z (t)\Phi _z^t  \hfill \\
  \,\,\,\,\,\,\,\,\,\,\,\,\,\,\,\,\,\,\,\,\,\,\,\,\,\,\,\,\,\,\,\,\,\,\,\,\,\,\,\,\,\,\,\,\,\,\,\,\,\,\,\,\,\,\,\,\,\,\,\,\,\,\,\,\,\,\,\,\,\,\,\, + \int\limits_0^1 {(1 - \xi ){\mathbf{U}}_z^\xi  {\text{[}}t + h,t{\text{]}}} \frac{{{\mathbf{Q}}_z {\text{[}}t + h,t{\text{]}}}}
{h}^2 \Phi _z^t d\xi.  \hfill \\ 
\end{gathered} 
\]
It follows that
\[
\left\| {\frac{{\left( {{\mathbf{U}}_z {\text{[}}t + h,t{\text{]}} - {\text{I}}} \right)}}
{h}\Phi _z^t  - {\mathcal{A}}_z (t)\Phi _z^t } \right\|_ \otimes   \leqslant \left\| {\frac{{{\mathbf{Q}}_z {\text{[}}t + h,t{\text{]}}}}
{h}\Phi _z^t  - {\mathcal{A}}_z (t)\Phi _z^t } \right\|_ \otimes   + \frac{1}
{2}\left\| {\frac{{{\mathbf{Q}}_z {\text{[}}t + h,t{\text{]}}}}
{h}^2 \Phi _z^t } \right\|_ \otimes.  
\]
The result now follows from Theorem 5.12, (2) and (3).  To prove (3), note that $
{\mathcal{A}}_z (t{\text{)}}\Phi  = {\mathcal{A}}{\text{(}}t{\text{)}}\left\{ {z{\mathbf{R}}(z, {\mathcal{A}}{\text{(}}t{\text{)}})} \right\}\Phi {\text{ = }}\left\{ {z{\mathbf{R}}(z, {\mathcal{A}}{\text{(}}t{\text{)}})} \right\} {\mathcal{A}}{\text{(}}t{\text{)}}\Phi, 
$
so that $
\left\{ {z{\mathbf{R}}(z, {\mathcal{A}}{\text{(}}t{\text{)}})} \right\}
$ commutes with ${\mathbf{U}}{\text{[}}t,a{\text{]}}$ and $ {\mathcal{A}}{\text{(}}t{\text{)}}$.  It is now easy to show that
\[
\begin{gathered}
  \left\| { {\mathcal{A}}_z {\text{(}}t{\text{)}}{\mathbf{U}}_z {\text{[}}t,a{\text{]}}\Phi  - {\mathcal{A}}_{z'} {\text{(}}t{\text{)}}{\mathbf{U}}_{z'} {\text{[}}t,a{\text{]}}\Phi } \right\| \hfill \\
  \;\;{\text{              }} \leqslant \left\| {{\mathbf{U}}_z {\text{[}}t,a{\text{]}}\left( { {\mathcal{A}}_z {\text{(}}t{\text{)}} - {\mathcal{A}}_{z'} {\text{(}}t{\text{)}}} \right)\Phi } \right\| + \left\| {z'{\mathbf{R}}(z', {\mathcal{A}}{\text{(}}t{\text{)}})\left[ {{\mathbf{U}}_z {\text{[}}t,a{\text{]}}\Phi  - {\mathbf{U}}_{z'} {\text{[}}t,a{\text{]}}} \right] {\mathcal{A}}{\text{(}}t{\text{)}}\Phi } \right\| \hfill \\
  \;\;{\text{         }} \leqslant M\left\| {\left( { {\mathcal{A}}_z {\text{(}}t{\text{)}} - {\mathcal{A}}_{z'} {\text{(}}t{\text{)}}} \right)\Phi } \right\| + M\left\| {\left[ {{\mathbf{U}}_z {\text{[}}t,a{\text{]}}\Phi  - {\mathbf{U}}_{z'} {\text{[}}t,a{\text{]}}} \right] {\mathcal{A}}{\text{(}}t{\text{)}}\Phi } \right\| \to 0,{\text{ }}z,z' \to \infty , \hfill \\ 
\end{gathered} 
\]
so that, for $\Phi  \in {\text{D(}}{\mathbf{Q}}{\text{[}}b,a{\text{])}}$, 
\[
{\mathcal{A}}_z {\text{(}}t{\text{)}}{\mathbf{U}}_z {\text{[}}t,a{\text{]}}\Phi  \to {\mathcal{A}}{\text{(}}t{\text{)}}{\mathbf{U}}{\text{[}}t,a{\text{]}}\Phi  = \frac{\partial }
{{\partial t}}{\mathbf{U}}{\text{[}}t,a{\text{]}}\Phi. 
\]
\end{proof} 
Since, as noted earlier, $exp\{Q[t,a]\}$ does not solve the initial-value problem, we restate the last part of the last theorem to emphasize the importance of this result, and the power of the constructive Feynman theory. 
\begin{thm} If $a < t < b$,
\[
\frac{\partial }
{{\partial t}}{\mathbf{U}}[t,a]\Phi  = {\mathcal{A}}(t){\mathbf{U}}[t,a]\Phi  = {\mathbf{U}}[t,a]{\mathcal{A}}(t)\Phi {\text{, }}\Phi  \in {\text{D}}_0  \subset {\text{D}}({\mathbf{Q}[b,a])}.
\]
\end{thm}
\subsection{{Application: Hyperbolic and Parabolic Evolution Equations}}\paragraph{}
We can now apply the previous results to show that the standard conditions imposed in the study of hyperbolic and parabolic evolution equations imply that the family of operators is strongly continuous (see Pazy [PZ]), so that our condition (\ref{CD: cond}) is automatically satisfied. Let us recall the specific assumptions traditionally assumed in the study of parabolic and hyperbolic evolution equations.  Without loss, we shift the spectrum of $A(t)$ at each $t$, if necessary, to obtain a uniformly bounded family of semigroups. 

\begin{flushleft}
{\bf Parabolic Case}
\end{flushleft}

In the abstract approach to parabolic evolution equations, it is assumed that: 
\begin{enumerate}
\item For each $t \in I,\;A(t)$ generates an analytic $C_0$-semigroup with domains $D(A(t)) = D$ independent of $t$.
\item For each $t \in I,\;R(\lambda ,A(t))$ exists for all $\lambda$ such that $\operatorname{Re} \lambda  \leqslant 0$, and there is an $M > 0$ such that: 
\[
\left\| {R(\lambda ,A(t))} \right\| \leqslant {M \mathord{\left/
 {\vphantom {M {\left[ {\left| \lambda  \right| + 1} \right]}}} \right.
 \kern-\nulldelimiterspace} {\left[ {\left| \lambda  \right| + 1} \right]}}.
\]
\item There exist constants $L$ and $0 < \alpha  \leqslant 1$ such that 
\[
\left\| {\left( {A(t) - A(s)} \right)A(\tau )^{ - 1} } \right\| \leqslant L\left| {t - s} \right|^\alpha\;  {\text{ for all}} \; t,\; s,\; \tau  \in I.
\]
\end{enumerate}
In this case, when (3) is satisfied and $\varphi  \in D$, we have
\[
\begin{gathered}
 \left\| {\left[ {A(t) - A(s)} \right]\varphi } \right\| = \left\| {\left[ {\left( {A(t) - A(s)} \right)A^{ - 1} (\tau )} \right]A(\tau )\varphi } \right\| \hfill \\
  {\text{                       }} \leqslant \left\| {\left( {A(t) - A(s)} \right)A^{ - 1} (\tau )} \right\|\left\| {A(\tau )\varphi } \right\| \leqslant L\left| {t - s} \right|^\alpha  \left\| {A(\tau )\varphi } \right\|, \hfill \\ 
\end{gathered} 
\]
so that the family $A(t),\;t \in I$, is strongly continuous on $D$.  It follows that the time-ordered family ${\mathcal{A}}(t),\;t \in I$, has a strong Riemann integral on ${\text{D}}_{\text{0}}$. 
\begin{flushleft}
{\bf Hyperbolic Case}
\end{flushleft}

In the abstract approach to hyperbolic evolution equations, it is assumed that: 
\begin{enumerate}
\item For each $t \in I,\;A(t)$ generates a $C_0$-semigroup.
\item  For each $t\in I, \;A(t)$ is stable with constants $M,\;0$ and the resolvent set $\rho (A(t)) \supset (0,\infty ),\;t \in I$, such that: 
\[
\left\| {\prod _{j = 1}^k \exp \{ \tau _j A(t_j )\} } \right\| \leqslant M.
\]
\item There exists a Hilbert space ${\mathcal{Y}}$ densely and continuously embedded in ${\mathcal{H}}$ such that, for each $t \in I$, $D(A(t)) \supset {\mathcal{Y}}$ and $A(t) \in L[{\mathcal{Y}},{\mathcal{H}}]$ (i.e., $A(t)$ is bounded as a mapping from ${\mathcal{Y}} \to {\mathcal{H}}$), and the function $ g(t) = \left\| {A(t)} \right\|_{{\mathcal{Y}} \to {\mathcal{H}}}$ is continuous.
\item The space ${\mathcal{Y}}$ is an invariant subspace for each semigroup $S_t (\tau ) = \exp \{ \tau A(t)\}$ and $S_t (\tau )$ is a stable $C_0$-semigroup on ${\mathcal{Y}}$ with the same stability constants.
\end{enumerate}
This case is not as easily analyzed as the parabolic case, so we need the following:
\begin{lem} Suppose conditions (3) and (4) above are satisfied with $
\left\| \varphi  \right\|_{\mathcal{H}}  \leqslant \left\| \varphi  \right\|_{\mathcal{Y}}$.  Then the family $A(t),\;t \in I$, is strongly continuous on ${\mathcal{H}}$ (a.e.) for $t \in I$.
\end{lem}
\begin{proof}  Let $\varepsilon  > 0$ be given and, without loss, assume that $\left\| \varphi  \right\|_{\mathcal{H}}  \leqslant 1$.  Set $c = {{\left\| \varphi  \right\|_{\mathcal{Y}} } \mathord{\left/
 {\vphantom {{\left\| \varphi  \right\|_{\mathcal{Y}} } {\left\| \varphi  \right\|_{\mathcal{H}} }}} \right.
 \kern-\nulldelimiterspace} {\left\| \varphi  \right\|_{\mathcal{H}} }}
$, so that $1 \leqslant c < \infty $.  Now
\[
\begin{gathered}
  \left\| {\left[ {A(t + h) - A(t)} \right]\varphi } \right\|_{\mathcal{H}}  \leqslant \left\{ {{{\left\| {\left[ {A(t + h) - A(t)} \right]\varphi } \right\|_{\mathcal{H}} } \mathord{\left/
 {\vphantom {{\left\| {\left[ {A(t + h) - A(t)} \right]\varphi } \right\|_{\mathcal{H}} } {\left\| \varphi  \right\|_{\mathcal{Y}} }}} \right.
 \kern-\nulldelimiterspace} {\left\| \varphi  \right\|_{\mathcal{Y}} }}} \right\}\left[ {{{\left\| \varphi  \right\|_{\mathcal{Y}} } \mathord{\left/
 {\vphantom {{\left\| \varphi  \right\|_{\mathcal{Y}} } {\left\| \varphi  \right\|_{\mathcal{H}} }}} \right.
 \kern-\nulldelimiterspace} {\left\| \varphi  \right\|_{\mathcal{H}} }}} \right] \hfill \\
  {\text{                                }} \leqslant c\left\| {A(t + h) - A(t)} \right\|_{{\mathcal{Y}} \to {\mathcal{H}}} . \hfill \\ 
\end{gathered} 
\]
Choose $\delta  > 0$ such that $\left| h \right| < \delta$ implies  $ \left\| {A(t + h) - A(t)} \right\|_{{\mathcal{Y}} \to {\mathcal{H}}}  < {\varepsilon  \mathord{\left/
 {\vphantom {\varepsilon  c}} \right.
 \kern-\nulldelimiterspace} c}$, which completes the proof.
\end{proof}
\section{\bf{Perturbation Theory}}
	In this section, we prove a few results  without attempting to be exhaustive. Because of Theorem 3.11(4), the general problem of perturbation theory can always be reduced to that of the strong limit of the bounded case.  Assume that, for each $t \in I$, $A_0 (t) $ and $A_1 (t) $ are generators of $C_0$-semigroups on ${\mathcal{H}}$. The (generalized) sum of $A_0 (t) $ and $A_1 (t) $, in its various forms, whenever it is defined (with dense domain), is denoted by $
A(t) = A_0 (t) \oplus A_1 (t) $ (see Kato \cite{KA}, and Pazy \cite{PZ}).    Let $
A_1^n (t) = n{A_1 (t)}R(n,A_1 (t)) $, be the Yosida approximator for $
A_1 (t) $ and set $
A_n (t) = A_0 (t) + A_1^n (t) $. 
\begin{thm} For each $n$, $A_0 (t) + A_1^n (t) $ (respectively ${\mathcal{A}}_0 (t) + {\mathcal{A}}_1^n (t) $) is the generator of a $C_0$-semigroup on ${\mathcal{H}}$  (respectively ${\mathcal{F}\mathcal{D}}_ \otimes ^2 $) and: 
\begin{enumerate}
\item If, for each $t \in I$, $A_0 (t) $ generates an analytic or contraction $C_0$-semigroup, then so does $A_n (t) $ and ${\mathcal{A}}_n (t) $. 
\item
 If, for each $t \in I$, $A(t) = A_0 (t) \oplus A_1 (t) $ generates an analytic or contraction $C_0$-semigroup, then so does ${\mathcal{A}}(t) = {\mathcal{A}}_0 (t) \oplus {\mathcal{A}}_1 (t) $ and $
\exp \{ \tau {\mathcal{A}}_n (t)\}  \to \exp \{ \tau {\mathcal{A}}(t)\} $  for $
\tau  \geqslant 0$.
\end{enumerate}
\end{thm}
\begin{proof}
The first two parts of (1) are standard (see Pazy [PZ], pp. 79, 81).  The third part (contraction) follows because $A_1^n (t) $  (respectively ${\mathcal{A}}_1^n (t) $) is a bounded m-dissipative operator.  The proof of (2) follows from Theorem 3.11(4), equation (5.3), and Theorem 4.12.  
\end{proof}
We now assume that $A_0 (t) $ and $A_1 (t) $ are weakly continuous generators of $C_0$-semigroups for each $t \in I$, and equation (\ref{CD: cond}) is satisfied.  Then, with the same notation, we have:
\begin{thm} If, for each $t \in I$, $A(t) = A_0 (t) \oplus A_1 (t) $ generates an analytic or contraction semigroup, then ${\mathbf{Q}}[t,a] $ generates an analytic or contraction semigroup and $\exp \{ {\mathbf{Q}}_n [t,a]\}  \to \exp \{ {\mathbf{Q}}[t,a]\} $.
\end{thm}
\begin{proof}
The proof follows from Theorem 5.9 and Theorem 5.14.
\end{proof}
At this point,  we should mention the Trotter product theorem (see Goldstein \cite{GS}, page 44 and references).   
\begin{thm}{(Trotter)} Suppose $A_0, \; A_1 \; {\rm and} \; A_0+A_1$ generates $C_0$-contraction semigroups $S(t), \ T(t), \ U(t) \; {\rm on} \; \mathcal H$.  Then 
\[
\mathop {\lim }\limits_{n \to \infty } \left\{ {S\left( {\tfrac{t}
{n}} \right)T\left( {\tfrac{t}
{n}} \right)} \right\}^n  = U(t).
\]
\end{thm}
\begin{rem} There are cases in which the above limit exists without the assumption that $A_0+A_1$ generates a $C_0$-contraction semigroup.  In fact, it is possible for the limit to exist while $D(A_0) \cap D(A_1)= \{ 0 \}$.  Goldstein \cite{GS} calls the generator $C$ of such a semigroup a generalized or Lie sum and writes it $C= A_0 {\oplus_L} A_1$(see page 57).  Kato \cite{KA1} proves that the limit can exist for an arbitrary pair of selfadjoint contraction semigroups.  The fundamental question is:  what are the general conditions that makes this possible?  
\end{rem}
\begin{thm} Suppose that $A_0 (t) $ and $A_1 (t) $ are weakly continuous generators of $C_0$-contraction semigroups for each $t \in I$, and that equation (5.4) is satisfied.  If ${\mathbf{Q}}_0[t,a]$ and ${\mathbf{Q}}_1[t,a] $ are the corresponding time-ordered generators of contraction semigroups, then 
\[
{\mathbf{Q}}[t,a]= {\mathbf{Q}}_0[t,a]  {\oplus_L} {\mathbf{Q}}_1[t,a] \; (a,s),
\]
is the generator of a contraction semigroup on ${\mathcal{FD}}_{\otimes}^2$.
\end{thm} 
\begin{proof}  Let  ${\mathbf{Q}}_{n,1}[t,a]$  be the Yosida approximator for ${\mathbf{Q}}_1[t,a]$.  It follows that, 
\[
{\mathbf{Q}}_n[t,a]= {\mathbf{Q}}_0[t,a]  + {\mathbf{Q}}_{n,1}[t,a]
\]
is the generator of a $C_0$-contraction semigroup for each $n$.  Furthermore, for any $m, \ n \in \N$ and $\Phi \in D_0$,
\[
\begin{gathered}
  \left\| {\left[ {\exp \{ \tau {\mathbf{Q}}_n [t,a]\}  - \exp \{ \tau {\mathbf{Q}}_m [t,a]\} } \right]\Phi } \right\|_ \otimes   \hfill \\
  = \left\| {\int_0^\tau  {\frac{d}
{{ds}}\left[ { \exp \{ (\tau  - s){\mathbf{Q}}_n [t,a]\} \exp \{ s{\mathbf{Q}}_m [t,a]\} } \right]} \Phi ds} \right\|_ \otimes   \hfill \\
 \le \int_0^\tau  {\left\| {\left[ { \exp \{ (\tau  - s){\mathbf{Q}}_n [t,a]\} \exp \{ s{\mathbf{Q}}_m [t,a]\} \left( {{\mathbf{Q}}_n [t,a] - {\mathbf{Q}}_m [t,a]} \right)  \Phi } \right]} \right\|} _ \otimes   \hfill \\
 \le \int_0^\tau  {\left\| {\left( {{\mathbf{Q}}_n [t,a] - {\mathbf{Q}}_m [t,a]} \right)\Phi } \right\|} _ \otimes ds \longrightarrow 0, \ n \rightarrow \iy.     \hfill \\
\end{gathered} 
\]
Thus, $\exp \{ \tau {\mathbf{Q}}_n {\text{[}}t,a{\text{]}}\} \Phi$ converges as $n \to \infty$ for each fixed $t \in I$; and the convergence is uniform on bounded $\tau$ intervals.  As $\left\| {\exp \{ {\mathbf{Q}}_n {\text{[}}t,a{\text{]}}\} } \right\|_ \otimes   \leqslant 1$, we have 
\[
\lim _{n \to \infty } \exp \{ \tau {\mathbf{Q}}_n [t,a]\} \Phi  = {\mathbf{S}}_t (\tau )\Phi ,{\text{ }}\Phi  \in {\mathcal{F}\mathcal{D}}_ \otimes ^2 .
\]
The limit is again uniform on bounded $\tau$ intervals.  It is easy to see that the limit $
{\mathbf{S}}_t (\tau ) $ satisfies the semigroup property, ${\mathbf{S}}_t (0) = I$, and 
$\left\| {{\mathbf{S}}_t (\tau )} \right\|_ \otimes   \leqslant 1$, so that $S_t(\tau)$ is a $C_0$-contraction semigroup.  Furthermore, as the uniform limit of continuous functions, we see that $\tau  \to {\mathbf{S}}_t (\tau )\Phi$ is continuous for $\tau  \geqslant 0$.  We are done if we show that ${\mathbf{Q}}[t,a] $ is the generator of ${\mathbf{S}}_t (\tau ) $.  For $\Phi  \in D_0 $, we have that
\[
\begin{gathered}
  {\mathbf{S}}_t (\tau )\Phi  - \Phi  = \lim _{n \to \infty } \exp \{ \tau {\mathbf{Q}}_n [t,a]\} \Phi  - \Phi  \hfill \\
  {\text{           }} = \lim _{n \to \infty } \int_0^\tau  {\exp \{ s{\mathbf{Q}}_n [t,a]\} } {\mathbf{Q}}_n [t,a]\Phi ds = \int_0^\tau  {{\mathbf{S}}_t (\tau )} {\mathbf{Q}}[t,a]\Phi ds,\ (a.s). \hfill \\ 
\end{gathered} 
\]
Our result now follows from the uniqueness of the generator, so that $ {\mathbf{Q}}[t,a]$ generates a $C_0$-contraction semigroup. 
\end{proof}
\begin{rem}
It clear that the above result does not depend on domain relationships, as observed by Goldstein, and extends to all contraction generators, in addition to the observation of Kato for selfadjoint operators.  Since a shift in spectrum and an equivalent norm can make any generator a contraction generator, we see that the above is a broad generalization of the Trotter Theorem.
\end{rem} 
\subsection{{Disentanglement and the Trotter-Kato Theory}}
In order to relate the results of the last section to the conventional approach, where the order of operators is determined by their position on paper, in this section we investigate the method of disentanglement suggested by Feynman to relate his theory to the standard theory.  As an application, we extend the conventional Trotter-Kato Theorem.

Since any closed densely defined linear operator may be replaced by its Yosida approximator, when convenient, without loss, we can restrict our study to bounded linear operators. We first need to establish some notation.   If  $\left\{ {A(t),\;t \in I} \right\}$ denotes an arbitrary family of operators in $L[\mathcal{H}]$, the operator $\prod _{t \in I} A(t)$, when defined, is understood in its natural order: $\prod _{b \geqslant t \geqslant a} A(t)$.  Let  $L[{\mathcal{FD}}_ \otimes ^2 ] \subset L^\#  [\mathcal{H}_ \otimes ^2]$ be the class of bounded linear operators on   ${\mathcal{FD}}_ \otimes ^2$. It is easy to see that every operator $\mathcal{A} \in L[{\mathcal{F}\mathcal{D}}_ \otimes ^2]$, which depends on a countable number of elements in $I$, may be written as:
\[
\mathcal{A} = \sum\nolimits_{i = 1}^\infty  {a_i \prod\limits_{k = 1}^{n_i } {\mathcal{A}}_i (t_k )}, 
\]
where 
\[
\mathcal{A}_i (t_k ) \in L[\mathcal{H}(t_k )],\;k = 1,\;2, \cdots ,\;n_i ,\;n_i  \in \mathbb{N}. 
\]
\begin{Def} The disentanglement morphism, $dT[ \, \cdot \, ]$, is a   mapping from $L[{\mathcal{F}\mathcal{D}}_ \otimes ^2]$  to $L[{\mathcal{H}}]$, such that:
\[
dT\left[ \mathcal{A} \right] = dT\left[ {\sum\nolimits_{i = 1}^\infty  {a_i \prod\limits_{k = 1}^{n_i } {\mathcal{A}_i (t_k )} } } \right] = \sum\nolimits_{i = 1}^\infty  {a_i \prod\limits_{n_i  \geqslant k \geqslant 1} {A_i (t_k )}}.
\]
\end{Def}
\begin{thm} 
The map $dT[ \, \cdot \, ]$  is a well-defined onto bounded linear mapping from $L[{\mathcal{F}\mathcal{D}}_ \otimes ^2]$ to $L[{\mathcal{H}}]$, which is not injective and $\left. {dT[ \, \cdot \,  ]} \right|_{L[\mathcal{H}(t)]}  = {\mathbf{T}}_\theta ^{- t}$, where ${\mathbf{T}}_\theta ^t  \circ {\mathbf{T}}_\theta ^{ - t}  = {\mathbf{T}}_\theta ^{ - t}  \circ {\mathbf{T}}_\theta ^t  = {\mathbf{I}}$.
\end{thm}
\begin{proof}  With the stated convention, it is easy to see that $dT[ \, \cdot \, ]$ is a well-defined bounded, surjective linear mapping.  To see that it is not injective, note that $dT[E[t,s]\mathcal{A}(s)] = dT[\mathcal{A}(s)]$, while  $E[t,s]\mathcal{A}(s) \in L[\mathcal{H}(t)]$ and $\mathcal{A}(s) \in L[\mathcal{H}(s)]$, so that these operators are not equal when $t \ne s$.
	To see that $\left. {dT[ \, \cdot \, ]} \right|_{L[{\mathcal{H}}(t)]}  = {\mathbf{T}}_\theta ^{- t}$, we need only show that  $dT[ \, \cdot \, ]$  is injective when restricted to $L[\mathcal{H}(t)]$.  If  $\mathcal{A}(t),\;\mathcal{B}(t) \in L[\mathcal{H}(t)]$ and $dT[  \mathcal{A}  (t)] = dT[\mathcal{B}(t)]$, then $A(t)=B(t)$, by definition of  $dT[ \, \cdot \, ]$, so that  ${\mathcal{A}}(t)= {\mathcal{B} }(t)$ by definition of  $L[\mathcal{H}(t)]$.
\end{proof}
\begin{Def} A Feynman-Dyson algebra (${\mathcal{FD}}$-algebra) over $L[\mathcal{H}(t)]$ for the parameter set $I$ is the quadruple $
\left( {\left\{ {{\mathbf{T}}_\theta ^t ,\;t \in I} \right\},\;L[\mathcal{H}],\;dT[ \, \cdot \, ], \;L[{\mathcal{F}\mathcal{D}}_ \otimes ^2 ]} \right)$. 
\end{Def}
We now show that the ${\mathcal{FD}}$-algebra is universal for time-ordering in the following sense.
\begin{thm} Let $\left\{ {A(t)\left. {} \right|t \in I} \right\} \in L[\mathcal{H}]$
 be any family of operators.  Then the following conditions hold:
\begin{enumerate}

\item The time-ordered operator  $\mathcal{A}(t) \in L[\mathcal{H}(t)]$ and $dT[\mathcal{A}(t)] = A(t),\;t \in I$.

\item  For any family $\left\{ {t_j \left| {} \right.1 \leqslant j \leqslant n,\;n \in \mathbb{N}} \right\}, \ t_j \in I$  (distinct) the map
$\mathop  \times \limits_{n = 1}^\infty  \left( {A(t_n ),\;A(t_{n - 1} ),\; \cdots ,\;A(t_1 )} \right) \to \sum\nolimits_{n = 1}^\infty  {a_n } \prod\limits_{n \geqslant j \geqslant 1} {A(t_j )}$ from  $\mathop  \times \limits_{n = 1}^\infty  \left\{ {\mathop  \times \limits_{j = 1}^n L[\mathcal{H}]} \right\} \to L[\mathcal{H}]$ has a unique factorization through  $L[{\mathcal{F}\mathcal{D}}_ \otimes ^2]$  so that $\sum\nolimits_{n = 1}^\infty  {a_n } \prod\limits_{n \geqslant j \geqslant 1} {A(t_j )}  \in L[\mathcal{H}]$, corresponds to $\sum\nolimits_{n = 1}^\infty  {a_n } \prod\limits_{j = 1}^n {\mathcal{A}(t_j )}$.  
\end{enumerate}
\end{thm}

\begin{proof} ${\mathcal{A}}(t) = {\mathbf{T}}_\theta ^t [{A}(t)]$
 and $dT[{\mathcal{A}}(t)] = {{A}}(t)$, gives (1).  

To prove (2), note that
\[
\Theta :\;\mathop  \times \limits_{n = 1}^\infty  \left\{ {\mathop  \times \limits_{j = 1}^n L[\mathcal{H}]} \right\} \to \mathop  \times \limits_{n = 1}^\infty  \left\{ {\mathop  \times \limits_{j = 1}^n L[\mathcal{H}(t_j )]} \right\},
\]
defined by 
\[
\Theta \left[ {\mathop  \times \limits_{n = 1}^\infty  \left( {A(t_n ),\;A(t_{n - 1} ),\; \cdots ,\;A(t_1 )} \right)} \right] = \mathop  \times \limits_{n = 1}^\infty  \left( {\mathcal{A}(t_n ),\;\mathcal{A}(t_{n - 1} ),\; \cdots ,\;\mathcal{A}(t_1 )} \right),
\]
is bijective and the mapping 

\[
\mathop  \times \limits_{n = 1}^\infty  \left( {\mathcal{A}(t_n ),\;\mathcal{A}(t_{n - 1} ),\; \cdots ,\;\mathcal{A}(t_1 )} \right) \to \sum\nolimits_{n = 1}^\infty  {a_n } \prod\limits_{j = 1}^n {\mathcal{A}(t_j )}
\]
factors though the tensor algebra  $ \oplus _{n = 1}^\infty  \left\{ { \otimes _{j = 1}^n L[\mathcal{H}(t_j )]} \right\}$ via the universal property of that object (see Hu \cite{HU}, pg. 19).  We now note that $ \oplus _{n = 1}^\infty  \left\{ { \otimes _{j = 1}^n L[\mathcal{H}(t_j )]} \right\} \subset L[{\mathcal{F}\mathcal{D}}_ \otimes ^2]$.  In diagram form we have:
\[
\begin{array}{*{20}c}
   {\mathop  \times \limits_{n = 1}^\infty  \left( {A(t_n ),\; \cdots ,\;A(t_1 )} \right) \in \mathop  \times \limits_{n = 1}^\infty  \left\{ {\mathop  \times \limits_{j = 1}^n L[\mathcal{H}]} \right\}} & {} & {\xrightarrow{f}} & {} & {\sum\nolimits_{n = 1}^\infty  {a_n } \prod\limits_{n \geqslant j \geqslant 1}^{} {A(t_j ) \in L[\mathcal{H}]} }  \\
   {} & {} & {} & {} & {}  \\
   {\Theta  \downarrow } & {} & {} & {} & { \uparrow dT}  \\
   {} & {} & {} & {} & {}  \\
   {\mathop  \times \limits_{n = 1}^\infty  \left( {\mathcal{A}(t_n ),\; \cdots ,\;\mathcal{A}(t_1 )} \right) \in \mathop  \times \limits_{n = 1}^\infty  \left\{ {\mathop  \times \limits_{j = 1}^n L[\mathcal{H}(t_j )]} \right\}} & {} & {\xrightarrow{{f_ \otimes  }}} & {} & {\sum\nolimits_{n = 1}^\infty  {a_n } \prod\limits_{j = 1}^n {\mathcal{A}(t_j ) \in L[{\mathcal{F}\mathcal{D}}_ \otimes ^2 ]} }  \\

 \end{array} 
\]
so that $dT \circ f_ \otimes   \circ \Theta  = f$.
\end{proof}

\begin{ex} If $A,B \in L[\mathcal{H}] \ {\rm and} \ s<t$, then $\mathcal{A}(t)\mathcal{B}(s) = \mathcal{B}(s)\mathcal{A}(t)$ and $dT[\mathcal{B}(s)\mathcal{A}(t)] =AB$ while $dT[\mathcal{B}(s)\mathcal{A}(t) -\mathcal{B}(t)\mathcal{A}(s)] =AB-BA$.
\end{ex}
\begin{ex}
Let  $\mathcal{A}(t) = {\mathbf{T}}_\theta ^t [A], \ \mathcal{B}(t) = {\mathbf{T}}_\theta ^t [B], \ {\rm with} \ I=[0,1]$, where $A, \ B$  are the operators in the last example.  Then
\[
\begin{gathered}
  \sum\limits_{k = 1}^n {\Delta t_k \left\| {A(s_k )e^i  - \left\langle {A(s_k )e^i ,e^i } \right\rangle e^i } \right\|^2 }  = (b - a)\left\| {Ae^i  - \left\langle {Ae^i ,e^i } \right\rangle e^i } \right\|^2 , \hfill \\
  \sum\limits_{k = 1}^n {\Delta t_k \left\| {B(s_k )e^i  - \left\langle {B(s_k )e^i ,e^i } \right\rangle e^i } \right\|^2 }  = (b - a)\left\| {Be^i  - \left\langle {Be^i ,e^i } \right\rangle e^i } \right\|^2 , \hfill \\ 
\end{gathered} 
\]
so that the operators are strongly continuous.  Hence, $\int_0^1 {\mathcal{A}(s)ds} ,\;\int_0^1 {\mathcal{B}(s)ds}$  both exist as strong integrals and 
\beqn
 e^{ \int_0^1 {[\mathcal{A}(s) + \mathcal{B}(s)]ds}}  = \exp \{ \int_0^1 {\mathcal{A}(s)ds} \} \exp \{ \int_0^1 {\mathcal{B}(s)ds} \} \ (a.s).
\eeqn
Expanding the right-hand side, we obtain:
\[
\begin{gathered}
  \exp \{ \int_0^1 {\mathcal{A}(s)ds} \} \exp \{ \int_0^1 {\mathcal{B}(s')ds'} \}  = \exp \{ \int_0^1 {\mathcal{A}(s)ds} \} \sum\nolimits_{n = 0}^\infty  {\frac{{\left[ {\int_0^1 {\mathcal{B}(s')ds'} } \right]^n }}
{{n!}}}  \hfill \\
   = \exp \{ \int_0^1 {\mathcal{A}(s)ds} \}  + \exp \{ \int_0^1 {\mathcal{A}(s)ds} \} \int_0^1 {\mathcal{B}(s')ds'}  \hfill \\
  + \tfrac{1}{2}\exp \{ \int_0^1 {\mathcal{A}(s)ds} \} \int_0^1 {\mathcal{B}(s')ds'} \int_0^1 {\mathcal{B}(s'')ds''}  +  \cdots  \hfill \\
   = \exp \{ \int_0^1 {\mathcal{A}(s)ds} \}  + \int_0^1  \exp \{ \int_0^1 {\mathcal{A}(s)ds} \} \mathcal{B}(s')ds' \hfill \\
   + \tfrac{1}
{2}\int_0^1  \int_0^1  \exp \{ \int_0^1 {\mathcal{A}(s)ds} \} \mathcal{B}(s')\mathcal{B}(s'')ds'ds'' +  \cdots . \hfill \\ 
\end{gathered}
\]
Restricting to the second term, we have 
\[
e^{ \int_0^1 {[\mathcal{A}(s) + \mathcal{B}(s)]ds}}  = \exp \{ \int_0^1 {\mathcal{A}(s)ds} \}  + \int_0^1  \exp \{ \int_0^{s'} {\mathcal{A}(s)ds} \} \mathcal{B}(s')\exp \{ \int_{s'}^1 {\mathcal{A}(s)ds} \} ds' +  \cdots .
\]
Thus, to second order, we have:
\[
\begin{gathered}
  \exp \{ A + B\}  = dT\left[ {\exp \{ \int_0^1 {[\mathcal{A}(s) + \mathcal{B}(s)]ds} \} } \right] \hfill \\
   = dT\left[ {\exp \{ \int_0^1 {\mathcal{A}(s)ds} \} } \right] + dT\left[ {\int_0^1  \exp \{ \int_{s'}^1 {\mathcal{A}(s)ds} \} \mathcal{B}(s')\exp \{ \int_0^{s'} {\mathcal{A}(s)ds} \} ds'} \right] +  \cdots  \hfill \\
   = \exp \{ A\}  + \int_0^1  \exp \{ (1 - s)A\} B\exp \{ sA\} ds +  \cdots . \hfill \\ 
\end{gathered} 
\]
\end{ex}
This last example was given by Feynman [F].
\begin{thm}{(Generalized Trotter-Kato)} Suppose $A, \; B \; {\rm and} \; C= A {\oplus_L} B$ generate $C_0$-contraction semigroups $S(t), \ T(t) \ {\rm and} \ U(t) \; {\rm on} \; \mathcal H$.  Then 
\[
\begin{gathered}
  dT\left[ {\exp \{ \int_0^t {\left[ {\mathcal{A}(s) + \mathcal{B}(s)} \right]ds} \} } \right]\mathop { = \lim }\limits_{n \to \infty } dT\left[ {\prod\nolimits_{j=1}^n {\exp \{ \tfrac{t}{n} (\mathcal{A}(\tfrac{{jt }}
{n}) + \mathcal{B}(\tfrac{{jt'_n }}
{n}) ) \} } } \right] \hfill \\
   = \mathop {\lim }\limits_{n \to \infty } dT\left[ {\prod\nolimits_1^n {\exp \{ \tfrac{t}{n}(\mathcal{A}(\tfrac{{jt }}
{n})\} \exp \{ \mathcal{B}(\tfrac{{jt'_n }}
{n}))\} } } \right] = \exp \{ t(A \oplus _L B)\},  \hfill \\ 
\end{gathered} 
\]
where $t'_n= t(1-\tfrac{1}{10^{10}}e^{-(n+1)^2})$.
\end{thm}
\subsection{Interaction Representation}\paragraph{}
The research related to this paper is part of a different point of departure in the investigation of the foundations of relativistic quantum theory (compared to axiomatic or constructive field theory approaches) and therefore considers different problems and questions (see \cite{GJ}  and also \cite{SW}).   However, within the framework of axiomatic field theory, an important theorem of Haag suggests that the interaction representation, used in theoretical physics, does not exist in a rigorous sense (see Streater and Wightman \cite{SW}, pg. 161).  Haag's theorem shows that the equal time commutation relations for the canonical variables of an interacting field are equivalent to those of a free field.  In trying to explain this unfortunate result, Streater and Wightman point out that (see [SW], p. 168) "... What is even more likely in physically interesting quantum field theories is that equal-time commutation relations will make no sense at all; the field might not be an operator unless smeared in time as well as space."  In this section, it is first shown that, if one assumes (as Haag did) that operators act in sharp time, then the interaction representation (essentially) does not exist.  

We know from elementary quantum theory that there is some overlapping of wave packets, so that it is natural to expect smearing in time.  In fact, striking results of a beautiful recent experiment of Lindner et al (see Horwitz \cite{HW} and references therein) clearly shows the effect of quantum interference in time for the wave function of a particle.   In this section, we also show that, if any time smearing is allowed, then the interaction representation is well-defined.

Let us assume that $A_0 (t) $ and $A_1 (t) $ are weakly continuous generators of  $C_0$-unitary groups for each $t \in I$, $A(t) = A_0 (t) \oplus A_1 (t) $ is densely defined and equation (\ref{CD: cond}) is satisfied.  Define ${\mathbf{U}}_n [t,a] $, ${\mathbf{U}}_0 [t,a] $ and ${\mathbf{\bar{U}}}_0^\sigma  [t,a] $ by:
\[
\begin{gathered}
  {\mathbf{U}}_n [t,a] = \exp \{ ( - {i \mathord{\left/
 {\vphantom {i \hbar }} \right.
 \kern-\nulldelimiterspace} \hbar })\int\limits_a^t {[ {\mathcal{A}}_0 (s)}  + {\mathcal{A}}_1^n (s)]ds\} , \hfill \\
  {\mathbf{U}}_0 [t,a] = \exp \{ ( - {i \mathord{\left/
 {\vphantom {i \hbar }} \right.
 \kern-\nulldelimiterspace} \hbar })\int\limits_a^t { {\mathcal{A}}_0 (s)} ds\} , \hfill \\
  {\mathbf{\bar U}}_0 [t,a] = \exp \{ ( - {i \mathord{\left/
 {\vphantom {i \hbar }} \right.
 \kern-\nulldelimiterspace} \hbar })\int\limits_a^t {{\mathbf{E}}[t,s] {\mathcal{A}}_0 (s)} ds\},  \hfill \\ 
\end{gathered} 
\]
where ${\mathbf{E}}[t,s] $ is the standard exchange operator (see Definition 5.4 and Theorem 5.5).  There are other possibilities. For example, we could replace $ {\mathbf{\bar{U}}}_0 [t,a] $ by ${\mathbf{\bar{U}}}_0^\sigma  [t,a] $, where
\[
\begin{gathered}
{\mathbf{\bar U}}_0^\sigma  [t,a] = \exp \{ ( - {i \mathord{\left/
 {\vphantom {i \hbar }} \right.
\kern-\nulldelimiterspace} \hbar })\int\limits_a^t { {\hat{\mathcal{A}}}_0^\sigma  (s)} ds\} , \hfill \\
{\mathcal{\hat{A}}}_0^\sigma  (t) = \int\limits_{ - \infty }^\infty  {\rho _\sigma  (t,s){\mathbf{E}}[t,s]{\mathcal{A}}_0 (s)} ds, \hfill \\ 
\end{gathered} 
\]
where $\rho _\sigma  (t,s) $ is a smearing density that may depend on a small parameter 
$\sigma $ with $\int_{ - \infty }^\infty  {\rho _\sigma  (t,s)ds = 1} $ (for example, $
\rho _\sigma  (t,s) = [{1 \mathord{\left/
 {\vphantom {1 {\sqrt {2\pi \sigma ^2 } }}} \right.
 \kern-\nulldelimiterspace} {\sqrt {2\pi \sigma ^2 } }}]\exp \{ {{-(t - s)^2 } \mathord{\left/
 {\vphantom {{(t - s)^2 } {2\sigma ^2 }}} \right.
 \kern-\nulldelimiterspace} {2\sigma ^2 }}\} 
$).   

In the first case, using ${\mathbf{U}}_0 [t,a] $, the interaction representation for $ {\mathcal{A}}_1^n (t) $ is given by:
\[
{\mathcal{A}}_{\mathbf{I}}^n (t) = {\mathbf{U}}_0 [a,t] {\mathcal{A}}_{\text{1}}^n (t){\mathbf{U}}_0 [t,a] = {\mathcal{A}}_{\text{1}}^n (t), (a.s)
\]
as ${\mathcal{A}}_1^n (t) $ commutes with ${\mathbf{U}}_0 [a,t] $ in sharp time.  Thus, the interaction representation does not exist.  In the first of the last two possibilities, we have 
\[
{\mathcal{A}}_{\mathbf{I}}^n (t) = {\mathbf{\bar{U}}}_0 [a,t] {\mathcal{A}}_{\text{1}}^n (t){\mathbf{\bar{U}}}_0 [t,a],
\]
and the terms do not commute.  If we set $
\Psi _n^{} (t) = {\mathbf{\bar U}}_0 [a,t]{\mathbf{U}}_n [t,a]\Phi 
$, we have
\[
\begin{gathered}
  \frac{\partial }
{{\partial t}}\Psi _n (t) = \frac{i}
{\hbar }{\mathbf{\bar{U}}}_0 [a,t] {\mathcal{A}}_0 (t){\mathbf{U}}_n [t,a]\Phi  - \frac{i}
{\hbar }{\mathbf{\bar{U}}}_0 [a,t]\left[ { {\mathcal{A}}_0 (t) + {\mathcal{A}}_1^n (t)} \right]{\mathbf{U}}_n [t,a]\Phi  \hfill \\
 {\text{so that}}\quad  \frac{\partial }
{{\partial t}}\Psi _n (t) = -\frac{i}
{\hbar }\{ {\mathbf{\bar{U}}}_0 [a,t] {\mathcal{A}}_1^n (t){\mathbf{\bar{U}}}_0 [t,a]\} {\mathbf{\bar{U}}}_0 [a,t]{\mathbf{U}}_n [t,a]\Phi  \hfill \\
{\text{and}}\quad  i\hbar \frac{\partial }
{{\partial t}}\Psi _n (t) = {\mathcal{A}}_{\mathbf{I}}^n (t)\Psi _n (t),\;\Psi _n (a) = \Phi . \hfill \\ 
\end{gathered} 
\]
With the same conditions as Theorem 7.2, we have
\begin{thm} If $Q_1 [t,a] = \int_a^t {A_1 (s)ds}$  generates a $C_0$-unitary group on $ {\mathcal{H}}$, then the time-ordered integral $
{\mathbf{Q}}_{\mathbf{I}} [t,a] = \int_a^t { {\mathcal{A}}_{\mathbf{I}} (s)ds}$, where ${\mathcal{A}}_{\mathbf{I}} (t) = {\mathbf{\bar{U}}}_0 [a,t] {\mathcal{A}}_1 (t){\mathbf{\bar{U}}}_0  [t,a] $ generates a $C_0$ unitary group on ${\mathcal{F}\mathcal{D}}_ \otimes ^2 $, and 
\[
\exp \{ ( - {i \mathord{\left/
 {\vphantom {i \hbar }} \right.
 \kern-\nulldelimiterspace} \hbar }){\mathbf{Q}}_{\mathbf{I}}^n [t,a]\}  \to \exp \{ ( - {i \mathord{\left/
 {\vphantom {i \hbar }} \right.
 \kern-\nulldelimiterspace} \hbar }){\mathbf{Q}}_{\mathbf{I}} [t,a]\}, 
\]
where $
{\mathbf{Q}}_{\mathbf{I}}^n [t,a] = \int_a^t { {\mathcal{A}}_{\mathbf{I}}^n (s)ds}$, and:
\[
i\hbar \frac{\partial }
{{\partial t}}\Psi (t) = {\mathcal{A}}_{\mathbf{I}} (t)\Psi (t),\;\Psi (a) = \Phi. 
\]
\end{thm}
\begin{proof} The result follows from an application of Theorem 7.2.
\end{proof}
\begin{Def}The evolution operator ${\mathbf{U}}^w {\text{[}}t,a{\text{] = exp}}\left\{ {w{\mathbf{Q}}[t,a]} \right\}$ is said to be asymptotic in the sense of Poincar\'{e} if, for each $n$ and each $
\Phi _a  \in D\left[ {\left( {{\mathbf{Q}}[t,a]} \right)^{n + 1} } \right] $, we have 
\beqn
\mathop {\lim }\limits_{w \to 0} w^{ - (n + 1)} \left\{ {{\mathbf{U}}_{}^w {\text{[}}t,a{\text{]}} - \sum\limits_{k = 1}^n {\frac{{\left( {w{\mathbf{Q}}[t,a]} \right)^k }}
{{k!}}} } \right\}\Phi _a  = \frac{{{\mathbf{Q}}[t,a]^{n + 1} }}
{{(n + 1)!}}\Phi _a .
\eeqn
	This is the operator version of an asymptotic expansion in the classical sense, but $
{\mathbf{Q}}[t,a]$ is now an unbounded operator. 
\end{Def}
\begin{thm} Suppose that ${\mathbf{Q}}[t,a] $  generates a contraction $C_0$-semigroup on 
${\mathcal{F}\mathcal{D}}_ \otimes ^2 $ for each $
t \in I$.  Then: 
\item the operator ${\mathbf{U}}^w [t,a] = exp\left\{ {w{\mathbf{Q}}[t,a]} \right\}
$ is asymptotic in the sense of Poincar\'{e}. 
\item For each n and each $\Phi _a  \in D\left[ {\left( {{\mathbf{Q}}[t,a]} \right)^{n + 1} } \right] $, we have 
\beqn
\begin{gathered}
  \Phi (t) = \Phi _a  + \sum\limits_{k = 1}^n {w^k \int\limits_a^t {ds_1 } \int\limits_a^{s_1 } {ds_2 }  \cdots \int\limits_a^{s_{k - 1} } {ds_k } {\mathcal{A}}(s_1 ) {\mathcal{A}}(s_2 ) \cdots {\mathcal{A}}(s_k )} \Phi _a  \hfill \\
  {\text{   }} + \int\limits_0^w {(w - \xi )^n d\xi } \int\limits_a^t {ds_1 } \int\limits_a^{s_1 } {ds_2 }  \cdots \int\limits_a^{s_n } {ds_{n + 1} } {\mathcal{A}}(s_1 ) {\mathcal{A}}(s_2 ) \cdots {\mathcal{A}}(s_{n + 1} ){\mathbf{U}}^\xi  {\text{[}}s_{n + 1} ,a{\text{]}}\Phi _a , \hfill \\ 
\end{gathered} 
\eeqn
where $\Phi (t) = {\mathbf{U}}^w [t,a]\Phi _a $.
\end{thm}
\begin{rem}
The above case includes all generators of $C_0$-unitary groups.  Thus,  the theorem provides a precise formulation and proof of Dyson's second conjecture for quantum electrodynamics that, in general, we can only expect the expansion to be asymptotic.  Actually, we prove more in that we produce the remainder term, so that the above perturbation expansion is exact for all finite $n$.
\end{rem} 
\begin{proof}  From Theorem 6.1, we have 
		\[
{\mathbf{U}}^w [t,a]\Phi  = \left\{ {\sum\limits_{k = 0}^n {\frac{{\left( {w{\mathbf{Q}}[t,a]} \right)^k }}
{{k!}}}  + \frac{1}
{{n!}}\int\limits_0^w {(w - \xi )^n } {\mathbf{Q}}[t,a]^{n + 1} {\mathbf{U}}_{}^\xi  [t,a]d\xi } \right\}\Phi ,
\]
so that 
\[
w^{ - (n + 1)} \left\{ {{\mathbf{U}}^w [t,a]\Phi _a  - \sum\limits_{k = 0}^n {\frac{{\left( {w{\mathbf{Q}}[t,a]} \right)^k }}
{{k!}}\Phi _a } } \right\} =  + \frac{{(n + 1)}}
{{(n + 1)!}}w^{ - (n + 1)} \int\limits_0^w {(w - \xi )^n d\xi } {\mathbf{U}}_{}^\xi  [t,a]{\mathbf{Q}}[t,a]^{n + 1} \Phi _a .
\]
Replace the right-hand side by 
\[
\begin{gathered}
  {\mathbf{I}} = {\frac{(n + 1)}{(n + 1)!}} w^{- (n + 1)}
 \int\limits_0^w 
 {
 (w - \xi )^n d\xi \left\{ 
 {
 {\mathbf{U}}_z^\xi  [t,a] +
  \left[ {{\mathbf{U}}^\xi} [t,a] - {\mathbf{U}}_z^\xi [t,a] \right]
  } 
  \right\}
  } 
  {\mathbf{Q}}[t,a]^{n + 1} \Phi _a  \hfill \\
  {\text{                             }} = {\mathbf{I}}_{1, z}  + {\mathbf{I}}_{2, z} , \hfill \\ 
\end{gathered} 
\]
where 
\[
{\mathbf{I}}_{1, z}  = \frac{{(n + 1)}}
{{(n + 1)!}}w^{ - (n + 1)} \int\limits_0^w {(w - \xi )^n d\xi {\mathbf{U}}_{z}^\xi [t,a] {\mathbf{Q}}[t,a]^{n + 1} \Phi _a}, 
\]
and
\[
{\mathbf{I}}_{2,{\text{z}}}  = \frac{{(n + 1)}}
{{(n + 1)!}}w^{ - (n + 1)} \int\limits_0^w {(w - \xi )^n d\xi \left[ {{\mathbf{U}}_{}^\xi  {\text{[}}t,a{\text{]}} - {\mathbf{U}}_{\text{z}}^\xi  {\text{[}}t,a{\text{]}}} \right]} {\mathbf{Q}}[t,a]^{n + 1} \Phi _a. 
\]
From the proof of Theorem 6.1, we see that $\lim _{z \to \infty } {\mathbf{I}}_{2, z}  = 0$.  Let $
\varepsilon  > 0$ be given and choose $Z$ such that $
{\text{z}} > {\text{Z}} \Rightarrow \left\| {{\mathbf{I}}_{2,{\text{z}}} } \right\| < \varepsilon. 
$
  Now, use 
\[
{\mathbf{U}}_{\text{z}}^\xi  {\text{[}}t,a{\text{] = I}}_ \otimes   + \sum\nolimits_{k = 1}^\infty  {\frac{{\xi ^k {\mathbf{Q}}_{\text{z}}^{\text{k}} {\text{[}}t,a{\text{]}}}}
{{k!}}} 
\]
  for the first term to get that 
\[
{\mathbf{I}}_{1, z}  = \frac{{(n + 1)}}
{{(n + 1)!}}w^{ - (n + 1)} \int\limits_0^w {(w - \xi )^n d\xi \left\{ {{\text{I}}_ \otimes   + \sum\nolimits_{k = 1}^\infty  {\frac{{\xi ^k {\mathbf{Q}}_{z}^{k} [t,a]}}
{{k!}}} } \right\}} {\mathbf{Q}}[t,a]^{n + 1} \Phi _a. 
\]
If we compute the elementary integrals, we get 
\[
\begin{gathered}
  {\mathbf{I}}_{1,{\text{z}}}  = \frac{1}
{{(n + 1)!}}{\mathbf{Q}}[t,a]^{n + 1} \Phi _a  \hfill \\
  {\text{                }} + \sum\nolimits_{{\text{k = 1}}}^\infty  {\frac{1}
{{k!n!}}\left\{ {\sum\nolimits_{l = 1}^n {\left( {\begin{array}{*{20}c}
   n  \\
   l  \\
\end{array} } \right)\frac{{w^k }}
{{\left( {n + k + 1 - l} \right)}}} } \right\}{\mathbf{Q}}_{\text{z}}^k {\text{[}}t,a{\text{]}}} {\mathbf{Q}}[t,a]^{n + 1} \Phi _a. \hfill \\ 
\end{gathered} 
\]
Then
\[
\begin{gathered}
  \left\| {{\mathbf{I}} - \frac{1}
{{(n + 1)!}}{\mathbf{Q}}[t,a]^{n + 1} \Phi _a } \right\| <  \hfill \\
  {\text{                    }}\left\| {\sum\nolimits_{{\text{k = 1}}}^\infty  {\frac{1}
{{k!n!}}\left\{ {\sum\nolimits_{l = 1}^n {\left( {\begin{array}{*{20}c}
   n  \\
   l  \\

 \end{array} } \right)\frac{{w^k }}
{{\left( {n + k + 1 - l} \right)}}} } \right\}{\mathbf{Q}}_{\text{z}}^k {\text{[}}t,a{\text{]}}} {\mathbf{Q}}[t,a]^{n + 1} \Phi _a } \right\| + \varepsilon . \hfill \\ 
\end{gathered} 
\]
Now let $w \to 0$ to get
\[
\left\| {{\mathbf{I}} - \frac{1}
{{(n + 1)!}}{\mathbf{Q}}[t,a]^{n + 1} \Phi _a } \right\| < \varepsilon. 
\]
Since $\varepsilon$ is arbitrary, $
{\mathbf{U}}{\text{[}}t,a{\text{] = exp}}\left\{ {{\mathbf{Q}}[t,a]} \right\}
$ is asymptotic in the sense of Poincar\'{e}.  

To prove (7.2), let $
\Phi _a  \in D\left[ {\left( {{\mathbf{Q}}[t,a]} \right)^{n + 1} } \right]
$ for each ${\text{k}} \leqslant {\text{n + 1,}} $ and use the fact that (Dollard and Friedman [DF])
\beqn
\begin{gathered}
  \left( {{\mathbf{Q}}_{\text{z}} [t,a]} \right)^k \Phi _a  = \left( {\int\limits_a^t {{\mathcal{A}}_{\text{z}} (s)} ds} \right)^k \Phi _a  \hfill \\
  {\text{      }} = (k!)\int\limits_a^t {ds_1 } \int\limits_a^{s_1 } {ds_2 }  \cdots \int\limits_a^{s_{k - 1} } {ds_n } {\mathcal{A}}_{\text{z}} (s_1 ){\mathcal{A}}_{\text{z}} (s_2 ) \cdots {\mathcal{A}}_{\text{z}} (s_k )\Phi _a . \hfill \\ 
\end{gathered} 
\eeqn
Letting $z \to \infty $ gives the result.
\end{proof}
	There are special cases in which the perturbation series may actually converge to the solution. It is known that, if $A_0 (t) $ is a nonnegative selfadjoint operator on $
{\mathcal{H}}$, then $\exp \{  - \tau A_0 (t)\} $ is an analytic $C_0$-contraction semigroup for $\operatorname{Re} \tau  > 0$  (see Kato [KA], pg. 491).  More generally, if $
\Delta  = \{ z \in {\mathbf{C}}\,:\;\varphi _1  < \arg z < \varphi _2 ,\;\varphi _1  < 0 < \varphi _2 \}$ and $z \in \Delta$, suppose that $T(z) $ is a bounded linear operator on ${\mathcal{H}}$.
\begin{Def} The family $T(w) $ is said to be an analytic semigroup on ${\mathcal{H}}$, for $w \in \Delta $, if 
\begin{enumerate}
\item $T(w)f$ is an analytic function of $w \in \Delta $ for each $f$ in ${\mathcal{H}}$,
\item $T(0) = I$ and $\lim _{w \to 0} T(w)f = f$ for every $f \in {\mathcal{H}}$,
\item $T(w_1  + w_2 ) = T(w_1 )T(w_2 ) $ for $w_1 ,\;w_2  \in \Delta $.
\end{enumerate}
\end{Def}
For a proof of the next theorem, see Pazy [PZ], page 61.
\begin{thm} Let $A_0 $ be a closed densely defined linear operator defined on ${\mathcal{H}}$, satisfying:
\begin{enumerate}
\item For some $0 < \delta  < {\pi  \mathord{\left/  {\vphantom {\pi  2}} \right.
 \kern-\nulldelimiterspace} 2}$, 
\[
\rho (A_0 ) \supset \Sigma _\delta   = \{ \lambda \,:\;\left| {\arg \lambda } \right| < {\pi  \mathord{\left/
 {\vphantom {\pi  2}} \right.
 \kern-\nulldelimiterspace} 2} + \delta \}  \cup \{ 0\}. 
\]
\item There is a constant $M$ such that,
\[
\left\| {R(\lambda \,:\;A_0 )} \right\| \leqslant {M \mathord{\left/
 {\vphantom {M {\left| \lambda  \right|}}} \right.
 \kern-\nulldelimiterspace} {\left| \lambda  \right|}}
\]
for $\lambda  \in \Sigma _\delta$,	$\lambda  \ne 0$.
\end{enumerate}
Then $A_0$ is the infinitesimal generator of a uniformly bounded analytic semigroup $T(w)$ for $w \in \bar{\Delta} _{\delta '}  = \{ w\,:\;\left| {\arg w} \right| \leqslant \delta ' < \delta \}$.  Furthermore, for $s > 0$ and $\left| {w - s} \right| \leqslant Cs$ for some constant $C$, 
\[
T(w + s) = T(s) + \sum\nolimits_{n = 1}^\infty  {{{(w^n } \mathord{\left/
 {\vphantom {{(w^n } {n!}}} \right.
 \kern-\nulldelimiterspace} {n!}})T^{(n)} (s)}, 
\]
and the series converges uniformly.	
\end{thm}
\begin{thm} Let 
${\mathbf{Q}}_{\text{0}} [t,a] = \int\limits_a^t { {\mathcal{A}}_{\text{0}} (s)} ds
$ and 
${\mathbf{Q}}_{\text{1}} [t,a] = \int\limits_a^t { {\mathcal{A}}_{\text{1}} (s)} ds
$ be nonnegative selfadjoint generators of analytic $C_0$-contraction semigroups  for $t \in (a,b] $.  Suppose $
{\text{D}}({\mathbf{Q}}_{\text{1}} [t,a]) \supseteq {\text{D}}({\mathbf{Q}}_{\text{0}} [t,a])$ and there are positive constants $\alpha ,\;\beta$ such that 
\beqn
\left\| {{\mathbf{Q}}_{\text{1}} [t,a]\Phi } \right\|_ \otimes   \leqslant \alpha \left\| {{\mathbf{Q}}_{\text{0}} [t,a]\Phi } \right\|_ \otimes   + \beta \left\| \Phi  \right\|_ \otimes  ,\;\Phi  \in {\text{D}}({\mathbf{Q}}_{\text{0}} [t,a]).
\eeqn
\begin{enumerate}
\item  Then ${\mathbf{Q}}[t,a] = {\mathbf{Q}}_{\text{0}} [t,a] + {\mathbf{Q}}_{\text{1}} [t,a] $ and ${\mathcal{A}}_{\mathbf{I}} (t) = {\mathbf{\bar U}}_0 [a,t] {\mathcal{A}}_1 (t){\mathbf{\bar{U}}}_0 [t,a]
$ both generate analytic $C_0$-contraction semigroups. 
\item For each $k$ and each $
\Phi _a  \in D\left[ {\left( {{\mathbf{Q}}_{\mathbf{I}} [t,a]} \right)^{k + 1} } \right] 
$, we have that  
\beqa
\begin{gathered}
  {\mathbf{U}}_{\mathbf{I}}^w [t,a]\Phi _a  = \Phi _a  + \sum\limits_{l = 1}^k {w^l \int\limits_a^t {ds_1 } \int\limits_a^{s_1 } {ds_2 }  \cdots \int\limits_a^{s_{k - 1} } {ds_k } {\mathcal{A}}_{\mathbf{I}} (s_1 ) {\mathcal{A}}_{\mathbf{I}} (s_2 ) \cdots {\mathcal{A}}_{\mathbf{I}} (s_k )} \Phi _a  \hfill \\
  {\text{     }} + \int\limits_0^w {(w - \xi )^k d\xi } \int\limits_a^t {ds_1 } \int\limits_a^{s_1 } {ds_2 }  \cdots \int\limits_a^{s_k } {ds_{k + 1} }  {\mathcal{A}}_{\mathbf{I}} (s_1 ) {\mathcal{A}}_{\mathbf{I}} (s_2 ) \cdots  {\mathcal{A}}_{\mathbf{I}} (s_{k + 1} ){\mathbf{U}}_{\mathbf{I}}^\xi  {\text{[}}s_{k + 1} ,a{\text{]}}\Phi _a . \hfill \\ 
\end{gathered} 
\eeqa
\item  If $
\Phi _a  \in  \cap _{k \geqslant 1} D\left[ {\left( {{\mathbf{Q}}_{\mathbf{I}} [t,a]} \right)^k } \right] $ and $w$ is small enough, we have  
\beqa
{\mathbf{U}}_{\mathbf{I}}^w [t,a]\Phi _a  = \Phi _a  + \sum\limits_{k = 1}^\infty  {w^l \int\limits_a^t {ds_1 } \int\limits_a^{s_1 } {ds_2 }  \cdots \int\limits_a^{s_{k - 1} } {ds_k } {\mathcal{A}}_{\mathbf{I}} (s_1 ) {\mathcal{A}}_{\mathbf{I}} (s_2 ) \cdots {\mathcal{A}}_{\mathbf{I}} (s_k )} \Phi _a .
\eeqa
\end{enumerate}
\end{thm}
\begin{proof} To prove (1), use the fact that $
{\mathbf{Q}}_{\text{0}} [t,a] $ generates an analytic $C_0$-contraction semigroup to find a sector $
\Sigma $ in the complex plane, with $\rho ({\mathbf{Q}}_{\text{0}} [t,a]) \supset \Sigma$  ($
\Sigma  = \{ \lambda :\;\left| {\arg \lambda } \right| < {\pi  \mathord{\left/
 {\vphantom {\pi  2}} \right.
 \kern-\nulldelimiterspace} 2} + \delta '\}$, for some $\delta ' > 0$, and for $\lambda  \in \Sigma $, \[
\left\| {R(\,\lambda \,:\,{\mathbf{Q}}_{\text{0}} [t,a])} \right\|_ \otimes   \leqslant \left| \lambda  \right|^{ - 1}. 
\]
From (7.4), ${\mathbf{Q}}_{\text{1}} [t,a]R(\,\lambda \,:\,{\mathbf{Q}}_{\text{0}} [t,a]) $ is a bounded operator and:
\[
\begin{gathered}
  \left\| {{\mathbf{Q}}_{\text{1}} [t,a]R(\,\lambda \,:\,{\mathbf{Q}}_{\text{0}} [t,a])\Phi } \right\|_ \otimes   \leqslant \alpha \left\| {{\mathbf{Q}}_{\text{0}} [t,a]R(\,\lambda \,:\,{\mathbf{Q}}_{\text{0}} [t,a])\Phi } \right\|_ \otimes   + \beta \left\| {R(\,\lambda \,:\,{\mathbf{Q}}_{\text{0}} [t,a])\Phi } \right\|_ \otimes   \hfill \\
  {\text{                                  }} \leqslant \alpha \left\| {\left[ {R(\,\lambda \,:\,{\mathbf{Q}}_{\text{0}} [t,a]) - {\mathbf{I}}} \right]\Phi } \right\|_ \otimes   + \beta \left| \lambda  \right|^{ - 1} \left\| \Phi  \right\|_ \otimes   \hfill \\
  {\text{                                  }} \leqslant 2\alpha \left\| \Phi  \right\|_ \otimes   + \beta \left| \lambda  \right|^{ - 1} \left\| \Phi  \right\|_ {\otimes}.   \hfill \\ 
\end{gathered} 
\]
Thus, if we set $
\alpha  = {1 \mathord{\left/
 {\vphantom {1 4}} \right.
 \kern-\nulldelimiterspace} 4}
$ and $\left| \lambda  \right| > 2\beta$, we have \[
\left\| {{\mathbf{Q}}_{\text{1}} [t,a]R(\,\lambda \,:\,{\mathbf{Q}}_{\text{0}} [t,a])} \right\|_ \otimes   < 1,
\]
 and it follows that the operator \[
{\mathbf{I}} - {\mathbf{Q}}_{\text{1}} [t,a]R(\,\lambda \,:\,{\mathbf{Q}}_{\text{0}} [t,a])
\]
 is invertible.  Now it is easy to see that: 
\beqa
\left( {\lambda {\mathbf{I}} - ({\mathbf{Q}}_{\text{0}} [t,a] + {\mathbf{Q}}_{\text{1}} [t,a])} \right)^{ - 1}  = R(\,\lambda \,:\,{\mathbf{Q}}_{\text{0}} [t,a])\left( {{\mathbf{I}} - {\mathbf{Q}}_{\text{1}} [t,a]R(\,\lambda \,:\,{\mathbf{Q}}_{\text{0}} [t,a])} \right)^{ - 1}. 
\eeqa
It follows that, using $\left| \lambda  \right| > 2\beta$, with $
\left| {\arg \lambda } \right| < {\pi  \mathord{\left/
 {\vphantom {\pi  2}} \right.
 \kern-\nulldelimiterspace} 2} + \delta ''
$ for some $
\delta '' > 0$, and the fact that ${\mathbf{Q}}_{\text{0}} [t,a] $ and $ {\mathbf{Q}}_{\text{1}} [t,a] $ are nonnegative generators,  
\[
\left\| {R(\,\lambda \,:\,{\mathbf{Q}}_{\text{0}} [t,a] + {\mathbf{Q}}_{\text{1}} [t,a])} \right\|_ \otimes   \leqslant \left| \lambda  \right|^{ - 1}. 
\]
Thus, ${\mathbf{Q}}_{\text{0}} [t,a] + {\mathbf{Q}}_{\text{1}} [t,a] $ generates an analytic $C_0$-contraction semigroup.  The proof of (2) follows from Theorem 7.5. Finally, if $w$ is such that $\left| {\arg w} \right| \leqslant \delta ' < \delta$ and $\left| {w - a} \right| \leqslant Ca$ for some constant $C$,  (3) follows from Theorem 7.16 (see Definition 7.15). 
\end{proof}
There are also cases where the series may diverge, but still respond to some summability method.  This phenomenon is well-known in classical analysis.  In field theory, things can be much more complicated.  The book by Glimm and Jaffe \cite{GJ} has a good discussion.

\section{\textbf{Path Integrals: Sum Over Paths}}
As noted earlier, Feynman stated in his book with Hibbs [FH] that the operator calculus is more general than the path integral, and includes it.  In this section, we show that his expectation was indeed warranted.  First we construct (what we call) the experimental evolution operator.  This allows us to rewrite our theory as a sum over paths.  We use a general argument so that the ideas apply to all cases.  Assume that the family $
\{ \tau _1 ,\tau _2 , \cdots ,\tau _n \}$ represents the time positions of $n$ possible measurements of a general system trajectory, as appears on a film of system history.  We assume that information is available beginning at time $T = 0$ and ends at time $T = t$.  Define ${\mathbf{Q}}_E [\tau _1 ,\tau _2 , \cdots ,\tau _n ] $ by 
\beqn
{\mathbf{Q}}_E [\tau _1 ,\tau _2 , \cdots ,\tau _n ] = \sum\limits_{j = 1}^n {\int_{t_{j - 1} }^{t_j } {E[\tau _j ,s]} {\mathcal{A}}(s)ds}. 
\eeqn
Here, $t_0  = \tau _0  = 0$, $t_j  = (1/2)[\tau _j  + \tau _{j + 1} ] $  (for $1 \leqslant j \leqslant n$), and $E[\tau _j ,s] $ is the exchange operator.  The effect of  $E[\tau _j ,s] $ is to concentrate all information contained in $[t_{j - 1} ,t_j ] $ at $\tau _j $, the midpoint of the time interval around $\tau _j $ relative to $\tau _{j - 1}$ and $\tau _{j + 1}$.  We can rewrite $
{\mathbf{Q}}_E [\tau _1 ,\tau _2 , \cdots ,\tau _n ] $
 as 
\beqn
{\mathbf{Q}}_E [\tau _1 ,\tau _2 , \cdots ,\tau _n ] = \sum\limits_{j = 1}^n {\Delta t_j \left[ {\frac{1}
{{\Delta t_j }}\int_{t_{j - 1} }^{t_j } {E[\tau _j ,s]} {\mathcal{A}}(s)ds} \right]}. 
\eeqn
Thus, we have an average over each adjacent interval, with information concentrated at the midpoint.  The evolution operator is given by
\[
U[\tau _1 ,\tau _2 , \cdots ,\tau _n ] = \exp \left\{ {\sum\limits_{j = 1}^n {\Delta t_j \left[ {\frac{1}
{{\Delta t_j }}\int_{t_{j - 1} }^{t_j } {E[\tau _j ,s]} {\mathcal{A}}(s)ds} \right]} } \right\}.
\]
For $\Phi  \in {\mathcal{F}\mathcal{D}}_ \otimes ^2 $, we define the function $
{\mathbf{U}}{\text{[}}N{\text{(}}t),0{\text{]}}\Phi $ by:
\beqn
{\mathbf{U}}{\text{[}}N{\text{(}}t),0{\text{]}}\Phi  = U[\tau _1 ,\tau _2 , \cdots ,\tau _n ]\Phi. 
\eeqn
${\mathbf{U}}{\text{[}}N{\text{(}}t),0{\text{]}}\Phi$ is a $
{\mathcal{F}\mathcal{D}}_ \otimes ^2 $-valued random variable which represents the distribution of the number of measurements, $N(t)$, that are possible up to time t.   In order to relate ${\mathbf{U}}{\text{[}}N{\text{(}}t),0{\text{]}}\Phi$ to actual experimental results, we must compute its expected value.  Let  $\lambda ^{ - 1} $
 denote the smallest time interval in which a measurement can be made, and define ${\mathbf{\bar U}}_\lambda  {\text{[}}t,0{\text{]}}\Phi$ by:
\beqa
{\mathbf{\bar U}}_\lambda  {\text{[}}t,0{\text{]}}\Phi  = {\mathcal{E}}\left[ {{\mathbf{U}}{\text{[}}N{\text{(}}t),0{\text{]}}\Phi } \right] = \sum\limits_{n = 0}^\infty  {\left. { {\mathcal{E}}\left\{ {{\mathbf{U}}{\text{[}}N{\text{(}}t),0{\text{]}}\Phi \left| {N{\text{(}}t) = n} \right.} \right.} \right\}} \Pr ob\left[ {N{\text{(}}t) = n} \right],
\eeqa
where
\[
\mathcal{E}\{ {\mathbf{U}}[N(t),0]\Phi \left| {N(t) = n} \right.\}  = \int_0^t {\frac{{d\tau _1 }}
{t}} \int_0^t {\frac{{d\tau _2 }}
{{t - \tau _1 }}}  \cdots \int_0^t {\frac{{d\tau _n }}
{{t - \tau _{n - 1} }}} {\mathbf{U}}[\tau _n , \cdots \tau _1 ]\Phi  = {\mathbf{\bar U}}_n [t,0]\Phi .
\]
We make the natural assumption that:  (see Gill and Zachary [GZ])
\[
\Pr ob\left[ {N{\text{(}}t) = n} \right] = (n!)^{ - 1} \left( {\lambda t} \right)^n \exp \{  - \lambda t\}. 
\]
	The expected value-integral is of theoretical use and is not easy to compute.  Since we are only interested in what happens when $\lambda  \to \infty$, and, as the mean number of possible measurements up to time t is $\lambda t$, we can take $
\tau _j  = (jt{\kern 1pt} /{\kern 1pt} n),{\text{ }}1 \leqslant j \leqslant n, $  ($
\Delta t_j  = t{\kern 1pt} /{\kern 1pt} n$ for each $n$).  We can now replace $
{\mathbf{\bar{U}}}_n {\text{[}}t{\text{,0]}}\Phi $ by $
{\mathbf{U}}_n {\text{[}}t{\text{,0]}}\Phi $ and, with this understanding, we continue to use $\tau _j $, so that
\beqn
{\mathbf{U}}_n {\text{[}}t{\text{,0]}}\Phi  = \exp \left\{ {\sum\limits_{j = 1}^n {\int_{t_{j - 1} }^{t_j } {E[\tau _j ,s]} {\mathcal{A}}(s)ds} } \right\}\Phi. 
\eeqn
	We define our experimental evolution operator ${\mathbf{U}}_\lambda  {\text{[}}t,0{\text{]}}\Phi$ by 
\beqn
{\mathbf{U}}_\lambda [t,0]\Phi  = \sum\limits_{n = 0}^{\left[\kern-0.15em\left[ {\lambda t} 
 \right]\kern-0.15em\right]
}  {\frac{{\left( {\lambda t} \right)^n }}
{{n!}}\exp \{  - \lambda t\} } {\mathbf{U}}_n {\text{[}}t{\text{,0]}}\Phi. 
\eeqn
We now have the following result, which is a consequence of the fact that Borel summability is regular.
\begin{thm}
Assume that the conditions for Theorem \ref{F: fun} are satisfied.  Then
\beqn
\mathop {\lim }\limits_{\lambda  \to \infty } {\mathbf{\bar U}}_\lambda [t,0]\Phi  = \mathop {\lim }\limits_{\lambda  \to \infty } {\mathbf{U}}_\lambda [t,0]\Phi  = {\mathbf{U}}[t,0]\Phi. 
\eeqn
\end{thm}
Since $\lambda  \to \infty$ implies  $ \lambda ^{ - 1}  \to 0$, this means that the average time between measurements is zero (in the limit) so that we get a continuous path.  It should be observed that this continuous path arises from averaging the sum over an infinite number of (discrete) paths.  The first term in (8.5) corresponds to the path of a system that created no information (i.e., the film is blank).  This event has probability $
\exp \{  - \lambda t\}$ (which approaches zero as $\lambda  \to \infty $).  The $n$-th term corresponds to the path that creates $n$ possible measurements, (with probability $[(\lambda t)^n /n!]\exp \{  - \lambda t\} $) etc.  	
 
 Let $U[t,a] $ be an evolution operator on $L^2 [{\R}^3 ] $, with time-dependent generator $A(t) $, which has a kernel $ {\mathbf{K}}[{\mathbf{x}}(t), t\, ; \, {\mathbf{x}}(s), s] $ such that:  
\[
\begin{gathered}
  {\mathbf{K}}\left[ {{\mathbf{x}}{\text{(}}t{\text{), }}t{\text{; }}{\mathbf{x}}{\text{(}}s{\text{), }}s} \right] = \int_{{\R}^3 }^{} {{\mathbf{K}}\left[ {{\mathbf{x}}{\text{(}}t{\text{), }}t{\text{; }}d{\mathbf{x}}(\tau ){\text{, }}\tau } \right]} {\mathbf{K}}\left[ {{\mathbf{x}}{\text{(}}\tau {\text{), }}\tau {\text{; }}{\mathbf{x}}{\text{(}}s{\text{), }}s} \right], \hfill \\
  U[t,s]\varphi (s) = \int_{{\R}^3 }^{} {{\mathbf{K}}\left[ {{\mathbf{x}}{\text{(}}t{\text{), }}t{\text{; }}d{\mathbf{x}}(s){\text{, }}s} \right]\varphi (s)} . \hfill \\ 
\end{gathered} 
\]
Now let ${\mathcal{H}} = \bf{KS}^2 [{\R}^n ] \supset \bf{L}^2 [{\R}^3 ] $ 
 in the construction of ${\mathcal{F}\mathcal{D}}_ \otimes ^2  \subset {\mathcal{H}}_ \otimes ^2$, let ${\mathbf{U}}{\text{[}}t{\text{,}}s{\text{]}}$ be the corresponding time-ordered version, with kernel $\mathbb{K}_{\mathbf{f}} \left[ {{\mathbf{x}}{\text{(}}t{\text{), }}t{\text{; }}{\mathbf{x}}{\text{(}}s{\text{), }}s} \right] $.  Since ${\mathbf{U}}{\text{[}}t{\text{,}}\tau {\text{]}}{\mathbf{U}}{\text{[}}\tau {\text{,}}s{\text{] = }}{\mathbf{U}}{\text{[}}t{\text{,}}s{\text{]}}$, we have:
\[
\mathbb{K}_{\mathbf{f}} \left[ {{\mathbf{x}}{\text{(}}t{\text{), }}t{\text{; }}{\mathbf{x}}{\text{(}}s{\text{), }}s} \right] = \int_{{\R}^3 }^{} {\mathbb{K}_{\mathbf{f}} \left[ {{\mathbf{x}}{\text{(}}t{\text{), }}t{\text{; }}d{\mathbf{x}}(\tau ){\text{, }}\tau } \right]} \mathbb{K}_{\mathbf{f}} \left[ {{\mathbf{x}}{\text{(}}\tau {\text{), }}\tau {\text{; }}{\mathbf{x}}{\text{(}}s{\text{), }}s} \right].
\]
From our sum over paths representation for ${\mathbf{U}}[t,s]$, we have:
\[
\begin{gathered}
  {\mathbf{U}}[t,s]\Phi (s) = \lim _{\lambda  \to \infty } {\mathbf{U}}_\lambda  [t,s]\Phi (s) \hfill \\
  {\text{        }} = {\text{lim}}_{\lambda  \to \infty } \operatorname{e} ^{ - \lambda \left( {t - s} \right)} \sum\limits_{k = 0}^{\left[\kern-0.15em\left[ {\lambda (t-s)} 
 \right]\kern-0.15em\right]
}  {\frac{{\mathop {\left[ {\lambda \left( {t - s} \right)} \right]}\nolimits^k }}
{{k!}}} {\mathbf{U}}_k [t,s]\Phi (s), \hfill \\ 
\end{gathered} 
\]
where
\[
{\mathbf{U}}_k [t,s]\Phi (s) = \exp \left\{ {( - i/\hbar )\sum\limits_{j = 1}^k {\int_{t_{j - 1} }^{t_j } {{\mathbf{E}}[{{(j} \mathord{\left/
 {\vphantom {{(j} \lambda }} \right.
 \kern-\nulldelimiterspace} \lambda }),\tau ]} {\mathcal{A}}(\tau )d\tau } } \right\}\Phi (s).
\]
As in Section 3.2, we define $\mathbb{K}_{\mathbf{f}} [{\mathcal{D}}_\lambda {\mathbf{x}}{\text{(}}\tau ){\text{ ; }}{\mathbf{x}}{\text{(s)}}] $ by:
\[
\begin{gathered}
  \int_{{\R}^{3[t,s]} } {\mathbb{K}_{\mathbf{f}} [{\mathcal{D}}_{\lambda}  {\mathbf{x}}{\text{(}}\tau ){\text{ ; }}{\mathbf{x}}{\text{(s)}}]}  \hfill \\
  {\text{         }} = :\operatorname{e} ^{ - \lambda \left( {t - s} \right)} \sum\limits_{k = 0}^{\left[\kern-0.15em\left[ {\lambda (t-s)} 
 \right]\kern-0.15em\right]
} {\frac{{\left[ {\lambda (t - s)} \right]^k }}
{{k!}}\left\{ {\prod\limits_{j = 1}^k {\int_{{\R}^3 } {\left. {\mathbb{K}_{\mathbf{f}} [t_j ,{\mathbf{x}}{\text{(}}t_j {\text{)}}\,;\,d{\mathbf{x}}{\text{(}}t_{j - 1} {\text{)}},t_{j - 1} ]} \right|^{({j \mathord{\left/
 {\vphantom {j \lambda }} \right.
 \kern-\nulldelimiterspace} \lambda })} } } } \right\}} , \hfill \\ 
\end{gathered} 
\]	
where $\left[\kern-0.15em\left[ {\lambda (t-s)} 
 \right]\kern-0.15em\right]
 $ is the greatest integer in $\lambda (t - s) $, and $\left. {} \right|^{^{({j \mathord{\left/ {\vphantom {j \lambda }} \right. \kern-\nulldelimiterspace} \lambda })} } $ denotes the fact that the integration is performed in time slot $({j \mathord{\left/ {\vphantom {j \lambda }} \right. \kern-\nulldelimiterspace} \lambda })$.
\begin{Def} We define the Feynman path integral associated with $ {\mathbf{U}}[t,s]$ by:
\[
{\mathbf{U}}[t,s] = \int_{{\R}^{3[t,s]} } {\mathbb{K}_{\mathbf{f}} [{\mathcal{D}}{\mathbf{x}}{\text{(}}\tau ){\text{ ; }}{\mathbf{x}}{\text{(s)}}]}  = \lim _{\lambda  \to \infty } \int_{{\R}^{3[t,s]} } {\mathbb{K}_{\mathbf{f}} [{\mathcal{D}}_{\lambda}  {\mathbf{x}}{\text{(}}\tau ){\text{ ; }}{\mathbf{x}}{\text{(s)}}]}. 
\]
\end{Def}
\begin{thm} For the time-ordered theory, whenever a kernel exists, we have that:
\[
\mathop {\lim }\limits_{\lambda  \to \infty } {\mathbf{U}}_\lambda  [t,s]\Phi (s) = {\mathbf{U}}[t,s]\Phi (s) = \int_{{\R}^{3[t,s]} } {\mathbb{K}_{\mathbf{f}} [{\mathcal{D}} {\mathbf{x}}{\text{(}}\tau ){\text{ ; }}{\mathbf{x}}{\text{(s)}}]\Phi [{\mathbf{x}}{\text{(}}s{\text{)}}]} ,
\]
and the limit is independent of the space of continuous functions. 
\end{thm}\paragraph{}Let us assume that $A_0 (t) $ and $A_1 (t) $ are strongly continuous generators of  $C_0$-contraction semigroups for each  $t \in E = [a,b] $, and let $
{\mathcal{A}}_{1,\rho } (t) = \rho {\mathcal{A}}_1 (t){\mathbf{R}}(\rho, {\mathcal{A}}_1 (t)) $ be the Yosida approximator for the time-ordered version of $
A_1 (t)$.  Define ${\mathbf{U}}^\rho  [t,a] $ and ${\mathbf{U}}^0 [t,a] $ by: 
\[
\begin{gathered}
  {\mathbf{U}}^\rho  [t,a] = \exp \{ ( - {i \mathord{\left/
 {\vphantom {i \hbar }} \right.
 \kern-\nulldelimiterspace} \hbar })\int\limits_a^t {[ {\mathcal{A}}_0 (s)}  + {\mathcal{A}}_{1,\rho } (s)]ds\} , \hfill \\
  {\mathbf{U}}^0 [t,a] = \exp \{ ( - {i \mathord{\left/
 {\vphantom {i \hbar }} \right.
 \kern-\nulldelimiterspace} \hbar })\int\limits_a^t { {\mathcal{A}}_0 (s)} ds\} . \hfill \\ 
\end{gathered} 
\]
Since ${\mathcal{A}}_{1,\rho } (s) $ is bounded, ${\mathcal{A}}_0 (s) + {\mathcal{A}}_{1,\rho } (s) $ is a generator of a $C_0$-contraction semigroup for $s \in E$ and finite $\rho $.  Now assume that ${\mathbf{U}}^0 [t,a] $ has an associated kernel, so that
${\mathbf{U}}^{0} [t,a] = \int_{{\R}^{3[t,s]}}{\mathbb{K}}_{\mathbf{f}} [{\mathcal{D}}{\mathbf{x}}(\tau) ; {\mathbf{x}}(a)]$. 
 We now have the following generalization of the Feynman-Kac Theorem, which is independent of the space of continuous functions.  
\begin{thm} (Feynman-Kac)* If ${\mathcal{A}}_0 (s) \oplus {\mathcal{A}}_1 (s) $ is a generator of a $C_0$-contraction semigroup, then
\beqa
\begin{gathered}
  \mathop {\lim }\limits_{\rho  \to \infty } {\mathbf{U}}^\rho  [t,a]\Phi (a) = {\mathbf{U}}[t,a]\Phi (a) \hfill \\
   = \int_{{\R}^{3[t,a]} } {\mathbb{K}_{\mathbf{f}} [{\mathcal{D}}{\mathbf{x}}{\text{(}}\tau ){\text{ ; }}{\mathbf{x}}{\text{(a)}}]\exp \{ ( - {i \mathord{\left/
 {\vphantom {i \hbar }} \right.
 \kern-\nulldelimiterspace} \hbar })\int\limits_a^\tau  { {\mathcal{A}}_1 (s)ds} ]\} \Phi [{\mathbf{x}}{\text{(}}a{\text{)}}]} . \hfill \\ 
\end{gathered} 
\eeqa
\end{thm}
\begin{proof} The fact that ${\mathbf{U}}^\rho  [t,a]\Phi (a) \to {\mathbf{U}}[t,a]\Phi (a) $, is clear.  To prove that 
\[
{\mathbf{U}}[t,a]\Phi (a) = \int_{{\R}^{3[t,a]} } {\mathbb{K}_{\mathbf{f}} [\mathcal{D}{\mathbf{x}}(\tau );{\mathbf{x}}(a)]} \exp \{ ( - i/\hbar )\int_a^t {\mathcal{A}_1 (s)ds} \},
\]
first note that, since the time-ordered integral exists and we are only interested in the limit, we can write for each $k$:
$$
U_k^\rho  [t,a]\Phi (a) = \exp \left\{ {( - i/\hbar )\sum\nolimits_{j = 1}^k {\int_{t_{j - 1} }^{t_j } {\left[ {{\mathbf{E}}[\tau _j ,s]\mathcal{A}_0 (s) + {\mathbf{E}}[\tau '_j ,s]\mathcal{A}_{1,\rho } (s)} \right]ds} } } \right\},
$$
where $\tau_j $ and $\tau'_j $ are distinct points in the interval $(t_{j-1}, t_j)$.  Thus, we can also write $U_k^\rho  [t,a] $ as 
\[
\begin{gathered}
  {\mathbf{U}}_k^\rho  [t,a]\Phi (a) \hfill \\
   = \exp \left\{ {( - i/\hbar )\sum\nolimits_{j = 1}^k {\int_{t_{j - 1} }^{t_j } {{\mathbf{E}}[\tau _j ,s]\mathcal{A}_0 (s)ds} } } \right\}\exp \left\{ {( - i/\hbar )\sum\nolimits_{j = 1}^k {\int_{t_{j - 1} }^{t_j } {\mathbf{E}}[\tau '_j ,s]\mathcal{A}_{1,\rho } (s)ds} } \right\} \hfill \\
   = \prod\nolimits_{j = 1}^k {\exp \left\{ {( - i/\hbar )\int_{t_{j - 1} }^{t_j } {{\mathbf{E}}[\tau _j ,s]\mathcal{A}_0 (s)ds} } \right\}} \exp \left\{ {( - i/\hbar )\sum\nolimits_{j = 1}^k {\int_{t_{j - 1} }^{t_j } {\mathbf{E}}[\tau '_j ,s]\mathcal{A}_{1,\rho } (s)ds} } \right\} \hfill \\
   = \prod\nolimits_{j = 1}^k {\int_{\R^3 } {\mathbb{K}_{\mathbf{f}} [t_j ,{\mathbf{x}}(t_j );t_{j - 1} ,d{\mathbf{x}}(t_{j - 1} )]} \left. {} \right|^{\tau _j } } \exp \left\{ {( - i/\hbar )\sum\nolimits_{j = 1}^k {\int_{t_{j - 1} }^{t_j } {\mathbf{E}}[\tau '_j ,s]\mathcal{A}_{1,\rho } (s)ds} } \right\}. \hfill \\ 
\end{gathered} 
\]
If we put this in our experimental evolution operator ${\mathbf{U}}_{\lambda}^{\rho} [t,a] \Phi (a) $ and compute the limit, we have:
\[
\begin{gathered}
  {\mathbf{U}}^\rho  [t,a]\Phi (a) \hfill \\
   = \int_{{\R}^{3[t,a]} } {\mathbb{K}_{\mathbf{f}} [\mathcal{D}{\mathbf{x}}(t);{\mathbf{x}}(a)]} \exp \left\{ {( - i/\hbar )\int_a^t {\mathcal{A}_{1,\rho } (s)ds} } \right\}\Phi (a). \hfill \\ 
\end{gathered} 
\]
Since the limit as $\rho  \to \infty $ on the left exists, it defines the limit on the right.
\end {proof}

\subsection{Examples}
In this section, we pause to discuss a few examples.  
Theorem 8.4 is rather abstract and it may not be clear as to its application.  Our first example is a direct application of this theorem, which covers all of nonrelativistic quantum theory.  

Let $\triangle$ be the Laplacian on ${\R}^n$ and let $V$ be any potential such that $A=(-\hbar^2/2)\triangle + V$ generates a unitary group.
Then the problem:
$$
(i\hbar) {\partial\psi({\bf{x}},t)}/{\partial t}= A\psi({\bf{x}},t),\;\;  \psi({\bf{x}},0)=\psi_{0}({\bf{x}}),
$$
has a solution with the Feynman-Kac representation. 

Our second example is more specific, and is due to Albeverio and Mazzucchi \cite{AM}.   Their paper provides an excellent view of the power of the approach first introduced by Albeverio and H${\varnothing}$egh-Krohn \cite{AH}.
Let $\mathbb{C}$ be a completely symmetric positive definite fourth-order covariant tensor on ${\R}^n $, let $\Omega $ be a symmetric positive definite $n \times n$
matrix and let $\lambda $ be a nonnegative constant.   It is known \cite{RS1} that the operator  
$$
\bar A =  - \tfrac{\hbar^2 }{2}\Delta  + \tfrac{1}
{2}{\mathbf{x}}\Omega ^2 {\mathbf{x}} + \lambda \mathbb{C}[\mathbf{x},\mathbf{x},\mathbf{x},\mathbf{x}]
$$
is a densely defined selfadjoint generator of a unitary group on ${\bf L}^2 [{\mathbb{R}}^n ]$.  Using a substantial amount of elegant analysis, Albeverio and Mazzucchi \cite{AM} prove that $\bar A$ has a path integral representation as the analytic continuation (in the parameter $\lambda $) of an infinite dimensional generalized oscillatory integral.  

Our approach to the same problem is both simple and direct using the results of the previous sections.  First, by Theorem 3.12, $\bar A$ has a closed densely defined selfadjoint extension $A$ to ${\bf{KS}}^2 [{\R}^n ]$, which also generates a unitary group by Theorem 3.13.  If we set $V = \tfrac{1}
{2}{\mathbf{x}}\Omega ^2 {\mathbf{x}} + \lambda \mathbb{C}[{\mathbf{x}},{\mathbf{x}},{\mathbf{x}},{\mathbf{x}}]$ and $V_\rho   = V(I + \rho V^2 )^{ - 1/2} ,{\text{ }}\rho  > 0$, it is easy to see that $V_\rho $ is a bounded selfadjoint operator which converges to $V$ on $D(V)$.   (This follows from the fact that  a bounded (selfadjoint) perturbation of an unbounded selfadjoint operator is selfadjoint.)   Now, since $ - \tfrac{\hbar^2 }{2}\Delta $ generates a unitary group, by Theorem 7.1 $A_\rho   =  - \tfrac{\hbar^2 }
{2}\Delta  + V_\rho  $ generates one also and converges to $A$ on $D(A)$.  Let
$$ 
\mathcal{A}(\tau) = (\mathop {\hat  \otimes }\limits_{t \geqslant s > \tau} {\text{I}}_s)  \otimes A \otimes (\mathop  \otimes \limits_{\tau > s \geqslant 0} {\text{I}}_s ),
$$
then, by Theorem 7.1, $ {\mathcal{A}}(t)$ generates a unitary group for each $t$ and ${\mathcal{A}}_\rho  (t)$ converges to ${\mathcal{A}}(t)$ on $D[\mathcal{A}(t)] \subset {\mathcal{F}\mathcal{D}}_ \otimes ^2 $.  We can now apply our Theorem 7.13 to get that:
\[
\begin{gathered}
  {\mathbf{U}}[t,a] = \int_{{\R}^{3[t,a]} } {\mathbb{K}_{\mathbf{f}} [{\mathcal{D}}{\mathbf{x}}(\tau )\,;\,{\mathbf{x}}(a)]} \exp \{  - ({i \mathord{\left/
 {\vphantom {i \hbar }} \right.
 \kern-\nulldelimiterspace} \hbar })\int_a^\tau  {V(s)ds} \}  \hfill \\
   = \lim _{\rho  \to 0} \int_{{\R}^{3[t,a]} } {\mathbb{K}_{\mathbf{f}} [{\mathcal{D}}{\mathbf{x}}(\tau )\,;\,{\mathbf{x}}(a)]} \exp \{  - ({i \mathord{\left/
 {\vphantom {i \hbar }} \right.
 \kern-\nulldelimiterspace} \hbar })\int_a^\tau  {V_\rho  (s)ds} \} , \hfill \\ 
\end{gathered} 
\]
where we have used the fact that  $1 \in {\bf{KS}}^2 [{\R}^n ]$ to obtain the propagator as Feynman would have wanted for this case.

Under additional assumptions, Albeverio and Mazzucchi are able to prove Borel summability of the solution in power series of the coupling constant.  With Theorem 66, we get the Dyson expansion to any order with remainder.   

The third example is taken from \cite{GZ2} and provides an example of a problem that cannot by solved using analytic continuation via a Gaussian kernel.  It is shown that, if the vector potential $\bf{A}$ is constant, $\mu =mc/\hbar$, and $\boldsymbol{\beta}$ is the standard beta matrix, then the solution to the equation for a spin $1/2$ particle in square-root form, 
\[
i\hbar {{\partial \psi ({\mathbf{x}},t)} \mathord{\left/
 {\vphantom {{\partial \psi ({\mathbf{x}},t)} {\partial t}}} \right.
 \kern-\nulldelimiterspace} {\partial t}} = \left\{ {{\boldsymbol{\beta}} \sqrt {c^2 \left( {{\mathbf{p}} - \tfrac{e}
{c}{\mathbf{A}}} \right)^2  + m^2 c^4 } } \right\}\psi ({\mathbf{x}},t),{\text{  }}\psi ({\mathbf{x}},0) = \psi _0 ({\mathbf{x}}),
\]
is  given by:
$$
\psi ({\mathbf{x}},t) = {\mathbf{U}}[t,0]\psi _0 ({\mathbf{x}}) = \int\limits_{{\R}^3 } {\exp \left\{ {\frac{{ie}}
{{2\hbar c}}\left( {{\mathbf{x}} - {\mathbf{y}}} \right) \cdot {\mathbf{A}}} \right\}{\mathbf{K}}\left[ {{\mathbf{x}},t\,;\,{\mathbf{y}},0} \right]\psi _0 ({\mathbf{y}})d{\mathbf{y}}},
$$
where
\[
{\mathbf{K}}\left[ {{\mathbf{x}},t\,;\,{\mathbf{y}},0} \right] = \frac{{ct\mu ^2 \beta }}
{{4\pi }}\left\{ {\begin{array}{*{20}c}
   {\tfrac{{ - H_2^{(1)} \left[ {\mu \left( {c^2 t^2  - ||{\kern 1pt} {\mathbf{x}} - {\mathbf{y}}{\kern 1pt} ||^2 } \right)^{1/2} } \right]}}
{{\left[ {c^2 t^2  - ||{\kern 1pt} {\mathbf{x}} - {\mathbf{y}}{\kern 1pt} ||^2 } \right]}}{\text{, }}\;ct <  - ||{\kern 1pt} {\mathbf{x}}{\kern 1pt} ||,}  \\
   {\tfrac{{ - 2iK_2 \left[ {\mu \left( {||{\kern 1pt} {\mathbf{x}} - {\mathbf{y}}{\kern 1pt} ||^2  - c^2 t^2 } \right)^{1/2} } \right]}}
{{\pi \left[ {||{\kern 1pt} {\mathbf{x}} - {\mathbf{y}}{\kern 1pt} ||^2  - c^2 t^2 } \right]}},{\text{ }}\;c\left| t \right| < \,||{\kern 1pt} {\mathbf{x}}{\kern 1pt} ||,}  \\
   {\tfrac{{H_2^{(2)} \left[ {\mu \left( {c^2 t^2  - ||{\kern 1pt} {\mathbf{x}} - {\mathbf{y}}{\kern 1pt} ||^2 } \right)^{1/2} } \right]}}
{{\left[ {c^2 t^2  - ||{\kern 1pt} {\mathbf{x}} - {\mathbf{y}}{\kern 1pt} ||^2 } \right]}},{\text{ }}\;ct > \,||{\kern 1pt} {\mathbf{x}}{\kern 1pt} ||.}  \\

 \end{array} } \right.
\]
The function $K_2 (\, \cdot \,)$ is a modified Bessel function of the third kind of second order, while $H_2^{(1)}, \; H_2^{(2)}$ are the Hankel functions (see Gradshteyn and Ryzhik \cite{GRRZ}). Thus, we have a kernel that is far from the standard form.   A direct application of the same theorems used in the previous example shows that any selfadjoint potential $V$ for which the generalized sum (of any type) generates a unitary group will lead to a path integral representation via the time-ordered operator calculus.

This example was first introduced in \cite{GZ2}, where we only considered the kernel for the Bessel function term.  In that case, it was shown that, under appropriate conditions, that term will reduce to the free-particle Feynman kernel and, if we set $\mu=0$, we get the kernel for a (spin $1/2$) massless particle.  In closing this section, we remark that the square-root operator is unitarily equivalent to the Dirac operator (in the case discussed).   

\subsection{The Kernel Problem}
Since any semigroup that has a kernel representation will generate a path integral via the operator calculus, a fundamental question is:  Under what general conditions can we expect a given (time-dependent) generator of a semigroup to have an associated kernel?  In this section we discuss a class of general conditions for unitary groups.  It will be clear that the results of this section carry over to semigroups with minor changes.

Let $A(\bf{x},\;\bf{p})$ denote a $k \times k$ matrix operator $[A_{ij}(\bf{x},\;\bf{p})]$, $i,\;j=1,2,\cdots,k$, whose components are pseudodifferential operators with symbols $a_{ij}(\bf{x},\; \boldsymbol{\eta}) \in \mathbb{C}^{\infty}(\R^{n} \times \R^{n})$ and we have, for any multi-indices $\alpha$ and   $\beta$, 
\beqn
\left| {a_{ij(\beta )}^{(\alpha )} ({\mathbf{x}}, \boldsymbol{\eta} )} \right| \leqslant C_{\alpha \beta } (1 + \left| \boldsymbol{\eta}  \right|)^{m - \xi \left| \alpha  \right| + \delta \left| \beta  \right|},
\eeqn
where
\beqa
a_{ij(\beta )}^{(\alpha )} ({\mathbf{x}}, \boldsymbol{\eta} ) = \partial ^\alpha  {\mathbf{p}}^\beta  a_{ij} ({\mathbf{x}},\boldsymbol{\eta} ),
\eeqa
with  $\partial _l  = {\partial  \mathord{\left/
 {\vphantom {\partial  {\partial \eta _l }}} \right.
 \kern-\nulldelimiterspace} {\partial \eta _l }}$, and 
$p _l  = (1/i)({\partial  \mathord{\left/
 {\vphantom {\partial  {\partial x _l }}} \right.
 \kern-\nulldelimiterspace} {\partial x _l }})$.  The multi-indices are defined in the usual manner by $\alpha=(\alpha_1,\cdots,\alpha_n)$ for integers $\alpha_j \ge 0$, and $\left| \alpha \right|=\sum_{j=1}^{n}\alpha_j$, with similar definitions for $\beta$.  The notation for derivatives is $\partial^{\alpha}=\partial_1^{\alpha_1} \cdots \partial_n^{\alpha_n}$ and ${\bf{p}}^{\beta}=p_1^{\beta_1}\cdots p_n^{\beta_n}$.  Here, $m,\; \beta$, and $\delta$ are real numbers satisfying $0 \le \delta < \xi $.  Equation (8.7) states that each $a_{ij}(\bf{x},\;\boldsymbol{\eta})$ belongs to the symbol class  $S_{\xi, \delta}^m$ (see \cite{SH}).

Let $a({\bf{x}}, {\boldsymbol{\eta}})=[a_{ij}(\bf{x}, \boldsymbol{\eta})]$ be the matrix-valued symbol for $A({\bf{x}}, {\boldsymbol{\eta}})$, and let $\lambda_1({\bf{x}}, {\boldsymbol{\eta}}) \cdots \lambda_k ({\bf{x}}, {\boldsymbol{\eta}})$ be its eigenvalues.  If $\left|\, \cdot\, \right|$ is the norm in the space of $k \times k$ matrices, we assume that the following conditions are satisfied by $a({\bf{x}}, {\boldsymbol{\eta}})$. 
For $0<c_0<\left|\, {\boldsymbol{\eta}} \right|$ and ${\bf{x}} \in {\mathbb{R}}^n$ we have 
\begin{enumerate}
\item $\left| a_{(\beta)}^{(\alpha)}({\bf{x}}, {\boldsymbol{\eta}}) \right| \le C_{\alpha \beta}\left| a({\bf{x}}, {\boldsymbol{\eta}}) \right| (1+\left| {\boldsymbol{\eta}}) \right|)^{-\xi \left| \alpha \right| + \delta \left| \beta \right|}$ (hypoellipticity),

\item $\lambda _0 ({\mathbf{x}},{\boldsymbol{\eta}} ) = \mathop {\max }\limits_{1 \leqslant j \leqslant k} \operatorname{Re} \lambda _j ({\mathbf{x}},{\boldsymbol{\eta}}) < 0$,

\item  $\frac{{\left| {a({\mathbf{x}},{\boldsymbol{\eta}} )} \right|}}
{{\left| {\lambda _0 ({\mathbf{x}},{\boldsymbol{\eta}} )} \right|}} = O\left[ {(1 + \left| {\boldsymbol{\eta}}  \right|)^{(\xi  - \delta )/(2k - \varepsilon )} } \right],\; \varepsilon >0$.
\end{enumerate}
We assume that $A({\mathbf{x}},{\bf{p}} )$ is a selfadjoint generator of a unitary group $U(t,0)$, so that
\beqa
U(t,0)\psi_0({\bf{x}})= exp[(i/\hbar)tA({\mathbf{x}},{\bf{p}})]\psi_0({\bf{x}})=\psi({\bf{x}},t)
\eeqa 
solves the Cauchy problem
\beqn
(i/\hbar) {{\partial \psi({\bf{x}},t)}/ {\partial t}} =A({\mathbf{x}},{\bf{p}} )\psi({\bf{x}},t), \quad \psi({\bf{x}},t)=\psi_0({\bf{x}}).
\eeqn
\begin{Def}
We say that $Q({\bf{x}},t,{\boldsymbol{\eta}},0)$ is a symbol for the Cauchy problem (8.8) if $\psi ({\bf{x}}, t)$ has a representation of the form
\beqn
\psi ({\bf{x}},t) = (2\pi )^{ - n/2} \int_{\mathbb{R}^n } {e^{i({\bf{x}},  {\boldsymbol{\eta}} )} Q({\bf{x}},t, {\boldsymbol{\eta}},0)\hat \psi _0 ({\boldsymbol{\eta}})d {\boldsymbol{\eta}}}. 
\eeqn
\end{Def}
It is sufficient that $\psi_0$ belongs to the Schwartz space ${\mathcal{S}}(\mathbb{R}^n )$, which is contained in the domain of $A({\mathbf{x}},{\bf{p}})$, in order that (8.9) makes sense.

Following Shishmarev \cite{SH}, and using the theory of Fourier integral operators, we can define an operator-valued kernel for $U(t,0)$ by
\beqa
K({\bf{x}},t\,;\,{\bf{y}},0) = (2\pi )^{ - n/2} \int_{\mathbb{R}^n } {e^{i({\bf{x-y}},  {\boldsymbol{\eta}} )} Q({\bf{x}},t, {\boldsymbol{\eta}},0)d {\boldsymbol{\eta}}}, 
\eeqa
so that
\beqn
 \quad \psi({\bf{x}},t)=U(t,0)\psi_0({\bf{x}})= (2\pi )^{ - n/2} \int_{\mathbb{R}^n } {K({\bf{x}},t\,;\,{\bf{y}},0)\psi_0({\bf{y}})d{\bf{y}}}. 
\eeqn
The following results are due to Shishmarev \cite{SH}.
\begin{thm}
If $A({\mathbf{x}},{\bf{p}})$ is a selfadjoint generator of a strongly continuous unitary group with domain $D$,  ${\mathcal{S}}(\mathbb{R}^n ) \subset D$ in $L^2({\mathbb{R}}^n)$, such that conditions (1)-(3) are satisfied, then there exists precisely one symbol $Q({\bf{x}},t, {\boldsymbol{\eta}},0)$ for the Cauchy problem (8.8).
\end{thm} 
\begin{thm}
If we replace condition (3) in Theorem 8.6 by the stronger condition

$ (3') \quad \frac{{\left| {a({\mathbf{x}},{\boldsymbol{\eta}} )} \right|}}
{{\left| {\lambda _0 ({\mathbf{x}},{\boldsymbol{\eta}} )} \right|}} = O\left[ {(1 + \left| {\boldsymbol{\eta}}  \right|)^{(\xi  - \delta )/(3k -1- \varepsilon )} } \right], \; \varepsilon >0, \left|{\boldsymbol{\eta}}\right|>c_0,
$

then the symbol $Q({\bf{x}},t, {\boldsymbol{\eta}},0)$ of the Cauchy problem (8.8) has the asymptotic behavior near $t=0$:
\beqa
{Q({\bf{x}},t,{\boldsymbol{\eta}},0)= 
exp[-(i/{\hbar})ta({\bf{x}}, 
\boldsymbol{\eta})] + o(1)},
\eeqa
uniformly for $\bf{x},\,\bf{y} \in \mathbb{R}^n$.
\end{thm} 
Now, using Theorem 8.7 we see that, under the stronger condition (3'), the kernel $K({\bf{x}},t\,;\,{\bf{y}},0)$ satisfies 
\beqa
 \begin{gathered}
  K({\mathbf{x}},t;{\mathbf{y}},0) = \int_{\mathbb{R}^n } {\exp [i({\mathbf{x}} - {\mathbf{y}},{\boldsymbol{\eta}} ) - (i/\hbar )ta({\mathbf{x}},{\boldsymbol{\eta}} )]\frac{{d{\boldsymbol{\eta}} }}
{{(2\pi )^{n/2} }}}  \hfill \\
\quad \quad \quad \quad \quad \quad \quad \quad + \int_{\mathbb{R}^n } {\exp [i({\mathbf{x}} - {\mathbf{y}},{\boldsymbol{\eta}} )]\frac{{d{\boldsymbol{\eta}} }}
{{(2\pi )^{n/2} }}}o(1).  \hfill \\ 
\end{gathered} 
\eeqa

In order to see the power of ${\bf{KS}}^2 [{\R}^n ]$, first note that $A({\mathbf{x}},{\bf{p}})$ has a selfadjoint extension to ${\bf{KS}}^2 [{\R}^n ]$, which also generates a unitary group (see Theorem 28).  This means that we can construct a path integral in the same (identical) way as was done for the free-particle propagator in Section 3.2 (i.e., for all Hamiltonians with symbols  in $\mathcal{S}_{\alpha,\delta}^m)$.  Furthermore, it follows that the same comment applies to any Hamiltonian that has a kernel representation, independent of its symbol class.  The important point of this discussion is that neither time-ordering nor initial data is required!  

On the other hand, if we want to consider perturbations of the above Hamiltonians with various potentials, the normal analytical problems arise.  In this case, it is easier to convert to the time-ordered theory and use Theorem 8.1.  The alternative is to resort to the limited number of Trotter-Kato type results that may apply (directly on ${\bf{KS}}^2 [{\R}^n ]$).
    
The results of Shishmarev have direct extensions  to time-dependent Hamiltonians.  However, in this case, the operators need not commute. Thus, in order to construct path integrals, we must use the full power of Sections 5,6 and 8.1.  

\section{\bf{Discussion}}
	The question of external forces requires discussion of the inhomogeneous problem.  Since the inhomogeneous problem is a special case of the semilinear problem, we provide a few remarks in that direction.  Since all of the standard results go through as in the conventional approach, we content ourselves with a brief description of a typical case.   Without loss in generality, we assume $\mathcal{H}$ has our standard basis.  With the conditions for the parabolic or hyperbolic problem in force, the typical semilinear problem can be represented on $\mathcal{H}$ as: 
\beqn
\frac{{\partial u(t)}}
{{\partial t}} = A(t)u(t) + f(t,u(t)),\;u(a) = u_a. 
\eeqn
We assume that $f$ is continuously differentiable with $u_a  \in \mathcal{H}$ in the parabolic or $
u_a  \in D$, the common dense domain, in the hyperbolic case.  These conditions are sufficient for $u(t)$ to be a classical solution (see Pazy \cite{PZ}, pg. 187).  The function $f$ has the representation $f(t,u(t)) = \sum\nolimits_{k = 1}^\infty  {f_k (t)} e^k$ in $\mathcal{H}$.  The corresponding function ${\mathbf{f}}$, in $\mathcal{F}\mathcal{D}_ \otimes ^2$, has the representation ${\mathbf{f}}(t,{\mathbf{u}}(t)) = \sum\nolimits_{k = 1}^\infty  {f_k (t)} E^k$, where ${\mathbf{u}}(t) $ is a classical solution to the time-ordered problem: 
\beqn
\frac{{\partial {\mathbf{u}}(t)}}
{{\partial t}} = {\mathbf{A}}(t){\mathbf{u}}(t) + {\mathbf{f}}(t,{\mathbf{u}}(t)),\;{\mathbf{u}}(a) = {\mathbf{u}}_a. 
\eeqn
This function ${\mathbf{u}}(t) $ also satisfies the integral equation (time-ordered mild solution): 
\beqa
{\mathbf{u}}(t) = {\mathbf{U}}(t,a){\mathbf{u}}_a  + \int_a^t {{\mathbf{U}}(t,s){\mathbf{f}}(s,{\mathbf{u}}(s))ds}. 
\eeqa
If $f$ does not depend on $u(t)$, we get the standard linear inhomogeneous problem.  It follows that all the basic results (and proofs) go through for the semilinear and linear inhomogeneous problem in the time-ordered case.  Similar statements apply to the problem of asymptotic behavior of solutions  (e.g., dynamical systems, attractors, etc). 
	
\section*{\textbf{Conclusion}}
In this paper we have shown how to construct a natural representation Hilbert space for Feynman's time-ordered operator calculus (in which operators acting at different times actually commute).  This space allows us to construct the time-ordered integral and evolution operator (propagator) under the weakest known conditions and extend all of semigroup theory to the time-ordered setting.  We have also constructed a new Hilbert space which contains the Feynman kernel and the delta function as norm bounded elements, and have shown that, on this space, we can rigorously construct the path integral in the manner originally intended by  Feynman. We have extended this path integral to very general interactions and have provided a substantial generalization of the Feynman-Kac formula. The approach is independent of the space of continuous paths and makes the apparent need for a measure more of a desire then a necessity.   In addition, we have also developed a general theory of perturbations for operators and have shown that all time-ordered evolution operators are asymptotic in the operator-valued sense of Poincar\'{e}.

A major problem envisioned by Feynman was the development of his disentanglement approach in order to relate his calculus to standard mathematical methods.  A number of researchers have made advances in this direction.   Work of Johnson and coworkers is of particular interest in this respect (see [JN], [JP] and [JJN]).  For additional important work on this approach, see the books by Jefferies [J], Johnson and Lapidus [JL], Maslov [M] and Nazaikminskii et al [NSS].

Our approach is different in that we have chosen to extend functional analysis so that the process of time-ordering has a natural place in mathematics.  This approach does not require disentanglement for its justification.

\acknowledgements
During the course of the development of this work, we have benefited from important critical remarks and encouragement from Professors  Brian DeFacio,  Louis Fishman, Jerome Goldstein, Brian Jefferies, Gerald Johnson, John Klauder, Michel Lapidus, Elliot Lieb, Lance Nielsen, Lawrence Schulman and Dave Skoug. Finally, we would like to thank Professor Pliczko for informing us of an error, that led to a substantial improvement of Section 3.

\end{document}